\documentclass[11pt,a4paper]{article}
\pdfoutput=1
\usepackage[top=25mm,bottom=30mm,left=29mm,right=29mm]{geometry}

\usepackage{tocloft}

\usepackage[nottoc,notlot,notlof]{tocbibind}

\usepackage[pdfencoding=auto]{hyperref}

\hypersetup{
    colorlinks,
    linkcolor={red!50!black},
    citecolor={blue!50!black},
    urlcolor={blue!80!black}
}

\usepackage[utf8]{inputenc}

\usepackage{amssymb,amsmath}

\usepackage{graphicx}
\usepackage{cite}

\usepackage{mathtools}
\usepackage{xcolor}
\usepackage{tikz}
\numberwithin{equation}{section}

\def\ket#1{|{#1}\rangle}

\newcommand{\one}{\mathbf{1}} 
\newcommand{\MC}{\mathcal{M}} 
\newcommand{\Hom}{\mathrm{Hom}} 

\DeclareMathOperator{\Tr}{Tr}
\DeclareMathOperator{\Res}{Res}

\begin{document} 
\thispagestyle{empty}
\phantom{.}
\vspace{.7cm}
\begin{center}{\Large \textbf{
Lattice models from CFT on surfaces with holes I:\\[.5em]
Torus partition function via two lattice cells
}}
\end{center}

\begin{center}
Enrico M. Brehm\textsuperscript{$\beta$},
Ingo Runkel\textsuperscript{$\mu$}
\end{center}

\begin{center}
	{ ${}^{\beta}$} Max Planck Institut für Gravitationsphysik,\\
	Albert-Einstein-Institut, \\
	Am Mühlenberg 1 \\
	14476 Potsdam-Golm,	Germany.\\[.5em]
{\small \sf brehm@aei.mpg.de}
\\[1em]
	{${}^{\mu}$} 
Fachbereich Mathematik,\\
Universität Hamburg,\\
Bundesstraße 55,\\
20146 Hamburg, Germany\\[.5em]
{\small \sf Ingo.Runkel@uni-hamburg.de}
\end{center}

\section*{Abstract}

We construct a one-parameter family of lattice models starting from a two-dimensional rational conformal field theory on a torus with a regular lattice of holes, each of which is equipped with a conformal boundary condition. The lattice model is obtained by cutting the surface into triangles with clipped-off edges using open channel factorisation. The parameter is given by the hole radius. At finite radius,
high energy states are suppressed and the model is effectively finite. In the zero-radius limit, it recovers the CFT amplitude exactly. In the touching hole limit, one obtains a topological field theory. 

If one chooses a special conformal boundary condition which we call ``cloaking boundary condition'', then for each value of the radius the fusion category of topological line defects of the CFT is contained in the lattice model. 
The fact that the full topological symmetry of the initial CFT is realised exactly is a key feature of our lattice models.

We provide an explicit recursive procedure to evaluate the interaction vertex on arbitrary states.  
As an example, we study the lattice model obtained from the Ising CFT on a torus with one hole, decomposed into two lattice cells. We numerically compare the truncated lattice model to the CFT expression obtained from expanding the boundary state in terms of the hole radius and we find good agreement at intermediate values of the radius.

\newpage
\setcounter{tocdepth}{2}
\tableofcontents

\newpage

\section{Introduction and overview}

The long-range behaviour of critical spin chains in one dimension or of critical statistical models in two dimensions is often captured by a two-dimensional conformal field theory (CFT). In this sense, CFTs describe universality classes of long-range critical behaviour.
Given a spin chain or lattice model, it is a difficult problem to verify if it is critical, and an even harder problem to identify its universality class in the form of a continuum CFT.

An important invariant of a CFT, and of a quantum field theory in any dimension for that matter, are its topological defects. In our situation of CFTs in $2d$, we are interested in topological line defects \cite{Petkova:2000ip,Frohlich:2006ch}. Suppose there is a finite subset $F$ of elementary topological line defects which closes under fusion, up to taking direct sums. If the CFT in question is unitary (or at least semisimple), the defects in $F$ provide a finer invariant, namely a \textit{fusion category} $\mathcal{F}$. Roughly speaking, $\mathcal{F}$ captures at the same time the fusion rules of the line defects in $F$ and the behaviour of their weight-zero junction fields \cite{Frohlich:2006ch}. The fusion category $\mathcal{F}$ is often referred to as a \textit{topological symmetry} of the CFT.

Topological symmetry is also realised in discrete models, and indeed often leads to criticality of the model in question. Conversely, identifying such a topological symmetry $\mathcal{F}$ in a discrete model helps in determining the corresponding universal CFT. In fact, constructions of spin chains 
\cite{PhysRevLett.98.160409,Pfeifer2012,Buican:2017rxc,belletete2020topological}, string net and tensor network models \cite{Levin:2004mi,Kitaev:2011dxc,Vanhove:2018wlb,Lootens:2020mso} and statistical lattice models \cite{Aasen:2016dop,Aasen:2020jwb} directly use $\mathcal{F}$ as an input to define the model itself, and thereby ensure that it realises the topological symmetry $\mathcal{F}$.

An emblematic example is the golden chain \cite{PhysRevLett.98.160409}, for which the topological symmetry is the Fibonacci category $\mathcal{F} = \mathrm{Fib}$. The resulting universal CFT is the tricritical Ising CFT with $c=\frac{7}{10}$. This CFT indeed realises $\mathrm{Fib}$ in its topological line defects, but its actual topological symmetry is larger (consisting of six elementary defects, rather than just the two contained in $\mathrm{Fib}$). The tricritical Ising CFT is, however, the smallest unitary CFT, in terms of central charge, which contains the topological symmetry $\mathrm{Fib}$. In the same spirit, the Ising CFT of $c = \frac12$ is the smallest unitary CFT with a $\mathbb{Z}_2$-symmetry.

Note, however, that if there is one CFT $C$ which contains $\mathcal{F}$ in its topological symmetry, then there are infinitely many, as one can simply consider tensor products $C \otimes C'$ with any other CFT $C'$. Outside of the Virasoro minimal models, i.e.\ for $c \ge 1$, the topological defects (are believed to) always form a continuum, and very little is known about their properties in general. The typical situation is to look at subsets of the full topological symmetry, e.g.\ at those line defects which commute with a larger set of conserved currents than just the stress-energy tensor. This is the approach taken in rational conformal field theory. 

\medskip

The above discussion raises the following questions: Given a topological symmetry $\mathcal{F}$ and a topological lattice model constructed from the data in $\mathcal{F}$, which of the infinitely many CFTs with $\mathcal{F}$ as a sub-symmetry describes the universality class of that model? And how can one modify the construction to change the universality class to a different CFT containing $\mathcal{F}$?

In this series of papers, we approach these problems from the opposite direction. Rather than starting from a topological symmetry, we start from the CFT itself and construct lattice models from the full data the CFT makes available. Concretely, we start from a rational CFT and produce a parameter-dependent lattice model with two key features:
\begin{enumerate}
\item The Boltzmann weights depend on a parameter that can be thought of as a high energy cutoff. In the high energy limit of that parameter, the lattice model \textit{reproduces the CFT amplitude exactly}, even for a finite number of lattice cells. On the other hand, in the low energy limit, the model becomes topological.

\item The lattice model \textit{realises the topological symmetry of the CFT} which preserves the rational chiral algebra. That is, at each value of the above parameter, the lattice model allows for topological line defects satisfying the same fusion relations as those of the rational CFT.
\end{enumerate}

\noindent
The lattice model initially has an infinite number of states at each site. However, for each finite value of the parameter, all but finitely many states are strongly suppressed.\footnote{
Actually, we produce a two-parameter family of models, all of which realise the topological symmetry exactly. Both parameters are energy cutoffs.
One induces a dampening for high-energy modes, while the other is a strict cutoff to a finite number of states, see Section~\ref{sec:truncated-sym}.
}

\medskip

In the remainder of the introduction, we outline the construction of these lattice models, the results of this paper, and the next steps in our program.

\subsection{Construction of the lattice model}\label{sec:lattice-model-def}

\begin{figure}[t]
	\centering
    \includegraphics[scale=1.2]{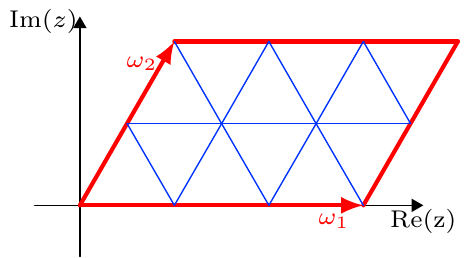}
    \includegraphics[scale=1.2]{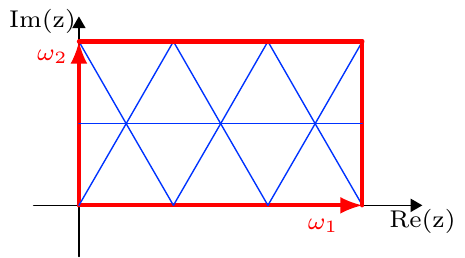}
	\caption{Two examples of decomposing a torus into $2MN$ equilateral triangles, here with $M=2$, $N=3$}
	\label{fig:intro-decompose-torus}
\end{figure}

\begin{figure}[t]
	\centering
	\begin{tikzpicture}[>=latex]
	\node (A) at (0,0)  {\includegraphics[width=.45\textwidth]{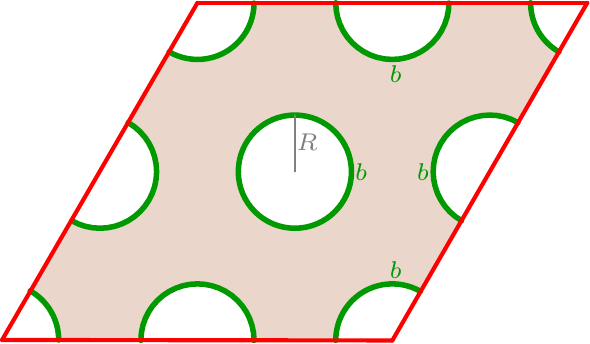}};
	\node (B) at (2,-6)  {\includegraphics[width=.45\textwidth]{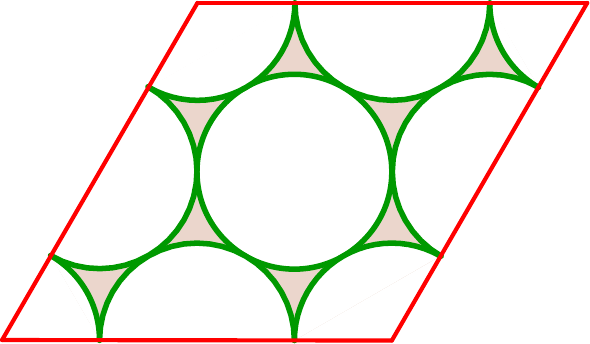}};
	\node (C) at (-4,-5)  {\includegraphics[width=.45\textwidth]{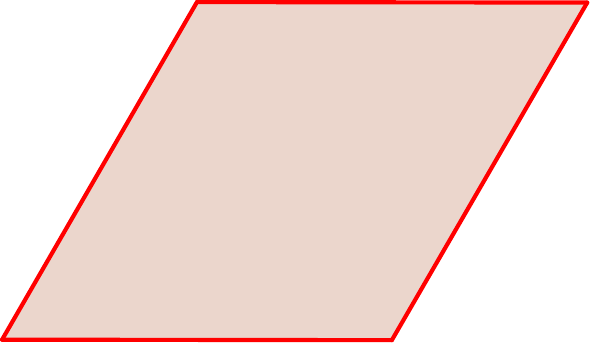}};
	\draw[->] (A) to[out=-45,in=80] node[right] {$R\to\frac d2$} (B);
	\draw[->] (A) to[out=-160,in=100] node[left] {$R\to0$} (C);
	\node at (-1.5,2) {a)};
	\node at (-5.5,-3) {b)};
	\node at (0.5,-4) {c)};
	\end{tikzpicture}
	
	\caption{(a) The torus (with $M=N=2$) with holes of radius $R$ cut out and conformal boundary condition $b$ at each hole. (b) The $R\to0$ limit recovers the torus without holes. (c) In the $R \to \frac d2$ limit the surface decomposes into a product of triangles with curved boundaries (which are conformally equivalent to discs).
	}
	\label{fig:intro-two-limits}
\end{figure}
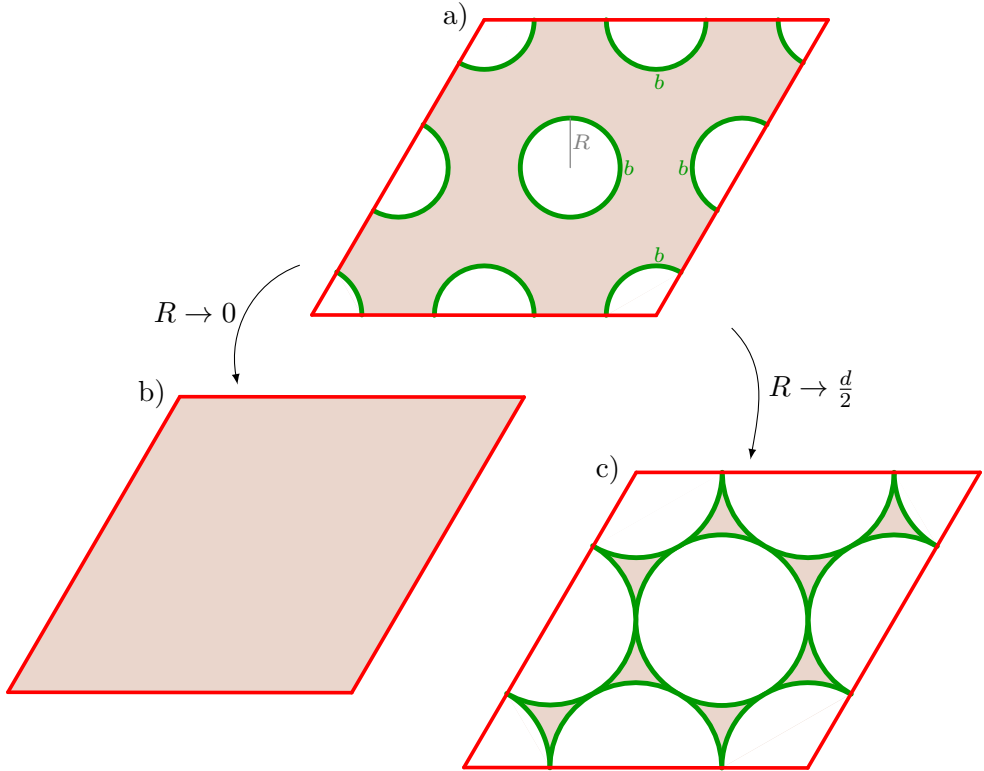

\begin{figure}[t]
	\centering
	\begin{tikzpicture}
	\node at (0,0) {\hspace{-.5cm}\includegraphics[scale=1.5]{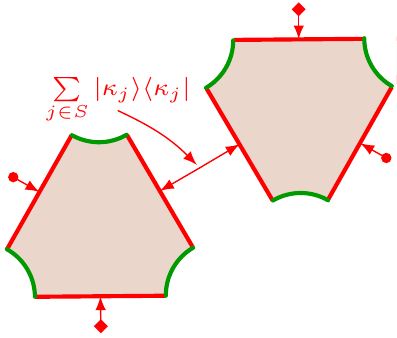}};
	\node at (6,0) {\includegraphics[scale=1.5]{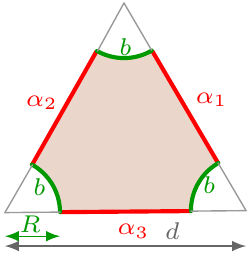}};
	\node at (-1,2) {a)};
	\node at (4.5,1.5) {b)};
	\end{tikzpicture}
	\caption{a) Inserting sums over intermediate states at all edges stretching between the holes. 		
	b) A clipped triangle with conformal boundary condition $b$ on each of its three boundary components and states $\alpha_1$, $\alpha_2$, $\alpha_3$ inserted at the open channel state boundaries.}
	\label{fig:intro-sum-over-states}
\end{figure}

Fix a unitary rational CFT $C$ which does allow for conformal boundary conditions. In particular, $C$ has the same holomorphic and anti-holomorphic central charge $c$. We will comment on non-unitary and non-rational models in Section~\ref{sec:nextsteps}.  
Pick a value $d>0$ (``the distance between lattice points'') and consider a complex torus with periods $\omega_1, \omega_2 \in \mathbb{C}$, which can be decomposed into equilateral triangles whose sides have length $d$. For example $\omega_1 = Nd$ and $\omega_2 = M d e^{\pi i/3}$ or $\omega_2 = i\frac{\sqrt{3}}{2} M d $ for $N,M \in \mathbb{N}$ (see Figure~\ref{fig:intro-decompose-torus}). Let 
\begin{equation}
Z(\omega_1,\omega_2)
\end{equation}
be the partition function of the CFT $C$ on this torus. Now fix a radius $0<R<\frac{d}2$ and cut out a hole of radius $R$ at each vertex of the triangles. Choose a conformal boundary condition $b$ for $C$ and place it on each of the circular boundaries  (see Figure~\ref{fig:intro-two-limits}\,a). Write
\begin{equation}\label{eq:ZalphaR-def}
Z_b(\omega_1,\omega_2;R)
\end{equation}
for the amplitude of the CFT $C$ on the torus with $MN$ holes and boundary condition $b$ on each boundary. In the limit $R\to 0$ (see Figure~\ref{fig:intro-two-limits}\,b), one recovers the torus partition function,
\begin{equation}\label{eq:Rto0limit}
Z(\omega_1,\omega_2) ~= \, \lim_{R\to 0} \, (a^b_1)^{-MN} \, R^{\frac{c}{6}MN} \, Z_b(\omega_1,\omega_2;R)   ~,
\end{equation}
see Section~\ref{sec:TorusBoundaryExpansion} for details on this limit and the normalisation factors. 
In the limit $R \to \frac{d}2$ (Figure~\ref{fig:intro-two-limits}\,c), only open channel vacuum states propagate
between the holes. If $b$ is an elementary boundary condition there is a unique open channel vacuum state, and -- up to a divergent factor $f(R)$ independent of $b$ -- 
the amplitude factorises into a product of disc amplitudes $D_b$ (without field insertions):
\begin{equation}\label{eq:Rtod2limit}
Z_b(\omega_1,\omega_2;R) ~\sim~ f(R) \, (D_b)^{2MN} \quad \text{for}~~ R\to \tfrac{d}2~.
\end{equation}
%
To obtain the lattice model, we fix a hole radius $0<R<\frac d2$ and factorise the torus amplitude into $2MN$ clipped triangles by inserting a sum over intermediate states at each of the edges stretching between the holes (Figure~\ref{fig:intro-sum-over-states}\,a). Let $\mathcal{H}_{bb}$ be the space of open channel states on a strip with boundary condition $b$ on either side, and let 
\begin{equation}
	T_R^b(\alpha_1, \alpha_2, \alpha_3)
\end{equation}
be the amplitude of the clipped triangle for three states $\alpha_1, \alpha_2, \alpha_3 \in \mathcal{H}_{bb}$ (Figure~\ref{fig:intro-sum-over-states}\,b). Denote by $\{ \kappa_s \}_{s \in S}$ an orthonormal basis of $\mathcal{H}_{bb}$, indexed by the (countably infinite) set $S$. Summing over the full basis at each edge allows one to express the amplitude of the CFT $C$ on the torus with $MN$ holes via clipped triangles:
\begin{equation}\label{eq:CFT-to-lattice}
Z_b(\omega_1,\omega_2;R) ~=
\sum_{\sigma \in S^{|E|}}
\prod_{f \in F} T_R^b\big(\,\kappa_{\sigma(f_1)},\,\kappa_{\sigma(f_2)},\,\kappa_{\sigma(f_3)}\,\big) ~.
\end{equation}
Here, $E$ is the set of edges of the triangle decomposition of the torus (there are $3MN$ of them). The summation variable $\sigma$ is an $|E|$-tuple of indices of $S$, or, equivalently, a function $\sigma : E \to S$. The product ranges over the set $F$ of triangles (``faces'', of which there are $2MN$). For each triangle $f$, its three edges are denoted $f_1,f_2,f_3$, and the clipped triangle is evaluated at the basis vectors assigned to these edges, i.e.\ at $\kappa_{\sigma(f_i)}$, $i=1,2,3$.

\begin{figure}[t]
	\centering
	\includegraphics[width=.7\textwidth]{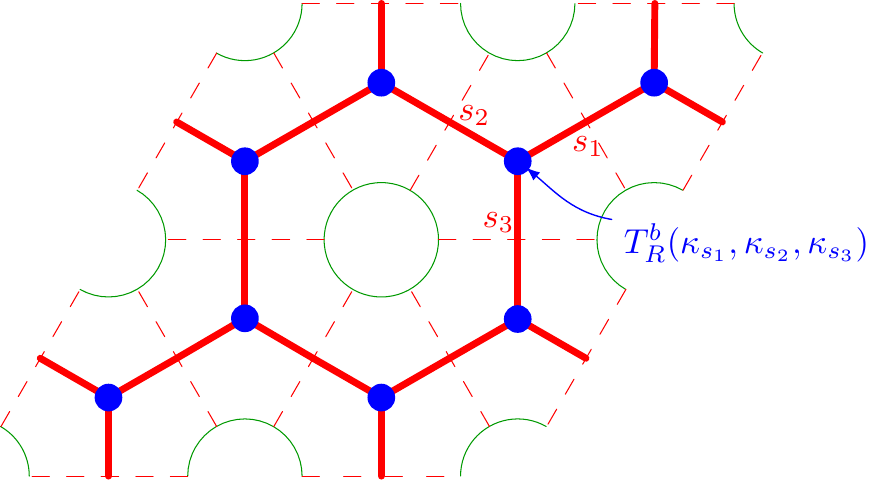}	
	\caption{Lattice model on the dual honeycomb lattice. The edges carry lattice spins $s_1, s_2, s_3,\dots \in S$. The Boltzmann weights 
	$T_R^b\big(\kappa_{s_1},\kappa_{s_2},\kappa_{s_3}\big)$, etc., are attached to the vertices.}
	\label{fig:intro-lattice-model}
\end{figure}

The lattice model is now obtained by a simple re-interpretation of \eqref{eq:CFT-to-lattice}. We replace the triangulation with the dual honeycomb lattice (Figure~\ref{fig:intro-lattice-model}). The degrees of freedom of the lattice model sit on the edges and are given by $S$, so there is an infinite number of lattice spin states at each edge. The Boltzmann weights reside on the vertices and are given by the value of $T_R^b$ on the basis vectors indexed by the value of the spin on the adjacent edges. The resulting lattice model has only nearest-neighbour interactions.
The partition function of this lattice model is given precisely by \eqref{eq:CFT-to-lattice}.

The limiting behaviours \eqref{eq:Rto0limit} and \eqref{eq:Rtod2limit} show that this lattice model by construction has property 1) stated at the beginning. We now turn to property 2).

\subsection{Topological symmetry and the cloaking boundary condition}\label{sec:top-sym-cloak-sym}

Let us suppose further that the unitary rational CFT $C$ has a diagonal modular invariant partition function, for example, $C$ could be an $A$-series unitary Virasoro minimal model. 
We comment on non-diagonal models in Section~\ref{sec:cloaking-bnd}.
Let $I$ be the finite set of irreducible representations of the chiral algebra (a rational vertex operator algebra). For diagonal models, $I$ labels both, the elementary conformal boundary conditions \cite{Cardy:1989ir} and the elementary topological line defects \cite{Petkova:2000ip,Frohlich:2006ch}. We will refer to the collection of topological line defects as \textit{topological symmetry} of the CFT $C$, even though they are typically not invertible under fusion.\footnote{ 
	In fact, topological line defects transparent to the holomorphic and anti-holomorphic copy of the chiral algebra define a fusion category $D_C$. For diagonal models, $D_C$ is given by the category $\mathrm{Rep}(V)$ of representations of the vertex operator algebra $V$ describing the chiral symmetry.}

\begin{figure}[t]
	\centering
	\begin{tikzpicture}
	\node at (0,0) {\includegraphics[width=.5\textwidth]{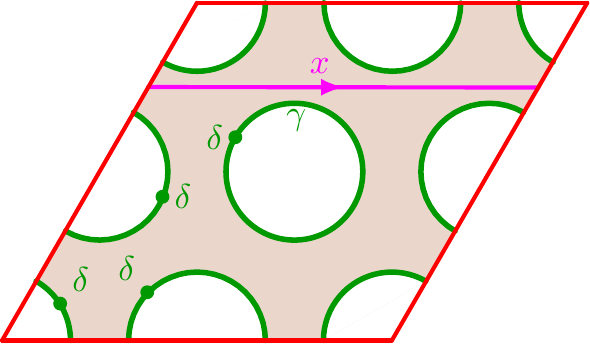}};
	\node at (6.5,0) {\includegraphics[width=.5\textwidth]{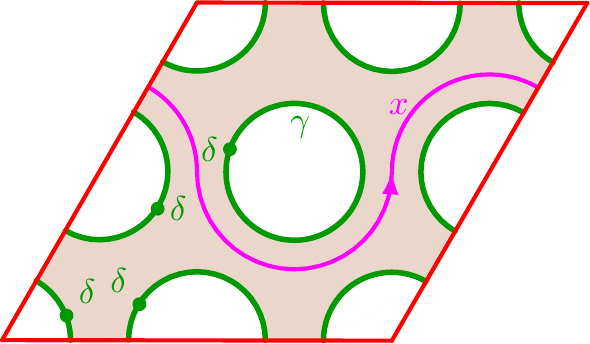}};
	\node at (0,-5) {\includegraphics[width=.5\textwidth]{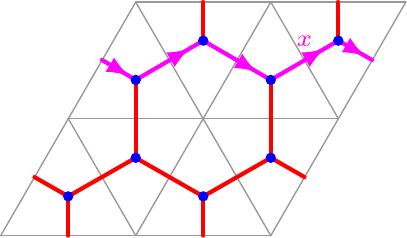}};
	\node at (6.5,-5) {\includegraphics[width=.5\textwidth]{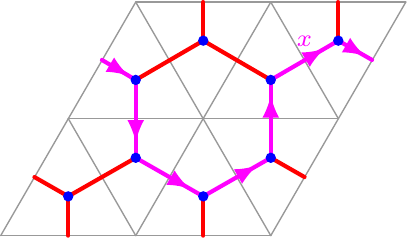}};
	\node at (-2.2,1.5) {a)};
	\node at (-2.2,-3.5) {b)};
	\node at (3.25,0) {$=$};
	\node at (3.25,-5) {$=$};
	\end{tikzpicture}		
	\caption{(a) Moving the line defect with topological defect condition $x \in I$ to the other side of a hole whose boundary is labelled by the cloaking boundary condition $\gamma(\delta)$ does not change the value of the CFT amplitude. (b) Representation of the identity in (a) in the lattice model.}
	\label{fig:intro-cloaking-bnd}
\end{figure}
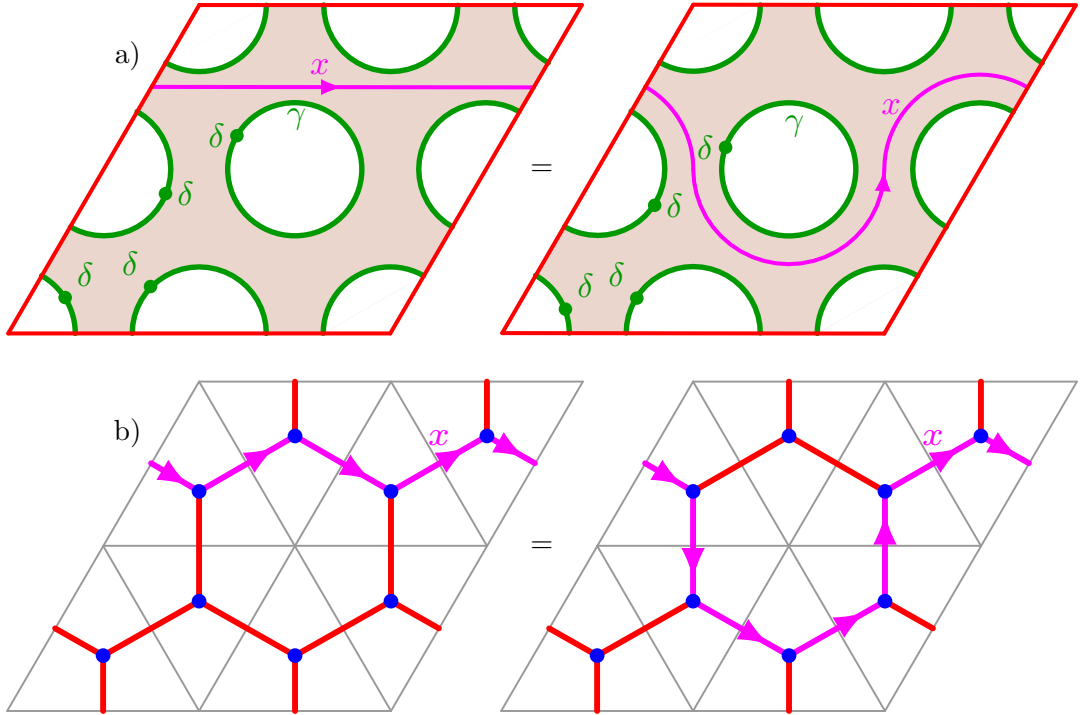

If we pick an elementary conformal boundary condition $b$ in the above construction, the resulting lattice model will not inherit the topological symmetry of the CFT $C$. However, there is a special superposition $\gamma$ of boundary conditions and a weight-zero field insertion $\delta$, such that a hole labelled in this way is transparent to all topological defects of $C$ (Figure~\ref{fig:intro-cloaking-bnd}\,a). This is described in detail in Section~\ref{sec:cloaking-bnd}. We refer to the conformal boundary condition $\gamma$ with field insertion $\delta$ as \textit{cloaking boundary condition} (as it makes the hole invisible to the topological defects) and denote it by $\gamma(\delta)$.

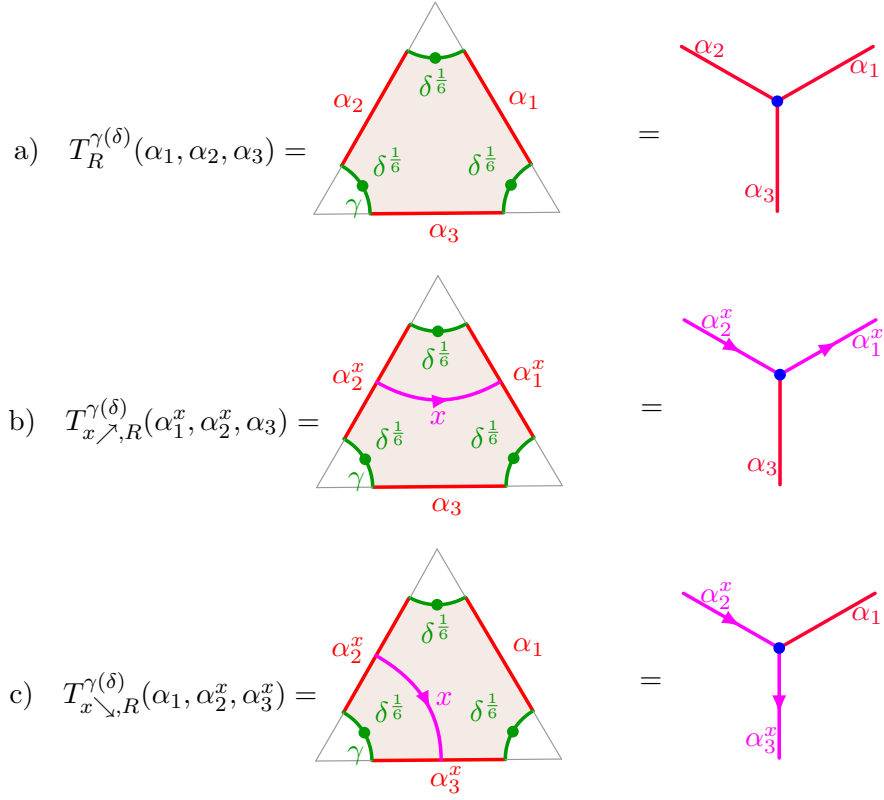
\begin{figure}[t]
	\centering
	\begin{tikzpicture}[scale=1]
	\node at (0,0) {a)~~ $ T_R^{\gamma(\delta)}(\alpha_1,\alpha_2,\alpha_3) = $};
	\node at (5.7,.3) {\includegraphics[scale=1.1]{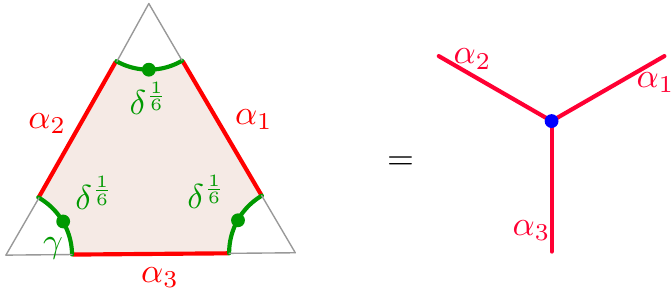}};
	\end{tikzpicture}
	\begin{tikzpicture}[scale=1]
	\node at (0,0) {b)~~ $T_{x\nearrow,R}^{\gamma(\delta)}(\alpha_1^x,\alpha_2^x,\alpha_3) = $};
	\node at (5.7,.3) {\includegraphics[scale=1.1]{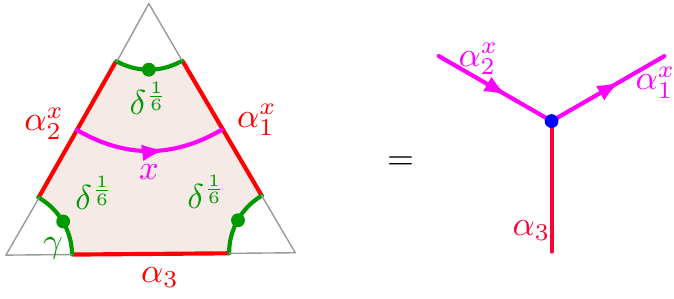}};
	\end{tikzpicture}
	\begin{tikzpicture}[scale=1]
	\node at (0,0) {c)~~ $T_{x\searrow,R}^{\gamma(\delta)}(\alpha_1,\alpha_2^x,\alpha_3^x) = $};
	\node at (5.7,.3) {\includegraphics[scale=1.1]{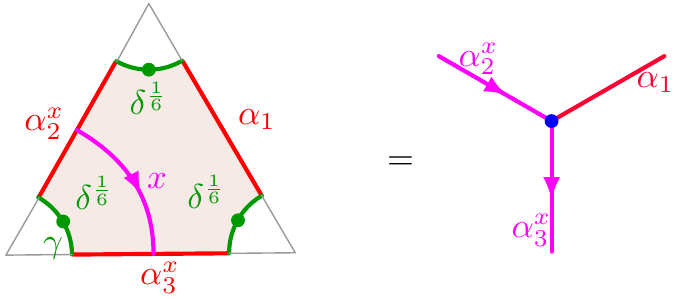}};
	\end{tikzpicture}
	\caption{(a) The clipped triangle and the interaction vertex for the cloaking boundary condition $\gamma(\delta)$. The clipped triangle carries an insertion of $\delta^{\frac16}$ on each of its three boundary components. 
	(b,c) The two possible clipped triangles with a topological line defect labelled $x \in I$ passing through them, and the corresponding interaction vertices. }
	\label{fig:intro-cloaking-vertex}
\end{figure}

To obtain a symmetric definition of the clipped triangle and of the interaction vertex, we need to take a sixth root of $\delta$ as in Figure~\ref{fig:intro-cloaking-vertex}\,a. After gluing the triangles together via sums over intermediate states as in Figure~\ref{fig:intro-sum-over-states}\,a, these insertions combine to one insertion of $\delta$ per boundary circle. 

To give the lattice representation of a line defect $x$, we need a new set of edge spins and a new interaction vertex. Denote by $\mathcal{H}^x_{\gamma\gamma}$ the space of open channel states on a strip with boundary condition $\gamma$ on either side and with a line defect $x$ running parallel to the boundary. Let $\{ \kappa_s^x \}_{s \in S^x}$ be an orthonormal basis of $\mathcal{H}^x_{\gamma\gamma}$, indexed by $S^x$. Then $S^x$ is the set of edge spins on an edge carrying the defect $x$.
The two clipped triangles defining the interaction vertex for two edges carrying a defect $x$ and one edge without defect are given in Figure~\ref{fig:intro-cloaking-vertex}\,b.

By construction, the lattice line defects are topological for each value of $R$ (Figure~\ref{fig:intro-cloaking-bnd}\,b) and they obey the same fusion rules as in the CFT $C$ (in fact they define the same fusion category). In this sense, the lattice model realises the topological symmetry of the CFT $C$ exactly for each value of $R$.
This explains property 2 stated at the beginning of the introduction.

\medskip

The cloaking boundary condition is special in another way. Write $|\gamma(\delta);R\rangle$ for the boundary state of the conformal boundary condition $\gamma$ placed on the boundary of a hole of radius $R$ with insertion of $\delta$. The insertion $\delta$ turns out to be such that the resulting weighted sum of Ishibashi states projects onto the vacuum Ishibashi state $|0\rangle\!\rangle$,
\begin{equation}\label{eq:only-vac-in-cloak}
|\gamma(\delta);R\rangle ~=~ R^{L_0+\overline L_0 - \frac c6} \, |0\rangle\!\rangle ~,
\end{equation}
where we use the standard radius-one Ishibashi states (see Sections~\ref{sec:BSdiag} 
for details). It is not hard to see that \eqref{eq:only-vac-in-cloak} 
is actually equivalent to the property of being transparent to all topological line defects (up to an overall constant which can multiply the right hand side).

We can now compare the leading contribution in the small-$R$ expansion of the boundary state $|b;R\rangle$ of an elementary boundary condition $b$ and of the cloaking boundary state $|\gamma(\delta);R\rangle$. Let $h_\mathrm{min}$ be the lowest $L_0$-weight above the vacuum in the closed channel spectrum of $C$, and let $\phi_\mathrm{min}$ be the leading summand in the corresponding Ishibashi state.\footnote{
	For Virasoro minimal models $\phi_\mathrm{min}$ is just the primary field of $(L_0,\overline L_0)$-weight $(h_\mathrm{min},h_\mathrm{min})$. For models with several	fields of weight $h_\text{min}$, $\phi_\mathrm{min}$ is an appropriate sum.}
Then
\begin{align}
\ket{b;R} ~&=~ R^{- \frac c6}\, a^b_1 \Big(  |0\rangle + R^{2 h_\mathrm{min}} \, a_2^b\,|\phi_\mathrm{min}\rangle + \text{(higher powers of $R$)} \Big)~,
\nonumber \\
\ket{\gamma(\delta);R} ~&=~ 
R^{- \frac c6} \Big( |0\rangle + R^{4} \,\tfrac2c\, L_{-2}\overline L_{-2}|0\rangle + \text{(higher powers of $R$)} \Big) ~,
\label{eq:bnd-state-leading-contrib}
\end{align}
where $a^b_1, a_2^b \in \mathbb{C}$ are prefactors as dictated by the boundary state.
Using this, one can write the first two leading terms in the small $R$-expansion of $Z_b(\omega_1,\omega_2;R)$ in \eqref{eq:ZalphaR-def} as follows. Let $\langle \phi(z) \rangle_{\omega_1,\omega_2}$ be the (normalised) expectation value of $\phi(z)$ on the torus with periods $\omega_1,\omega_2$, let $\Lambda$ be the set of positions of the $MN$ vertices on the torus, and abbreviate $g(R) = R^{-\frac{c}{6}MN} Z(\omega_1,\omega_2)$. Then, using \eqref{eq:Rto0limit},
\begin{align}
Z_b(\omega_1,\omega_2;R) ~&=~ g(R) (a^b_1)^{MN} \Big( \langle 1 \rangle_{\omega_1,\omega_2} + R^{2 h_\mathrm{min}} \, a_2^b \sum_{z \in \Lambda} \langle \phi_\mathrm{min}(z) \rangle_{\omega_1,\omega_2} + \cdots \Big) ~,
\nonumber \\
Z_{\gamma(\delta)}(\omega_1,\omega_2;R) ~&=~ g(R) \Big( \langle 1 \rangle_{\omega_1,\omega_2} + R^{4} \, \tfrac2c \sum_{z \in \Lambda} \langle T\overline T(z) \rangle_{\omega_1,\omega_2} + \cdots \Big)~.
\label{eq:Z-hole-smallR-expansion}
\end{align}

\noindent
If we chose $M,N$ large, the sum over $\Lambda$ is well approximated by an integral over the torus $T^2$ of periods $\omega_1,\omega_2$. Namely, with $d = |\omega_2|/N$ the area of a lattice cell is $\sqrt{3} \, d^2/2$ and we get
\begin{equation}\label{eq:sum-to-integral}
\sum_{z \in \Lambda} \phi(z)
~\approx~
\frac{2}{\sqrt{3}\,d^2}
\int_{T^2} \phi(z) \, dz ~.
\end{equation}
(In fact, for one-point functions in \eqref{eq:Z-hole-smallR-expansion}, the sum and integral are actually equal as the one-point functions are position-independent.)
Therefore, the leading terms in the expansion \eqref{eq:Z-hole-smallR-expansion} agree with those one obtains when perturbing the CFT $C$ on $T^2$ by the fields
\begin{align}
\text{bnd. cond. } b  \quad:\quad & 
\text{(const)} \,\tfrac{R^2}{d^2} R^{2 h_\mathrm{min}-2} \cdot \phi_\mathrm{min}(z) ~,
\nonumber \\
\text{bnd. cond. } \gamma(\delta) \quad:\quad & 
\text{(const)} \,\tfrac{R^2}{d^2} R^{2} \cdot T\overline T(z) ~.
\end{align}
Consider the torus of periods $\omega_1,\omega_2$ with a finer and finer lattice of holes, but such that the lattice cells keep their shape up to rescaling (i.e.\ $R/d$ is constant). Then the radius $R$ will become smaller and smaller. For $2h_\mathrm{min}<2$ this will increase the strength of the perturbation by $\phi_\mathrm{min}$ in the first line, but it will decrease the strength of the perturbation by $T\overline T$ in the second line.
From a more field theoretical perspective,
for $2h_\mathrm{min}<2$ the field $\phi_\mathrm{min}(z)$ creates a relevant perturbation which changes the IR-fixed point. On the other hand, $4>2$ and so $T\overline T(z)$ is irrelevant and preserves the IR-fixed point. 

We take this as an indication that among the lattice models we construct, the one obtained from the cloaking boundary condition is the best candidate to reproduce the original CFT $C$ in the continuum limit.
To support this argument further, one should not reply on the approximation \eqref{eq:sum-to-integral} and instead express the lattice model systematically as a perturbation of the continuum model by a suitable (infinite) collection of fields, but we will not attempt this here.

\medskip
The observation that there seems to be a relation between being critical and preserving a topological symmetry has been made in the context of lattice models and tensor-network models e.g.\ in \cite{PhysRevB.82.115126,Aasen:2016dop,Aasen:2020jwb,Vanhove:2018wlb}.

\subsection{Results in this paper}
\label{sec:intro-results}

We start with a detailed discussion of the cloaking boundary condition $\gamma(\delta)$, see Section~\ref{sec:cloaking-bnd}. We show how to construct this boundary condition both in diagonal and non-diagonal rational conformal field theories.  
In Section \ref{sec:TFT}, we compute the $R \to \frac d2$ limit of the lattice model obtained from the interaction vertex $T_R^{\gamma(\delta)}$ in a diagonal unitary rational CFT $C$. It is given by a 2d state-sum TFT $C_\mathrm{top}$ which realises the topological symmetry of the original CFT $C$.

\medskip

For the remainder of this paper, we focus on the complex-analytic aspects of the interaction vertex $T_R^b$. We will restrict ourselves to elementary boundary conditions $b$, leaving a closer investigation of the more involved vertex for the cloaking boundary condition for a future part of this series. The main new technical ingredient we present in this paper is an exact uniformisation formula for the clipped triangle at arbitrary $0<R<\frac d2$ (Section~\ref{sec:uniform}). This makes it possible to evaluate $T_R^b$ exactly for each triple of open channel states as an explicit function of $R$. If one disregards the Liouville factor arising from the conformal anomaly, the dependence is only on $R/d$. 

\begin{figure}[t]
	\centering	
	\includegraphics{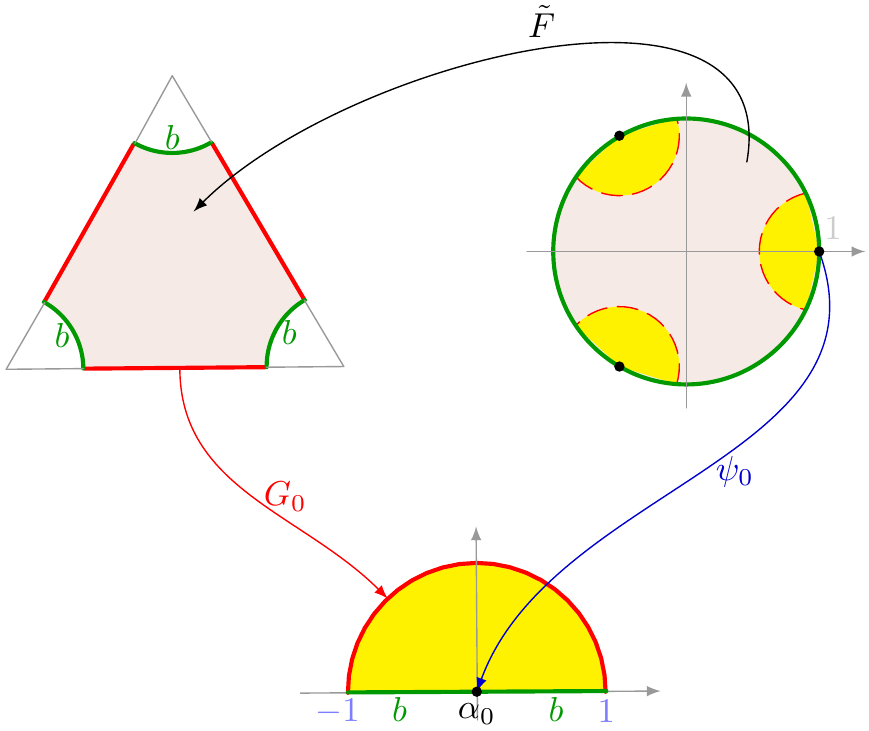}
	\caption{Uniformisation and local coordinates. The open channel states for the clipped triangle are described by glueing three half-discs with boundary fields $\alpha_j$, $j=0,\pm 1$ inserted at $0$ via the maps $G_j$. The resulting surface is a disc and is uniformised by $\tilde F$. The disc has field insertions at $e^{ 2 \pi i j/3}$ with local coordinate charts $\psi_j$.}
	\label{fig:glue-open-uniformise}
\end{figure}

The open channel states are described by gluing half-discs with a boundary field insertion at $0$ (with standard local coordinates) to the three open channel boundaries. The resulting surface is then mapped to the unit disc with boundary field insertions at $1, e^{\pm 2 \pi i / 3}$, but with complicated local coordinates expressed via exponential and hypergeometric functions (see Figure~\ref{fig:glue-open-uniformise} and Section~\ref{sec:uniform}).

We describe the action of the coordinate transformation on descendant fields and 
give a recursion formula for the disc amplitudes of descendant fields. The first main result of this paper is an algorithm to reduce $T^b_R$ on three arbitrary fields to the disc amplitude of the three corresponding primary fields (Section~\ref{sec:disc-recursion-and-vertex}).

\medskip

In order to get a lattice model that can be studied numerically, we fix a cutoff energy $h_\mathrm{max}$ and truncate the sum over open channel states in Figure~\ref{fig:intro-sum-over-states}\,a to those $s \in S$ where $\kappa_s \in \mathcal{H}_{bb}$ has $L_0$ weight $h_s \le h_\mathrm{max}$. We stress that also this cutoff model still exactly realises the full topological symmetry of the CFT. This is explained in Section~\ref{sec:truncated-sym}.
The error of the truncated sum relative to the exact CFT expression decreases for $R \to \frac d2$, where the high energy states are more and more suppressed. 
But the truncated sum will be an increasingly worse approximation as $R \to 0$, where all open channel states contribute.

We study the Ising CFT for different values of the cutoff $h_\mathrm{max}$ in the extreme case $M=N=1$ in Figure~\ref{fig:intro-decompose-torus}\,a where the torus has a single hole and is decomposed into two triangles. In order to do this, we need to address two problems:
\begin{enumerate}
\item The value of $T^b_R$ on three primary fields is multiplied by a universal Liouville factor which arises from the conformal anomaly and depends only on the geometry. We do not compute the Liouville factor in this paper, and so we can only determine $Z_b(\omega_1,\omega_2;R)$ up to an $R$-dependent factor.
\item The CFT amplitude for a torus with a hole is not known exactly, and so the question arises to what we should compare our truncated expressions in order to judge its accuracy.
\end{enumerate}
We dispense with problem 1 simply by looking at ratios of CFT amplitudes,
\begin{equation}\label{eq:ZZ-ratio}
	\frac{Z_b(\omega_1,\omega_2;R)}{Z_{b'}(\omega_1,\omega_2;R)}~,
\end{equation}
for two different conformal boundary conditions $b,b'$. The Liouville factor then cancels from this expression. 

Problem 2 is more work. We give a second numerical approximation scheme for the partition function $Z_b(\omega_1,\omega_2;R)$ which becomes exact for $R \to 0$ but loses accuracy as $R \to \frac d2$. In a region for $R$ where both schemes approximately agree, we believe that they are also a good approximation to the exact CFT amplitude. 

\begin{figure}[t]
	\centering
	\includegraphics[width=.49\textwidth]{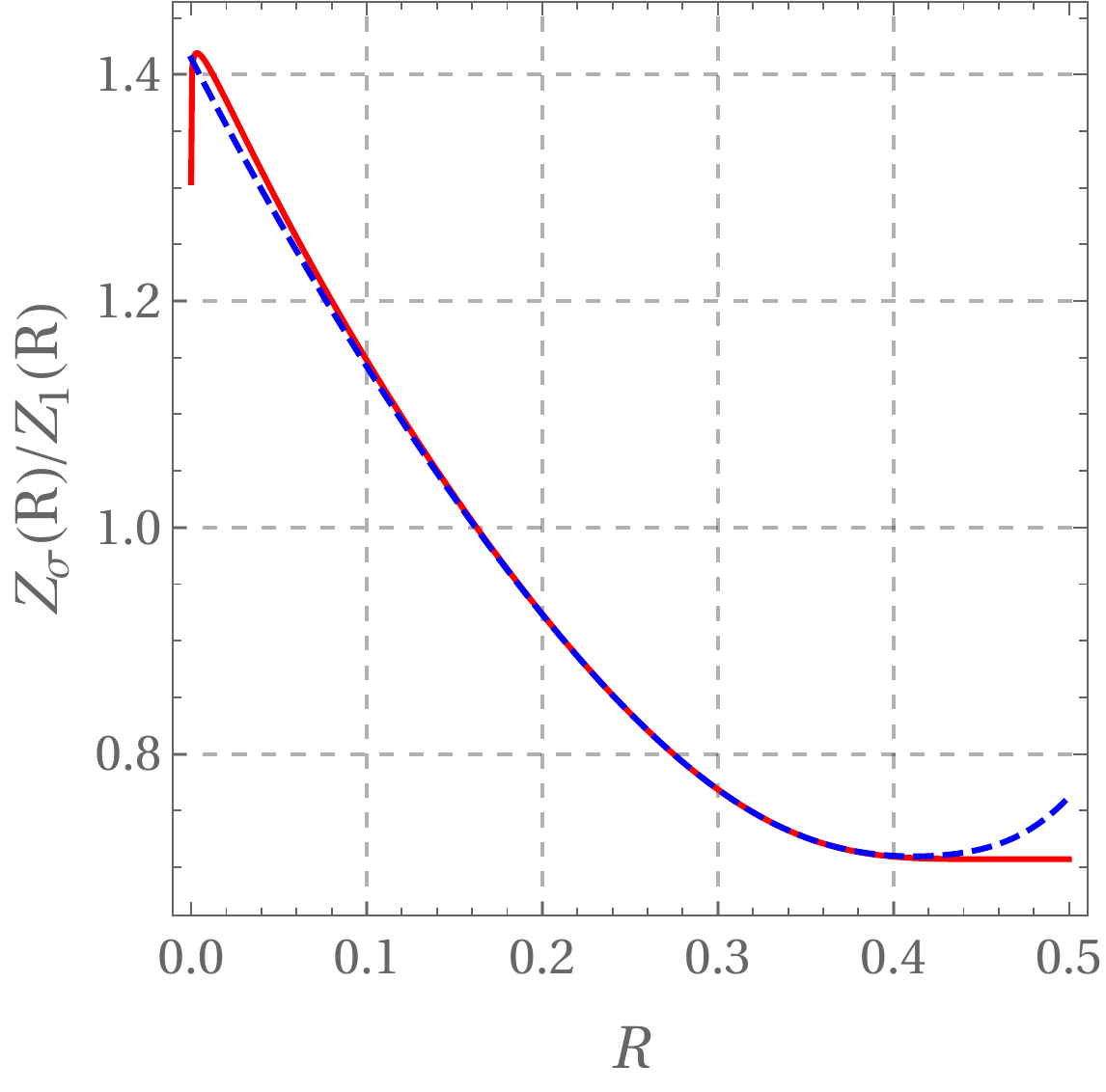}
	\includegraphics[width=.49\textwidth]{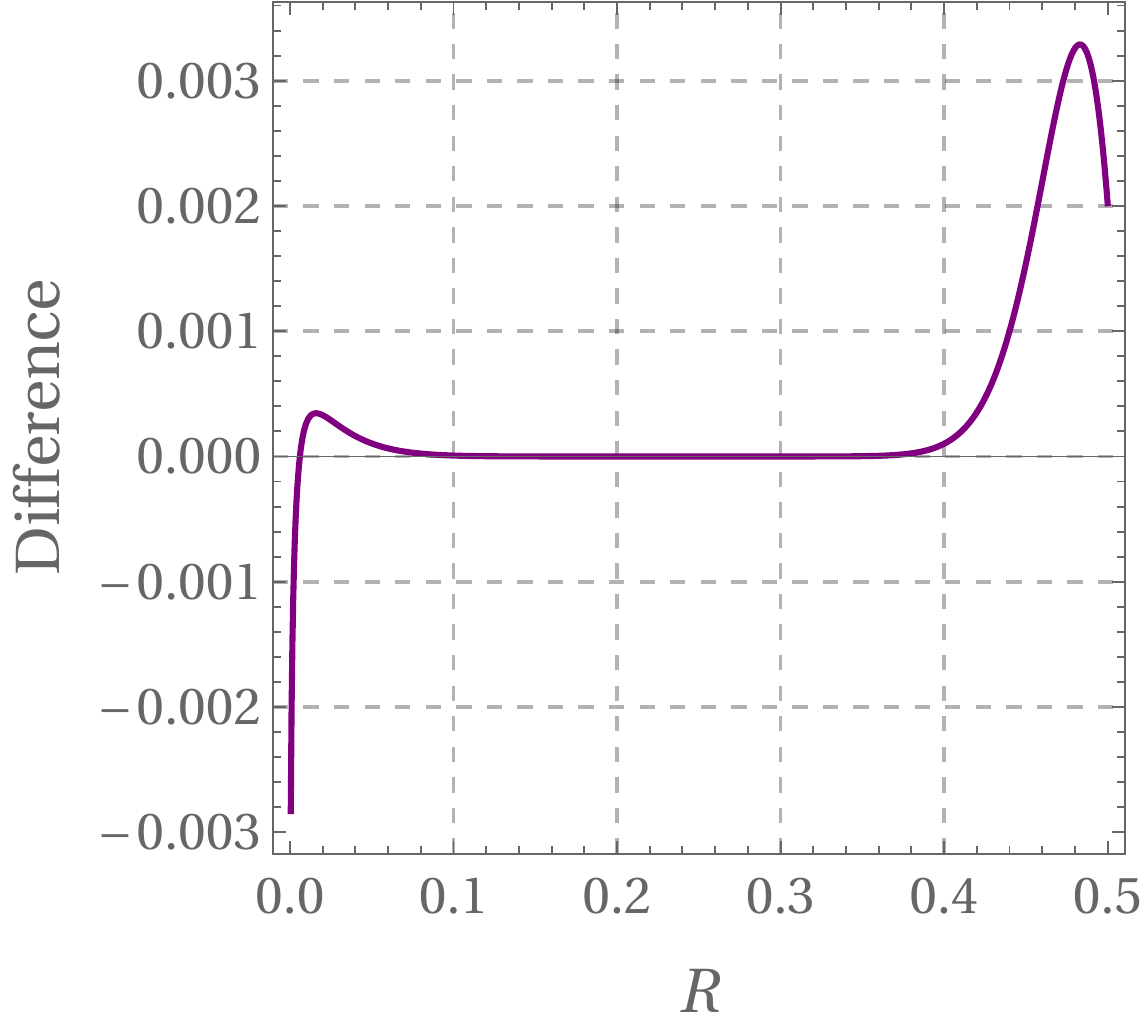}	
	\caption{Results for the Ising CFT.
	(a) $Z_\sigma/Z_\one$ in the two approximation schemes: 
	the red solid curve gives the result for the open state sum truncated at $h_\mathrm{max} = 2$ and is exact at $R=\frac12$, the blue dotted curve gives the ratio for the boundary state truncated at $\Delta_\mathrm{max} = 20$ and is exact at $R=0$. (b) The difference of the results in the open and closed channel for $h_\mathrm{max} =5$ and $\Delta_\mathrm{max} = 28$.}
	\label{fig:intro-Ising-plot}
\end{figure}

The second approximation scheme works in the closed channel and truncates the boundary state at some cutoff energy $\Delta_\mathrm{max}$. To compute the resulting CFT amplitude, we give a recursive formula to evaluate one-point functions of descendant fields on the torus, analogous to the procedures derived in \cite{Eguchi:1986sb,Zhu:1996,Gaberdiel:2012yb}, see Section~\ref{sec:recusion-one-point}. 
This explicit recursion formula is the second main result of our paper. 

\medskip

Combining all these ingredients, we can finally study the ratio \eqref{eq:ZZ-ratio} for the Ising CFT in the open and closed approximation schemes. 
We take $b = \sigma$, the free boundary condition, and $b' = \one$, the fixed spin-up boundary condition. 
The torus has a single hole of radius $R$, and we take $d=1$, $M=N=1$, and $\omega_1=1$, $\omega_2 = e^{\pi i/3}$. In Figure~\ref{fig:intro-Ising-plot} we plot the ratio in both approximation schemes, as well as their difference. 
This shows good agreement in a surprisingly (for us) large region of $R$, even at the moderate truncation levels used in the open channel.
More detailed numerical results are given in Section~\ref{sec:explicit_results_Ising}.

\subsection{Next steps}\label{sec:nextsteps}

In future parts of this series, we will derive the various structure constants and normalisations of disc amplitudes needed to compute $T^{\gamma(\delta)}_R$, the interaction vertex for the cloaking boundary condition. If one restricts $T^{\gamma(\delta)}_R$ to only primary open channel states, the resulting expressions bear some resemblance to the tensor network approach to strange correlators \cite{you2014wave,Vanhove:2018wlb}, and we intend to investigate this connection in the future.

While the cloaking boundary condition makes the computation of $T^{\gamma(\delta)}_R$ more involved, it actually makes the computation in the truncated boundary state approximation scheme much simpler as one only has to account for the vacuum Ishibashi state. This can be computed solely in terms of derivatives of the torus partition function (without field insertions). We will repeat the comparison of the two approximation schemes on the torus with one hole for the cloaking boundary condition in several minimal models.

Starting from a CFT $C$, our construction produces a two-parameter family of lattice models. The parameters are $R$, the size of the holes, and $h_\mathrm{max}$, the cutoff in the sum over open channel states. The lattice models are described by the correspondingly truncated set of edge spin states $S(h_\mathrm{max})$, and by the interaction vertex $T^{\gamma(\delta)}_R$ restricted to these edge states. As will be noted in Section~\ref{sec:truncated-sym}, each model in this 2-parameter family realises the full topological symmetry of the initial CFT $C$.

Further parts of this series will be concerned with the investigation of the lattice models produced from this construction, in particular by studying transfer matrices for values $N>1$. 
For simple rational CFTs, such as the Ising CFT, and small open channel cutoffs $h_\mathrm{max}$ we will compare to well-established lattice constructions. In other cases, we will do numerical investigations. 

\medskip

Let us conclude the outline of this research program with some speculations about hoped-for properties of the parameter-dependent lattice model
we construct from a given CFT. We aim to lend support to these ideas in future parts of this series.

\subsubsection*{Realise different CFTs with the same topological symmetry}

The topological symmetry of a rational CFT $C$ is described by a fusion category $D_C$. 
To be more precise $D_C = D_{C,V}$ consists of those topological line defects that commute with the holomorphic and anti-holomorphic copy of the rational VOA $V$ in the space of bulk fields of $C$. So if $W \subset V$ is a sub-VOA, then more line defects may exist, $D_{C,W} \supset D_{C,V}$.
    
The assignment $C \mapsto D_C$ is many-to-one (it is not known if it is surjective). As a trivial example of this, take the holomorphic WZW CFT $(E_8)_1$. The diagonal modular invariant has a single sector $H := (E_8)_1 \boxtimes \overline{(E_8)_1}$, and its topological symmetry is just $D_H = \mathrm{vect}$, the category of vector spaces. This means that $H$ does not possess non-trivial elementary topological line defects which are transparent to the chiral and anti-chiral copy of the 248 weight-one currents of $(E_8)_1$.
Then the product theories $C \otimes H^{\otimes n}$ all produce the same fusion category of topological symmetries,
\begin{equation}
	D_C ~\cong~ D_{C \otimes H^{\otimes n}} ~.
\end{equation}
On the other hand, in the context of tensor networks and strange correlators, one uses the fusion category $D_C$ as input datum. It is then clear that these models have to be supplemented by additional data in order to produce the infinite tower of CFTs with the same topological symmetry. One may hope that including enough descendant fields in the vertex $T^{\gamma(\delta)}_R$ provides this missing information. 

This leads us to expect the following picture:
Let us take $R$ to be the fundamental parameter of the lattice model because it has a clear geometric interpretation in the CFT. After fixing $R$, one can pick an open channel cutoff $h_\mathrm{max}$ large enough to obtain a good approximation of the CFT amplitude. Since each of these models realises the topological symmetry of the CFT, one may expect that the lattice model is critical and defines a CFT in the continuum limit. 

\begin{figure}[t]
	\centering
	\includegraphics{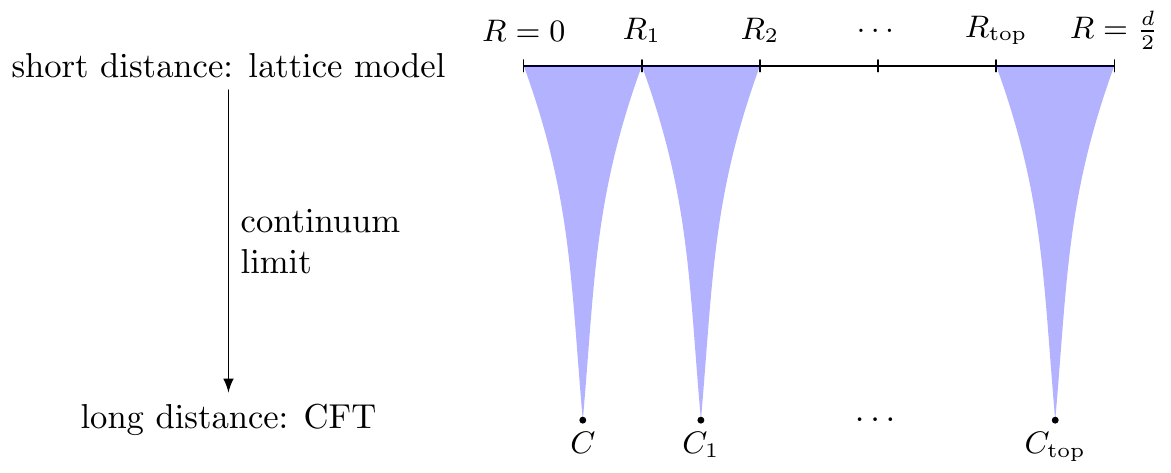}	
	\caption{Highly speculative phase diagram for the dependence of the continuum limit (the CFT describing the long distance behaviour) on the parameter $R$ in the short distance theory (the lattice model).}
	\label{fig:speculative-phase-diagram}
\end{figure}

For $R=0$ and $h_\mathrm{max} = \infty$ one obtains the original CFT. One can then speculate that this stays the case up to some critical value $R_1$, after which the universality class changes to another CFT $C_1$. The CFT $C_1$ will then realise the topological symmetry of $C$, at least as part of its full symmetry, $D_{C,V} \subset D_{C_1,V_1}$. This behaviour could possibly repeat several times, leading to CFTs $C_2,C_3,\dots$ in the continuum limit, up to some final transition at $R=R_\mathrm{top}$ after which the model becomes the topological model $C_\mathrm{top}$, see Figure~\ref{fig:speculative-phase-diagram}. 
It may well happen, that $R_\mathrm{top}=\frac d2$, this remains to be investigated.

As we will see in Section~\ref{sec:TFT}, the 2d\,TFT  $C_\mathrm{top}$ obtained from a diagonal unitary rational CFT $C$ has one vacuum state for each topological line defect of $C$. That is, the topological symmetry is maximally broken in the vacuum. It would be interesting to see if in the (hypothetical) chain of CFTs $C_1,C_2,\dots$ the topological symmetry gets partially broken below some value of $R$, so that the number of vacuum states increases in with decreasing $R$, up to the maximum degeneracy present in $C_\mathrm{top}$.

Related ideas on the behaviour of topological symmetries under RG-flow between continuum theories (rather than lattice and continuum theories) have recently been studied in \cite{Chang:2018iay,Kikuchi:2021qxz}.

\subsubsection*{Lattice models for non-unitray and non-rational CFTs}

For non-unitary rational CFTs, such as non-unitary minimal models, the main change is that the state of weight $(h,\overline h) = (0,0)$ is no longer the vacuum of the theory, i.e.\ no longer the state of lowest energy. This modifies the $R \to \frac d2$ behaviour of the lattice model. Otherwise, the construction is unaffected, in particular the cloaking boundary condition exists just the same and the resulting lattice model still realises the topological symmetry of the CFT.

Now suppose the CFT in question is still unitary and has a discrete spectrum, but is no longer rational, i.e.\ it possesses an infinite number of sectors.  In these theories, one can again write the lattice model in the same way for an elementary boundary condition $b$. However, when trying to construct the cloaking boundary condition, one encounters a problem:
Consider, as an example, $su(2)_1$ which has $c=1$, but take as chiral algebra only the Virasoro algebra, which is not rational at $c=1$. The conformal boundary conditions and topological defects which respect the Virasoro algebra are known and parametrised by $SU(2)$ and $(SU(2) \times SU(2))/\mathbb{Z}_2$, respectively \cite{Gaberdiel:2001xm,Fuchs:2007tx}. One can try to take an integral over all boundary conditions, but this leads to a continuous spectrum in the open channel, and hence to a lattice model with an infinite number of spin states, irrespective of the cutoff. Looking at what happens if one considers the full chiral algebra of $su(2)_1$, which is rational, leads one to find finite subsets of topological defects which close under fusion \cite{Thorngren:2021yso} (finite subgroups of $(SU(2) \times SU(2))/\mathbb{Z}_2$ in this example). One can then build a ``partially cloaking boundary condition'' which is a finite superposition of elementary conformal boundary conditions but is invisible only to a finite subalgebra of topological defects.

The final class of non-rational CFTs we would like to discuss are logarithmic CFTs which are still finite in the sense that the corresponding vertex operator algebra is $C_2$-cofinite (plus some further conditions). In particular, there are only a finite number of distinct irreducible representations. Such CFTs are increasingly well understood on surfaces with boundaries \cite{Runkel:2012rp,Creutzig:2016fms,Fuchs:2017unc}. Again, the same construction for the lattice model applies if one picks a fixed conformal boundary condition for the holes. It is not clear to us if a cloaking boundary condition exists, but we expect that it is at least possible to construct partially cloaking boundaries as above. 

\bigskip

\subsection*{Organisation of this paper}

We start in Section~\ref{sec:cloaking-bnd} with the derivation of the cloaking boundary condition in diagonal and non-diagonal models, and we show that the topological symmetry of the CFT is still realised if an open-channel energy cutoff is introduced.

In Section~\ref{sec:TFT} we compute the $R \to \frac d2$ (touching hole) limit of the lattice model and show that it results in a state sum 2d\,TFT whose topological line defects include those of the initial CFT.

In Section~\ref{sec:TorusDescCorr} we give a recursive formula to compute correlators of descendant bulk fields on the torus.
This recursion is used in Section~\ref{sec:torus-closed} to compute the closed channel approximation to the partition function of the torus with a hole as an expansion around $R=0$.

Section~\ref{sec:openstring} contains the computation of the interaction vertex and forms the technical core of this paper. We first describe the uniformisation of the clipped triangle to the unit disc, then give the transformation on descendant fields resulting from these coordinates, and finally provide a recursive procedure to evaluate the disc correlators. 
This is the input for Section~\ref{sec:open-channel-approx}, where the open channel approximation is computed via the truncated sum over open channel states.

In Section~\ref{sec:Ising-compare} we compare the two approximation schemes in the Ising CFT and find good agreement.

Finally, two appendices contain conventions and technical details on the analytic functions we use.

\subsection*{Acknowledgements}

We would like to thank Sukhwinder Singh for many helpful discussions, for useful comments on a draft of this paper, and for collaboration in the early stages of this project.
IR is partially supported by the Cluster of Excellence EXC 2121 ``Quantum Universe'' - 390833306. 

\newpage
\section{The cloaking boundary condition}
\label{sec:cloaking-bnd}

In this section, we first give the details for the construction of the cloaking boundary condition both in diagonal and non-diagonal models. As a tool, we introduce a ``cloaking line defect''. 
We also explain why truncating the open channel state spaces still leads to lattice models which realise the full topological symmetry of the initial CFT.

\subsection{Diagonal models}\label{sec:cloak-diagonal}

In this section, we assume that $C$ is a rational CFT (not necessarily unitary) with diagonal modular invariant.\footnote{
		There could be several distinct CFTs with diagonal modular invariant.
		To be precise, we mean the Cardy-case CFT, i.e.\ the CFT described by the Frobenius algebra $\one$ in the formalism of \cite{Fuchs:2002cm}.}
Thus $C$ has the same holomorphic and anti-holomorphic chiral algebra, given by a rational vertex operator algebra $V$. Let $I$ be the set of distinct irreducible representations of $V$. We write $\one \in I$ for the vacuum representation. 

For diagonal models, $I$ labels both, the elementary conformal boundary conditions and elementary topological line defects (both preserving $V$) \cite{Cardy:1989ir,Petkova:2000ip,Fuchs:2002cm,Frohlich:2006ch}. 
The weight zero fields on junctions of line defects and on junctions of boundaries with line defects can be described in terms of Hom-spaces of the modular fusion category $\MC = \mathrm{Rep}(V)$. For example, $\varphi \in \Hom_\MC(x \otimes y,z)$ labels the weight-zero fields on a junction with incoming line defects $x,y$ and outgoing line defect $z$, and also the junction of a line defect $x$ joining boundary conditions $y$ and $z$:
\begin{equation}
\begin{tikzpicture}
\node at (0,0) {\includegraphics[scale=1.3]{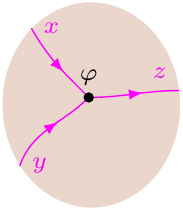} };
\node at (5,0) {\includegraphics[scale=1.3]{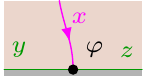} };
\end{tikzpicture}
\end{equation}
For each $x,y,z \in I$ such that $\Hom_\MC(x \otimes y,z) \neq \{0\}$ we choose a basis $\{ \alpha \}$ in this hom space, as well as a dual basis $\{\overline\alpha\}$ in $\Hom_\MC(z,x \otimes y)$, in the sense that $\alpha \circ \overline \beta = \delta_{\alpha,\beta} id_z$. In terms of line defects, this gives the identities
\begin{align}
\includegraphics[scale=1.2]{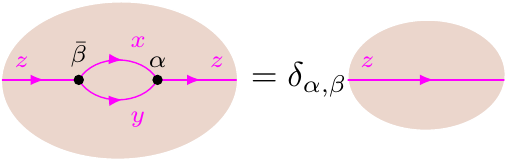} \quad ,\\
\includegraphics[scale=1.2]{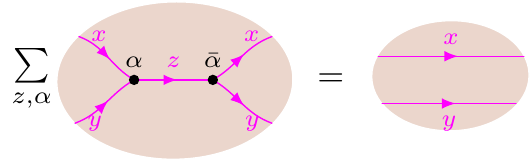} ~~.
\label{eq:defect-fusion-same-orientation}
\end{align}

\noindent
These and the following identities for topological line defects and conformal boundary conditions in rational CFT are derived in \cite{Frohlich:2006ch} using the 3d\,TFT approach, see also 
e.g.\ \cite{Runkel:2008gr,Runkel:2010ym,Kojita:2016jwe,Bhardwaj:2017xup,Chang:2018iay,Konechny:2019wff,Thorngren:2019iar,Thorngren:2021yso} 
for discussions on how to compute with topological line defects.
We will need the following facts:
\begin{enumerate}
\item A circular line defect labelled $x$ evaluates to the quantum dimension $\dim(x)$ of $x$:
\begin{equation}
\includegraphics[scale=1]{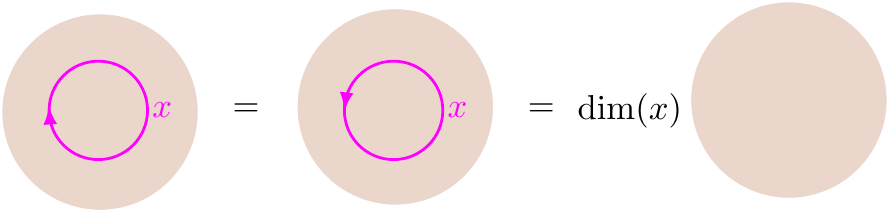}
\end{equation}

\noindent
The quantum dimension is given by $\dim(x) = \frac{S_{\one x}}{S_{\one\one}}$,
where $S_{xy}$ is the modular $S$-matrix giving the transformation of characters under $\tau \mapsto -\frac1\tau$.
\item Fusing a line defect labelled $x \in I$ to the boundary condition labelled $\one \in I$ gives the boundary condition labelled $x$:
\begin{equation}
\includegraphics[scale=1]{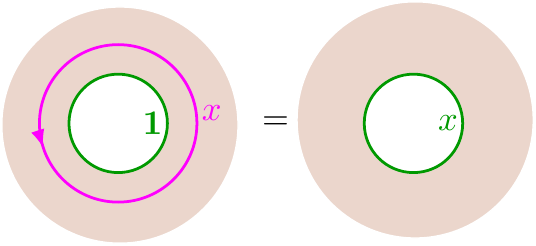}
\label{eq:fuse-defect-to-bnd}
\end{equation}
\item Locally fusing two line defects with opposite orientation incurs an extra quantum dimension factor as compared to \eqref{eq:defect-fusion-same-orientation}:
\begin{equation}
\includegraphics[scale=1.3]{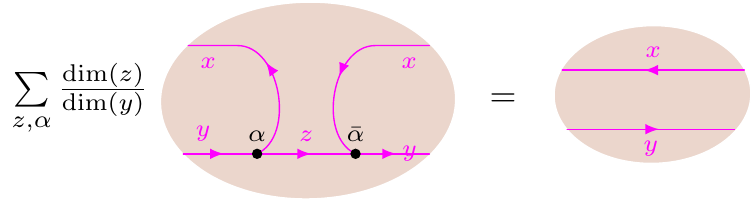}
\label{eq:defect-fusion-opposite-orientation}
\end{equation}
\end{enumerate}

\begin{figure}[t]
	\centering
    \includegraphics{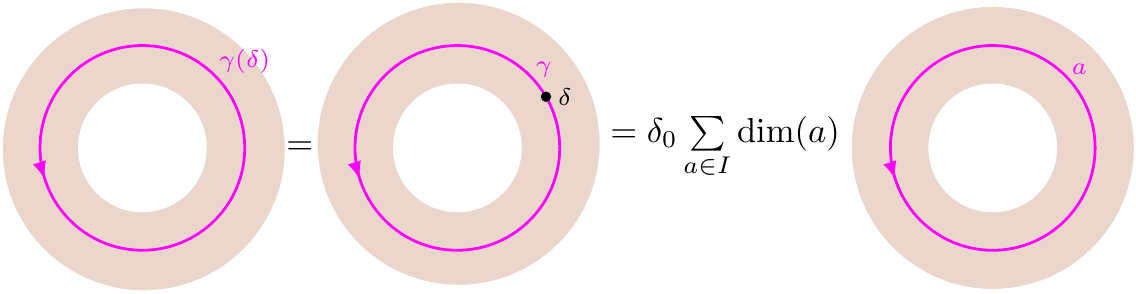}
	\caption{An annular region of the CFT surface with a cloaking defect loop $\gamma(\delta)$, and its definition in terms of the topological defect $\gamma$ with insertion of the weight-zero field $\delta$, as well as its expansion in terms of elementary defects.}
	\label{fig:cloaking-defect-loop}
\end{figure}

\noindent
We now come to the main ingredient, the \textit{cloaking defect loop}, which we also call $\gamma(\delta)$ (we will see in a moment why this is consistent with Section~\ref{sec:top-sym-cloak-sym}). It is given by a circular defect labelled by the superposition $\bigoplus_{a \in I} a$ of all elementary defects and carries the weight-zero field insertion
\begin{equation}
\delta ~:=~
\delta_0 \sum_{a \in I} \dim(a) \,1_a ~,
\end{equation}
where $\delta_0 \in \mathbb{C}^\times$ is a normalisation constant which will be fixed at the end of Section~\ref{sec:BSdiag},
and $1_a$ is the identity field on the defect $a$, see Figure~\ref{fig:cloaking-defect-loop}.

Consider the situation where a cloaking defect loop wraps around some region of the surface that can contain boundary components, field insertions or higher genus parts, and is marked ``anything'' in the identities below. We want to show that one can replace a defect line $x$ passing just above the cloaking defect loop by the same defect line passing just below it without changing the CFT correlator:
\begin{equation}
\includegraphics[scale=.95]{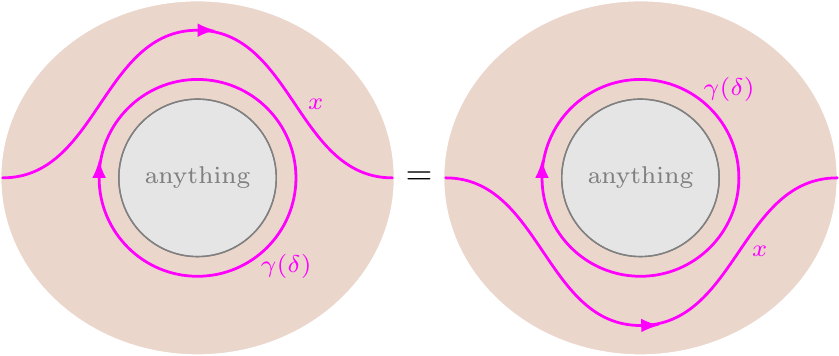}
\label{eq:move-defect-past-cloaking-defect}
\end{equation}
The computation is exactly the same as used 
in Reshetikhin-Turaev TFT to show the handle-slide property of the Kirby colour, and as in string-net models when one wants to hide a puncture, see \cite{kirillov2011string} (it is in this context that the qualifier ``cloaking'' appeared \cite{Goosen2018Oriented1V}).
The details for showing \eqref{eq:move-defect-past-cloaking-defect} are:
\begin{equation}
\includegraphics[width=\textwidth]{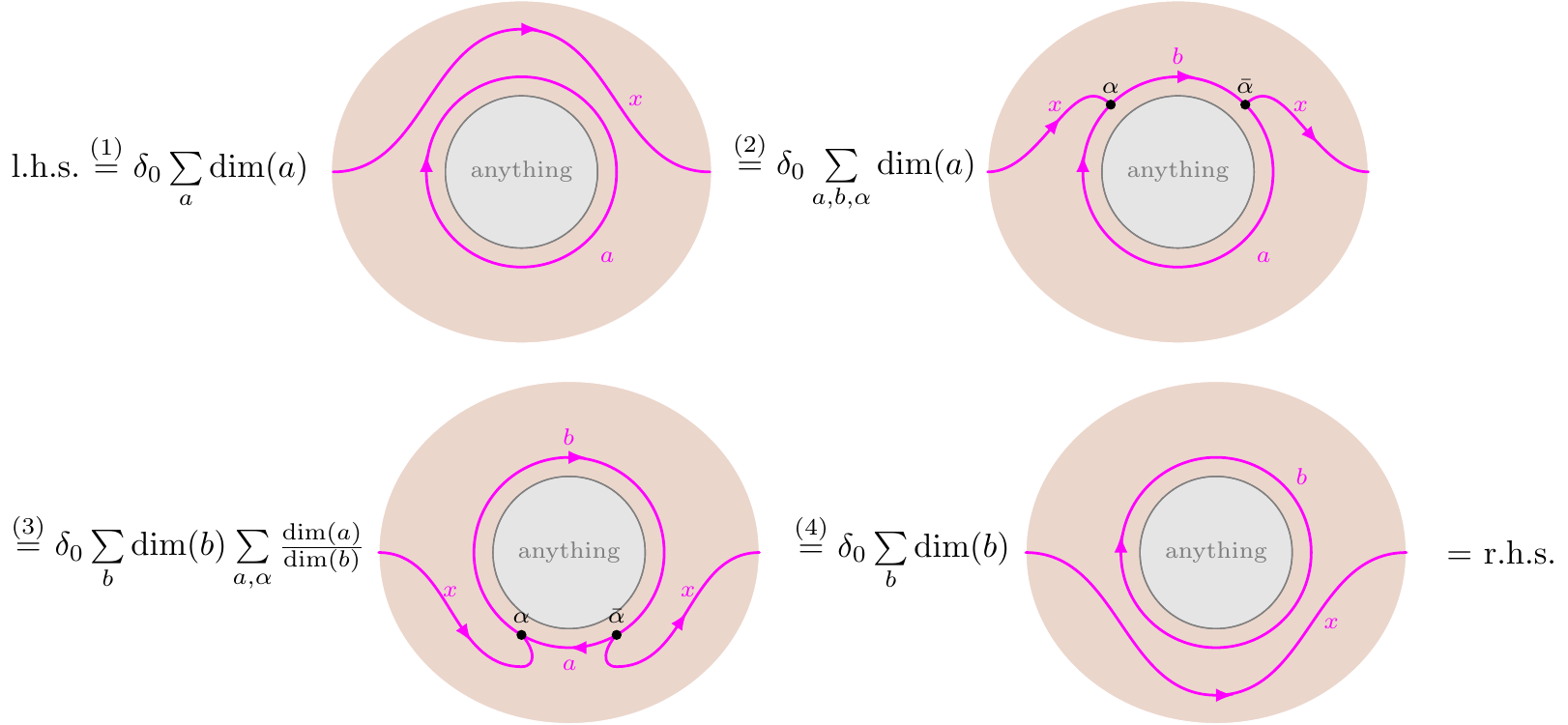}
\label{eq:move-defect-past-cloaking-defect-calc}
\end{equation}

\noindent
In step 1 we substituted the definition of the cloaking defect as in Figure~\ref{fig:cloaking-defect-loop}. Step 2 is the identity \eqref{eq:defect-fusion-same-orientation}. In step 3 we have multiplied by $\dim(b)/\dim(b)$ and deformed the defect lines. In step 4 we use \eqref{eq:defect-fusion-opposite-orientation}.

To get the cloaking boundary condition, we simply fuse the cloaking defect to the boundary condition with label $\one$:
\begin{equation}\label{eq:cloaking-def-cloaking-bc}
\includegraphics[scale=1]{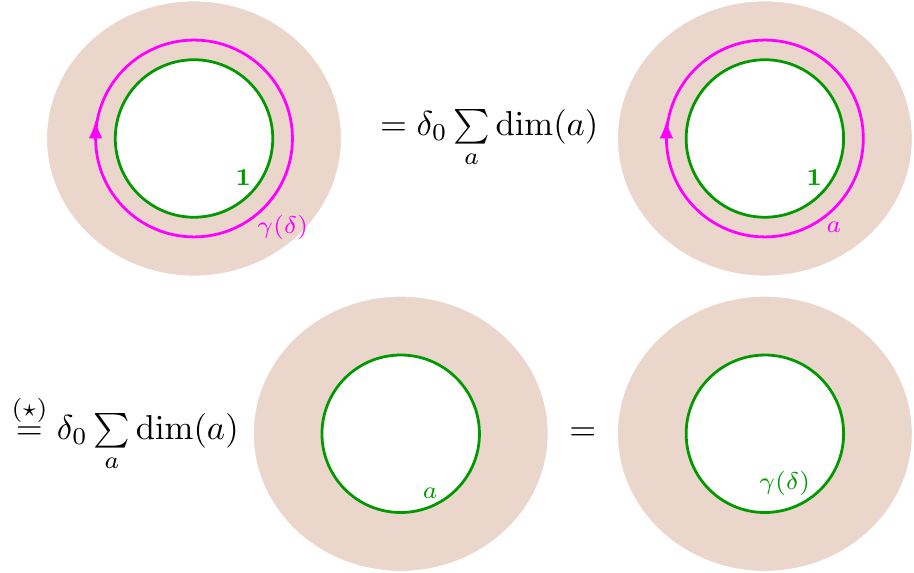},
\end{equation}
where in $(\star)$ we used \eqref{eq:fuse-defect-to-bnd}.
It now follows from \eqref{eq:move-defect-past-cloaking-defect} that the cloaking boundary condition commutes with all topological line defects. That is, for all $x \in I$ the following identity holds inside correlators (only the relevant portion of the surface is shown, the rest does not change):
\begin{equation}
\includegraphics[scale=.9]{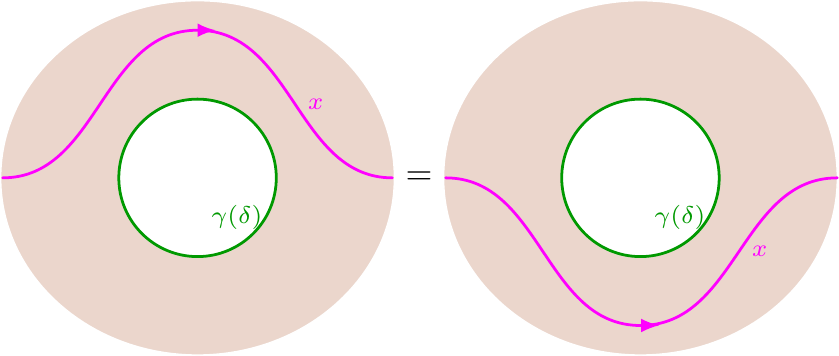}
\end{equation}

\subsection{Non-diagonal models}\label{sec:non-diag-cloak}

The above derivation of the cloaking boundary condition may seem specific to diagonal models because it uses that topological defects and conformal boundary conditions are labelled by the same set $I$. In this section, we want to briefly explain how -- after one extra step -- the same argument applies to non-diagonal models.

In the TFT-approach to CFT correlators \cite{Fuchs:2002cm}, the diagonal model is described by the simple special symmetric Frobenius algebra $\one \in \MC$. Let us denote the diagonal CFT by $C_\one$. All other CFTs which contain the holomorphic and anti-holomorphic symmetry $V$ (and unique vacuum and non-degenerate two-point functions) are described by a choice of simple special symmetric Frobenius algebra $A \in \MC$ \cite{Fjelstad:2006aw,Kong:2009inh}. A useful interpretation is that every CFT with this chiral symmetry is a generalised orbifold of the diagonal model \cite{Frohlich:2009gb}. Denote the resulting CFT by $C_A$. A correlator of $C_A$ is given by embedding a 3-valent network of $A$-defects in the surface and evaluating with $C_\one$. 

\begin{figure}[ht]
\begin{center}
\begin{tikzpicture}[scale=.95]
\node (A) at (0,0) {\includegraphics[width=.25\textwidth]{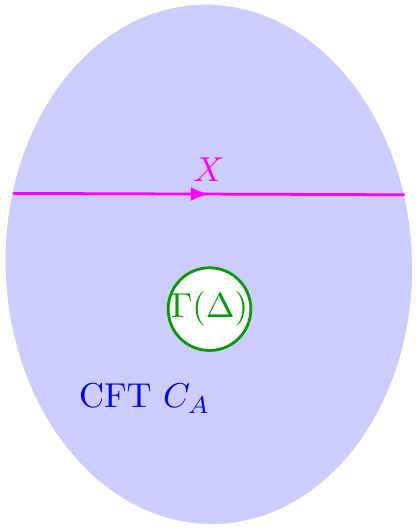}};
\node (B) at (3.7,-2.5) {\includegraphics[width=.25\textwidth]{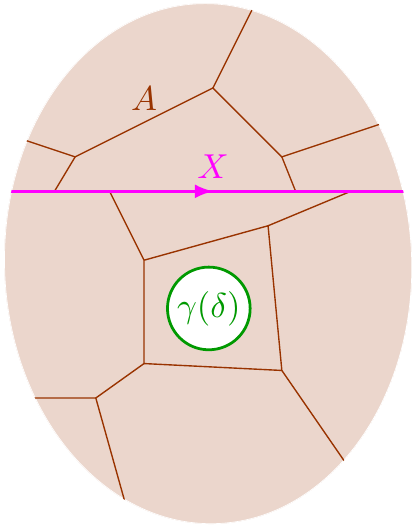}};
\node (C) at (7.4,0) {\includegraphics[width=.25\textwidth]{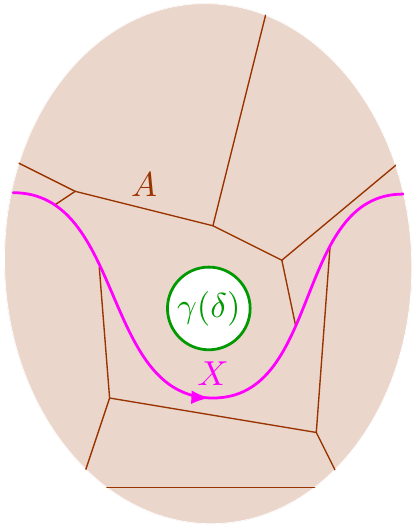}};
\node (D) at (11.1,-2.5) {\includegraphics[width=.25\textwidth]{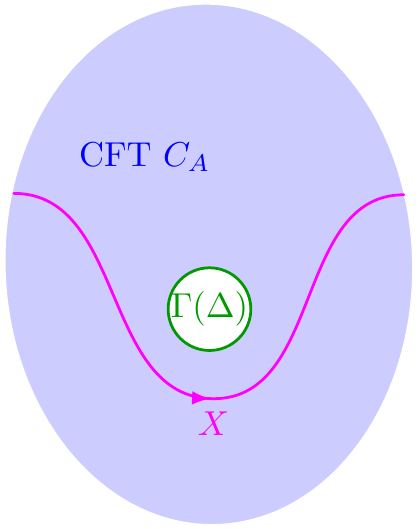}};
\node[rotate=-30] at (1.85,-1.2) {$\stackrel{(1)}{=}$};
\node[rotate=30] at (5.55,-1.2) {$\stackrel{(2)}{=}$};
\node[rotate=-30] at (9.25,-1.2) {$\stackrel{(3)}{=}$};
\end{tikzpicture}
\end{center}
\caption{Moving a topological defect labelled by the $A$-$A$-bimodule $X$ past the cloaking boundary condition $\Gamma(\Delta)$: steps 1 and 3 amount to the definition of correlators of the CFT $C_A$ in terms of those of $C_\one$ in the presence of a network of $A$-defects. In step 2 the cloaking boundary condition $\gamma(\delta)$ for $C_\one$ is used to move the defect line $X$ and a portion of the network of $A$-lines past the boundary.
}
\label{fig:cloaking-bc-nondiag}
\end{figure}
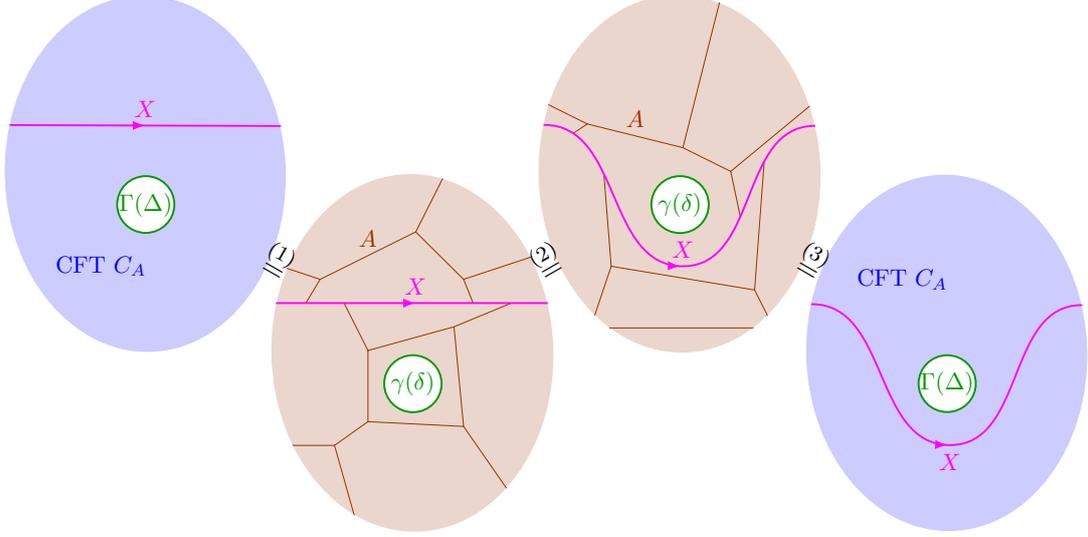

Boundary conditions of $C_A$ are labelled by $A$-modules in $\MC$ and topological defects by $A$-$A$-bimodules in $\MC$ \cite{Frohlich:2006ch}. 
The cloaking boundary condition for $C_A$ is the induced module $\Gamma := A \otimes \gamma$, where $\gamma = \bigoplus_{a \in I} a$ is the cloaking boundary condition of $C_\one$. The weight-zero field insertion on the boundary is $\Delta := id_A \otimes \delta$. The computation in Figure~\ref{fig:cloaking-bc-nondiag} shows that $\Gamma(\Delta)$ commutes with all topological defects of $C_A$.

\subsection{Topological symmetry with truncated state space}\label{sec:truncated-sym}

As explained in Section~\ref{sec:top-sym-cloak-sym}, 
the cloaking boundary condition produces a lattice model that realises the topological line defects of the original CFT (cf.\ Figure~\ref{fig:intro-cloaking-bnd}). A priori, this statement holds for the lattice model with the full infinite set of states indexed by a basis of the open channel state space of the CFT as in \eqref{eq:CFT-to-lattice}. In numerical computations, one needs to truncate this to a finite sum. In this section we show that the truncated model still carries the topological symmetry of the original CFT (though the truncated model may allow for additional topological defects). 
An extreme example of this will be given in the next section, where the model is truncated to only ground states, leading to a 2d\,TFT.

\medskip

The first point to note is that by definition, topological defects are transparent to insertions of the stress tensor $T(z)$ or $\bar T(z)$. Thus we can still move topological defects freely on a surface with cloaking boundary conditions and with an arbitrary number of stress tensor insertions, e.g.\ as in Figure~\ref{fig:intro-cloaking-bnd} but with additional insertion $T(z_i)$ and $\bar T(w_j)$, $i=1,\dots,m$, $j=1,\dots,n$. The topological invariance of the line defects is still preserved if we integrate over the insertion points $z_i$, $w_j$. In particular, we can integrate copies of $T$ and $\bar T$ along closely spaced contours parallel to the gluing edges where the sums over intermediate open states are inserted (cf.\ Figure~\ref{fig:intro-sum-over-states}). Each integral of $T+\bar T$ amounts to inserting an operator which acts as $L_0$ on the corresponding open state (up to an overall factor). Altogether, on each edge of each clipped triangle we can insert an arbitrary number of copies of $L_0$ while still maintaining topological invariance of the topological defect lines, as well as the ability to move them past cloaking boundaries as in Figure~\ref{fig:intro-cloaking-bnd}.

The next step in the argument is that using polynomials in $L_0$, one can approximate spectral projections to arbitrary precision. Thus, for each edge $e$ of each clipped triangle, one can pick a separate projection $P_e$ to part of the $L_0$-spectrum of the open state space $\mathcal{H}_{\gamma\gamma}$ and still maintain invariance with respect to moving topological defect lines. We will restrict ourselves to the situation that we truncate $\mathcal{H}_{\gamma\gamma}$ at some $L_0$-weight $h_\mathrm{max}$.

\medskip

\begin{figure}[t]
	\centering
	\vspace{1em}
	\includegraphics[scale=1]{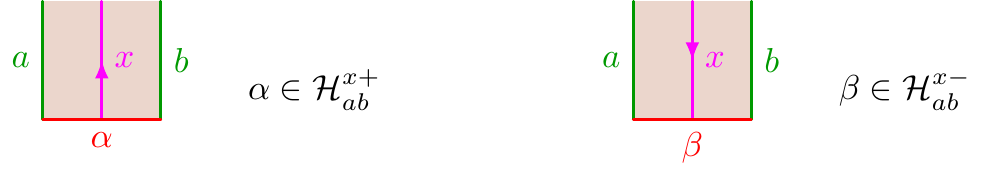}	
	\vspace{-1em}
	\caption{Notation for the open channel state spaces in the presence of a line defect $x$ for the two possible orientations of $x$.}
	\label{fig:open-states-with-def}
\end{figure}

Altogether we arrive at the following lattice model: Let $S$ index an $L_0$-homogeneous ON-basis of $\mathcal{H}_{\gamma\gamma}$ as in \eqref{eq:CFT-to-lattice}, and denote by
\begin{equation}
    S(h_\mathrm{max}) \subset S
\end{equation}
the subset of basis vectors of $L_0$-weight $\le h_\mathrm{max}$.
Write $\mathcal{H}^{x,\pm}_{ab}$ for the open channel states on a strip with boundary conditions $a$ and $b$, depending on the orientation of $x$, see Figure~\ref{fig:open-states-with-def}, and
let $S^{x}$ index an $L_0$-homogeneous basis of $\mathcal{H}^{x+}_{\gamma\gamma}$
(this determines a corresponding dual basis in $\mathcal{H}^{x-}_{\gamma\gamma}$). Finally, let $S^{x}(h_\mathrm{max}) \subset S^{x}$ be the subset with $L_0$-weight $\le h_\mathrm{max}$.

In the depiction of the lattice model via vertices and edges as in Figure~\ref{fig:intro-cloaking-bnd}\,(b), for the truncated model the states on an edge without line defect are now labelled by $S(h_\mathrm{max})$, and states on an edge carrying the line defect $x$ are labelled by $S^{x}(h_\mathrm{max})$. Both these sets are finite so that at this point we have a finite lattice model. The interaction vertices are the same as before (cf.\ Figure~\ref{fig:intro-cloaking-vertex}), the only change is that the set of states has been restricted. For example, in the partition function \eqref{eq:CFT-to-lattice}, the $\sigma$-sum is now only over $\sigma \in S(h_\mathrm{max})^{|E|}$.

By construction, this model satisfies the topological invariance condition in Figure~\ref{fig:intro-cloaking-bnd}\,(b) for any choice of $h_\mathrm{max}$. 
We thus have a two-parameter family of lattice models which realise the topological symmetry given by topological line defects of the original CFT: One parameter is the radius $R$ of the holes, and the other is the cutoff energy $h_\mathrm{max}$ in the open channel state space.

\section{Topological field theory in the touching hole limit}\label{sec:TFT}

In this section, we return to the initial assumption made in the introduction, namely that $C$ is a diagonal unitary rational CFT. We will study the $R \to \frac d2$ limit of $Z_b(\omega_1,\omega_2;R)$ from \eqref{eq:ZalphaR-def}. This limit contains a divergent Liouville factor independent of the boundary condition $b$. We will not treat this factor in the present paper and cancel it by considering the ratio \begin{equation}\label{eq:ZZ-ratio-gives-TFT}
Y(R) 
~:=~ \frac{Z_{\gamma(\delta)}(\omega_1,\omega_2;R)}{\delta_0^{\,MN}\,Z_{\one}(\omega_1,\omega_2;R)} ~.
\end{equation}
The factor of $\delta_0$ to the number of holes is included in the denominator to cancel the corresponding normalisation factor in $\delta$ in the numerator.
The $R \to \frac d2$ limit of $Y(R)$ exists and we will explain in this section that it is given by evaluating a 2d state-sum TFT $C_\mathrm{top}$. We will verify that $C_\mathrm{top}$ realises the topological symmetry of $C$.

\subsection{State-sum TFT from the lattice model}

In the $R \to \frac d2$ limit, only the weight-zero states in the open channel state spaces $\mathcal{H}_{\gamma\gamma}$ (for the numerator of $Y(R)$) and $\mathcal{H}_{\one\one}$ (for the denominator) contribute.\footnote{
	Here we use the unitarity assumption. The leading contribution will come from the $V$-representation of smallest $L_0$-weight, which may be negative in non-unitary theories. The simplest example where this happens is the Lee-Yang minimal model.}
For $\mathcal{H}_{\one\one}$
the subspace $(\mathcal{H}_{\one\one})_0$ of weight-zero fields is one-dimensional and spanned by the identity field $1_\one$ on the $\one$-boundary. For the $\gamma$-boundary,  we have
\begin{equation}
	(\mathcal{H}_{\gamma\gamma})_0
	~=~
	\mathrm{span}\{ \,1_a \,|\, a \in I \,\}
	~=~
	\bigoplus_{a \in I} \Hom_\MC(a,a) ~=:~ A ~,
\end{equation}
where $1_a$ is the identity field on the $a$-boundary (which is a summand in the superposition $\gamma$). For the second equality we used the description of weight-zero fields in terms of morphisms of $\MC$ as  in Section~\ref{sec:cloak-diagonal}.

The vertex $T^{\gamma(\delta)}_R(1_a,1_b,1_c)$ is non-zero only for $a=b=c$. Using the representation of the cloaking boundary condition via the cloaking defect in \eqref{eq:cloaking-def-cloaking-bc}, one obtains:
\begin{equation}
\begin{tikzpicture}
\node at (0,0) {$T^{\gamma(\delta)}_R(1_a,1_b,1_c) =$};
\node at (4,.5) { \includegraphics[scale=.9]{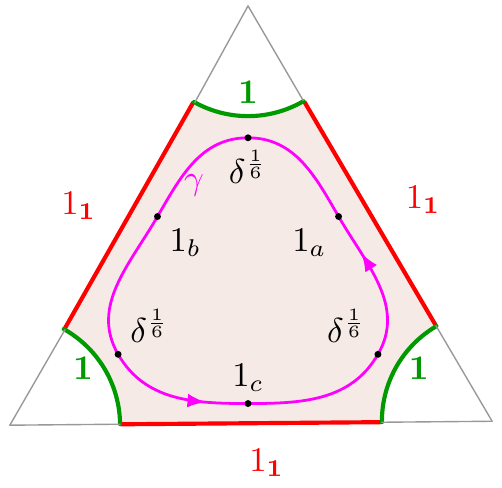}};
\node at (9.5,0) {$= \delta_{a=b=c}\, (\dim a)^{\frac32} \, \delta_0^{\,\frac12}\,T^\one_R(1_\one,1_\one,1_\one) ~.$};
\end{tikzpicture}
\end{equation}
In the ratio $Y(R)$, the factor $\delta_0^{\,\frac12}\,T^\one_R(1_\one,1_\one,1_\one)$ cancels, and the state-sum TFT $C_\mathrm{top}$ assigns the map
\begin{equation}
A^{\otimes 3} \longrightarrow \mathbb{C} \quad , \quad
1_a \otimes 1_b \otimes 1_c \longmapsto \delta_{a=b=c}\, (\dim a)^{\frac32}
\end{equation}
to a triangle.
The value on an edge is an element in $A \otimes A$ which describes the sum over orthonormal states. It is given by:
\begin{equation}
(\mathcal{H}_{\one\one})_0 ~:~
(S_{\one\one})^{-\frac12} 1_\one \otimes 1_\one
\quad , \quad
(\mathcal{H}_{\gamma\gamma})_0 ~:~
(S_{\one\one})^{-\frac12}
\sum_{a \in I}
\frac{1}{\dim a} 
1_a \otimes 1_a ~.
\end{equation}

\noindent 
The factor $(S_{\one\one})^{-\frac12}$ arises from the unnormalised disc-correlator (see Section~\ref{sec:intermediate_state_sum}), but in any case cancels from the ratio \eqref{eq:ZZ-ratio-gives-TFT}.
The factor $(\dim a)^{-1}$ arises from the following identity for line defects passing through an edge of the triangulation,
\begin{equation}
\begin{tikzpicture}
\node at (0,0) {\includegraphics{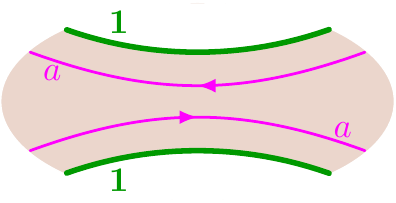}};
\node at (2.8,0) {$=\frac{1}{\text{dim}a}$};
\node at (5.6,0) {\includegraphics{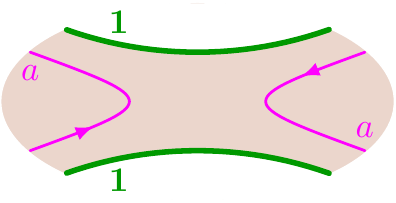}};
\node at (3.5,-2.2) {$+\sum\limits_{z\neq \one} \sum\limits_\alpha\frac{\text{dim}z}{\text{dim}a}$};\
\node at (6.7,-2) {\includegraphics{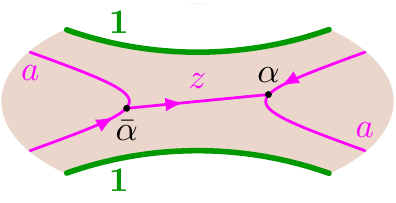}};
\end{tikzpicture}
\end{equation}
where we used \eqref{eq:defect-fusion-opposite-orientation}. In the $R \to \frac d2$ limit, only the first summand on the right hand side survives.
Altogether, to an edge $C_\mathrm{top}$ assigns the copairing
\begin{equation}
b ~=~ \sum_{a \in I} \frac{1}{\dim a} 1_a \otimes 1_a ~ \in A \otimes A~. 
\end{equation}
We can use $b$ to turn the map $A^{\otimes 3} \to \mathbb{C}$ into a map $A^{\otimes 2} \to A$ to obtain the multiplication $\mu$ on $A$. From $b$ one can also read off the appropriate Frobenius counit $\varepsilon : A \to \mathbb{C}$. One finds
\begin{equation}
	\mu(1_a \otimes 1_b) ~=~ \delta_{a,b}\, (\dim a)^{\frac 12}\, 1_a
	\quad , \quad
	\varepsilon(1_a) = (\dim a)^{\frac12}~.
\end{equation}
This is a semisimple Frobenius algebra and it hence defines a state-sum TFT $C_\mathrm{top}$ \cite{Bachas:1992cr,Fukuma:1993hy} (we will use the conventions in \cite{Davydov:2011kb}).
By construction, we have
\begin{equation}
\lim_{R \to \frac d2} Y(R) ~=~C_\mathrm{top}(T^2) ~,
\end{equation}
where $C_\mathrm{top}(T^2)$ is the value of the TFT on a torus. The torus consists of $2MN$ triangles and $3MN$ edges. The multiplication forces all boundary labels to be the same, and we get
\begin{equation}
C_\mathrm{top}(T^2) ~=~ \sum_{a \in I} (\dim a)^{\frac32 \cdot 2MN} (\dim a)^{-3MN} ~=~|I|~.
\end{equation}
This is of course expected as evaluating a TFT on a torus gives the dimension of its state space. For a state sum TFT this is the centre of $A$, but $A$ is already commutative, and $\dim(A)=|I|$.

\subsection{TFT with defect lines}

Next, we turn to the situation where the torus contains defect loops. 
As before, we need to determine the weight-zero part of the open channel state spaces, the interaction vertices, and the normalisation of the sum over intermediate states.

Recall the notation $\mathcal{H}^{x,\pm}_{ab}$ for the open channel states on a strip with boundary conditions $a$ and $b$, cf.\ Figure~\ref{fig:open-states-with-def}. 
We have
\begin{equation}
(\mathcal{H}^{x+}_{ab})_0
~=~
\Hom_\MC(a,x \otimes b)
\quad , \quad
(\mathcal{H}^{x-}_{ab})_0
~=~
\Hom_\MC(x \otimes a,b)~.
\end{equation}
Recall the dual bases we had chosen for these spaces above \eqref{eq:defect-fusion-same-orientation}. To keep track of the indices we write $\alpha_{xa}^b \in (\mathcal{H}^{x-}_{ab})_0$ and $\overline\alpha_a^{xb} \in (\mathcal{H}^{x+}_{ab})_0$. Similar to before, for the two clipped triangles in Figure~\ref{fig:intro-cloaking-vertex}\,b we compute
\begin{equation}
\begin{tikzpicture}
\node at (0,0) {$T^{\gamma(\delta)}_{x \nearrow, R}(\beta_{xb}^a,\overline \alpha_a^{xb},1_b) = 
\delta_0^{\frac12}\,(\dim a)^{\frac16}\, (\dim b)^{\frac 26}$};
\node at (5.5,.5) {\includegraphics[scale=.8]{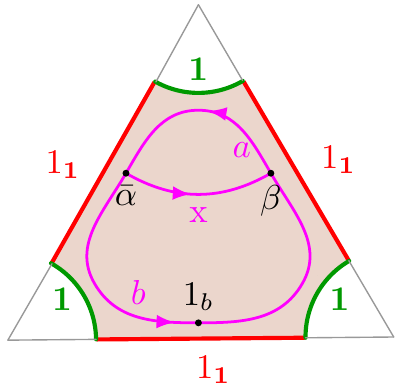}};
\node at (3.15,-1.8) {$= \delta_{\alpha,\beta}\,\delta_0^{\frac12}\,(\dim a)^{\frac76}\, (\dim b)^{\frac 26}
 T^\one_R(1_\one,1_\one,1_\one) $};
\end{tikzpicture}
\end{equation}
and analogously
\begin{equation}
T^{\gamma(\delta)}_{x \searrow, R}(1_a,\overline \alpha_a^{xb},\beta_{xb}^a) =
\delta_{\alpha,\beta}\, 
\delta_0^{\frac12}\,(\dim a)^{\frac76}\, (\dim b)^{\frac 26}
T^\one_R(1_\one,1_\one,1_\one)  ~.
\end{equation}
The weight-zero part of the sum over intermediate states is given by
\begin{equation}\label{eq:TFT-with-defect-copairing}
(S_{\one\one})^{-\frac12}
\sum_{a,b \in I} \frac{1}{\dim a} \alpha^a_{xb} \otimes \overline \alpha_a^{xb}
~\in~ (\mathcal{H}^{x-}_{ab})_0 \otimes (\mathcal{H}^{x+}_{ab})_0~.
\end{equation}
The $(S_{\one\one})^{-\frac12}$ factor will again cancel as before, and the factor $(\dim a)^{-1}$ is obtained from
\begin{equation}
\begin{tikzpicture}
\node at (0,0) {\includegraphics{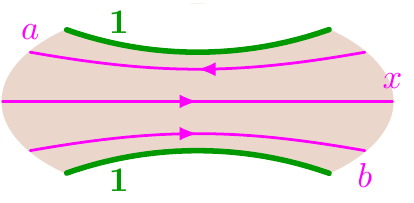}};
\node at (2.8,-.15) {$=\sum\limits_{c,\alpha}$};
\node at (5.5,0) {\includegraphics{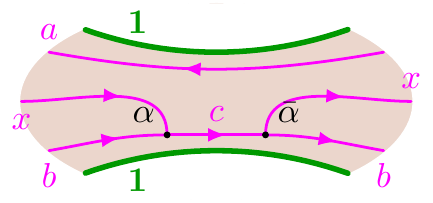}};
\node at (3.5,-2.15) {$=\frac{1}{\text{dim}a}\sum\limits_\alpha$};
\node at (6.5,-2) {\includegraphics{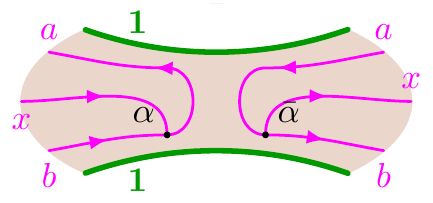}};
\node at (4.3,-4.15) {$+\sum\limits_{c,\alpha}\sum\limits_{z\neq1,\beta}\frac{\text{dim}z}{\text{dim}c}$};
\node at (7.7,-4) {\includegraphics{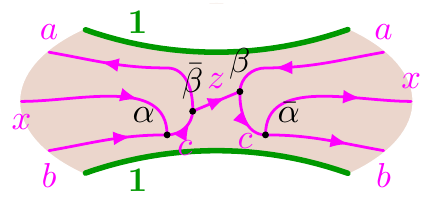}};
\end{tikzpicture}
\end{equation}
where we first used \eqref{eq:defect-fusion-same-orientation} and then \eqref{eq:defect-fusion-opposite-orientation}. The terms with $z \neq \one$ do not survive the $R \to \frac d2$ limit.

These data define particular line defects in the state sum TFT $C_\mathrm{top}$ we obtained above. To see this we use the state sum construction of 2d TFTs with line defects \cite[Sec.\,3]{Davydov:2011kb}. 
Line defects in $C_\mathrm{top}$ are given by finite-dimensional $A$-$A$-bimodules $M$. The relevant bimodule for a line defect labelled $x \in I$ is
\begin{equation}
M_x = \bigoplus_{a,b \in I} \Hom_\MC(x \otimes b,a) ~.
\end{equation}
with left and right action obtained from the corresponding clipped triangles. For $y \in M_x$ one finds $1_a.y.1_b = \text{(const)} \cdot id_a \circ y \circ (id_x \otimes id_b)$, where the identity morphisms project to the corresponding summand of $M_x$.
Computing the proportionality constant involves identifying $\Hom_\MC(a,x \otimes b)$ with $\Hom_\MC(x \otimes b,a)^*$ using the copairing \eqref{eq:TFT-with-defect-copairing}. We skip the details as we do not need the explicit factor here.

We can now verify that the line defects $M_x$, $x \in I$ of $C_\mathrm{top}$ satisfy the same fusion rules as the line defects $x$ in the original CFT $C$: For $x,y \in I$
\begin{align}
M_x \otimes_A M_y 
&= \bigoplus_{a,b,c,d \in I} \Hom_\MC(x \otimes b,a) \otimes_A \Hom_\MC(y \otimes d,c) 
\nonumber\\ &
\cong \bigoplus_{a,b,d \in I} \Hom_\MC(x \otimes b,a) \otimes_{\mathbb{C}} \Hom_\MC(y \otimes d,b) 
\nonumber\\ &
\cong \bigoplus_{a,d \in I} \Hom_\MC(x \otimes y \otimes d,a) 
\nonumber\\ &
\cong \bigoplus_{z \in I} N_{xy}^{~z} \bigoplus_{a,d \in I} \Hom_\MC(z \otimes d,a)
= \bigoplus_{z \in I} N_{xy}^{~z}\, M_z~,
\end{align}
where $N_{xy}^{~z}$ are the fusion rules of $\MC$. 

Thus, $C_\mathrm{top}$ does indeed contain the topological symmetry of $C$. The full topological symmetry of $C_\mathrm{top}$ is $A\text{-bimod}$, the category of finite-dimensional $A$-$A$-bimodules, and we have
\begin{equation}
D_C \hookrightarrow D_{C_\mathrm{top}} = A\text{-bimod}~.
\end{equation}
This is an embedding of fusion categories. However, if $\MC \neq \mathrm{vect}$, i.e.\ if $|I|>1$, this embedding is neither full nor
essentially surjective. Indeed, not every bimodule is a direct sum of the $M_x$, and the $M_x$ are not simple as $A$-$A$-bimodules. In particular we see that an elementary defect $x$ for $C$ is no longer an elementary defect for $C_\mathrm{top}$.

\medskip

This discussion lends some support to the speculative phase diagram in Figure~\ref{fig:speculative-phase-diagram}. In the limit $R \to \frac d2$ we do indeed obtain a TFT $C_\mathrm{top}$ which realises the topological symmetry of $C$. We also saw that in general the topological symmetry of $C_\mathrm{top}$ is strictly larger than that of $C$. 
Note that when interpreted as a CFT, the vacuum of $C_\mathrm{top}$ is $|I|$-fold degenerate, so that it is a superposition of $|I|$ many one-dimensional CFTs. The topological defects connect different summands in this superposition.

\section{Torus correlation functions of descendant fields} \label{sec:TorusDescCorr}

In this and the next section, we describe the closed channel approximation scheme for the partition function of the torus with one hole. 
This scheme is significantly less complicated than the open channel one in Sections~\ref{sec:openstring} and \ref{sec:open-channel-approx} and serves as our reference case. We will give the details in Section~\ref{sec:torus-closed} after reviewing some preliminaries in this section.

We recall some basic definitions and properties of elliptic functions in Section~\ref{sec:elliptic-functions}. This is applied in Section~\ref{sec:torus-correlators} to obtain a recursive formula to express an $n$-point function of descendant fields as a linear differential equation acting on the $n$-point function of the corresponding primary fields. Applying this method to null-vectors results in a differential equation for correlators of primary fields.

\subsection{Elliptic functions}\label{sec:elliptic-functions}

Consider the complex torus 
$T^2 = \mathbb{C}/\Lambda$, where
\begin{equation}
\Lambda  = \Lambda_{\omega_1,\omega_2} = \big\{\,n \omega_1 +m \omega_2\,\big|\, n,m\in \mathbb{Z} \,\big\}
\end{equation}
is a lattice with primitive periods $\{\omega_1,\omega_2\}$ (the \textit{lattice periods}). 
Giving a meromorphic function on $T^2$ is equivalent to giving a meromorphic function on $\mathbb{C}$ which is periodic with respect to $\omega_1$, $\omega_2$. Such functions are called \textit{elliptic}. We will need the first two of Liouville's three theorems on elliptic functions. Let $f : \mathbb{C} \to \overline{\mathbb{C}} = \mathbb{C} \cup \{\infty\}$ be elliptic. Then:
\begin{enumerate}
\item If $f$ has no poles, it is constant.
\item $f$ only has finitely many poles modulo $\Lambda_{\omega_1,\omega_2}$ and the sum of their residues is zero:
\begin{equation}\label{eq:ResIsZero}
\sum_{p\in P} \text{Res}_p(f) ~=~ 0~.
\end{equation}
Here, $P$ is a choice of representatives for the poles of $f$ modulo $\Lambda$.
\end{enumerate}

\subsubsection{Weierstrass functions}

We introduce some special functions that we will use heavily in what follows. 
The elliptic Weierstrass $\wp$ function is defined as
\begin{equation}\label{eq:wp1}
    \wp(z;\Lambda) = \frac{1}{z^2} + \sum_{w\in\Lambda\backslash\{0\}}\left(\frac{1}{(z-w)^2}-\frac{1}{w^2}\right)\,.
\end{equation}

\noindent
In a neighborhood of the origin, the Laurent series expansion of $\wp$ is 
\begin{equation}\label{eq:wp2}
    \wp(z;\Lambda) = \frac{1}{z^2} + \sum_{k=1}^\infty a^{(0)}_k z^{2k}\,.
\end{equation}

\noindent 
The coefficients $a_i^{(0)}$ are polynomials in the so-called \textit{invariants} $g_2$ and $g_3$ of the torus and are given in Appendix~\ref{app:Weierstr}. 
Any derivative $\wp^{(n)}(z) \equiv \partial_z^n \wp(z)$ is elliptic, too. 

The Weierstrass $\zeta$ function is defined through
\begin{equation}\label{eq:zeta-def}
    \frac{d \zeta(z)}{dz} = -\wp(z)\,, \quad \lim_{z\to0}\left(\zeta(z)-\frac{1}{z}\right) =0\,.
\end{equation}

\noindent 
Integrating \eqref{eq:wp1} term by term gives
\begin{equation}\label{eq:wz1}
    \zeta(z;\Lambda) = \frac{1}{z} + \sum_{w\in \Lambda\backslash\{0\}} \left(\frac{1}{z-w} +\frac{1}{w} + \frac{z}{w^2}\right)\,.
\end{equation}

\noindent
The function has a single first order pole on the torus and, hence, cannot be elliptic. However, it is quasi-periodic with
\begin{equation}\label{eq:ZetaPeriod}
    \zeta(z+\omega_i) = \zeta(z) + 
    2 \eta_i \,.
\end{equation}
where $\eta_i$ denotes the $\zeta$-values at half-periods,
\begin{equation}\label{eq:eta-i-def}
    \eta_i = \zeta(\omega_i/2) \,.
\end{equation}

\noindent
Next, we have the Weierstrass $\sigma$ function which is defined by
\begin{equation}
    \frac{\sigma'(z)}{\sigma(z)} = \frac{d \log\sigma}{dz} = \zeta(z)\,,\quad 
    \lim_{z\to0} \frac{\sigma(z)}{z} =1\,.
\end{equation}

\noindent
Integrating \eqref{eq:wz1} term by term gives
\begin{equation}
    \sigma(z;\Lambda) = z\prod_{w\in\Lambda\backslash\{0\}} \left[ \left(1-\frac{z}{w}\right) \exp\!\left(\frac{z}{w}+\frac{z^2}{2\,w^2}\right) \right]\,.
\end{equation}
The periodicity properties of the Weierstrass $\sigma$ function are
\begin{equation}\label{eq:sigmaPeriod}
    \sigma(z+\omega_i) = -e^{2 (z+\omega_i) \eta_i } \sigma(z)\,.
\end{equation}

\subsubsection{Elliptic functions in terms of Weierstrass functions}

Let $f : \mathbb{C} \to \overline{\mathbb{C}}$ be elliptic, and let $P = \{ z_i \,|\, i = 1,\dots,n\}$ be representatives of its poles modulo $\Lambda$. Let $r_i$ be the pole order at $z_i$ and 
\begin{equation}\label{eq:f1}
f(z) = \frac{c_{i,r_i}}{(z-z_i)^{r_i}} + \dots + \frac{c_{i,2}}{(z-z_i)^{2}} + \frac{c_{i,1}}{z-z_i} + \mathcal{O}\!\left(\left(z-z_i\right)^0\right)
\end{equation}
the Laurant series around $z_i$. From the second Liouville theorem we know that $\sum_i c_{i,1} =0$. We claim that
\begin{align}
f(z) &= C + \sum_{i=1}^n \left(c_{i,1} \,\zeta(z-z_i) + c_{i,2} \,\wp(z-z_i) + \sum_{m=2}^{r_i-1} c_{i,m+1} \frac{\zeta^{(m)}(z-z_i)}{(-1)^{m}\,m!} \right)
\nonumber\\
&\equiv C+g(z)
\label{eq:f2}
\end{align}
for some constant $C$. To see this, first note that the Laurent expansion around zero of the $n$-th derivative of the Weierstrass $\zeta$ function is
\begin{equation}
\frac{d^n\zeta(z)}{dz^n} \equiv \zeta^{(n)}(z) = (-1)^n \frac{n!}{z^{n+1}} + \mathcal{O}(z^0)\,.
\end{equation}
The sum in \eqref{eq:f2} is  elliptic because 
    $g(z+\omega_k) - g(z) = - 2 \eta_k  \sum_i c_{i,1} =0$\,, 
and by construction the difference $f(z) - g(z)$ has no poles.
Hence by the first Liouville theorem, the difference is constant.

The constant $C$ can be given in terms of the integral of $f$ along one of the periods.
Let
\begin{equation}
    I_k := \int_u^{u+\omega_k} \hspace{-.5em} f(z)\, dz ~,
\end{equation}
where the integration path is a straight line in $\mathbb{C}$, and where $u$ is chosen such that the contour does not pass through any of the poles of $f$. We have the following integrals for the Weierstrass $\zeta$ and $\wp$ functions:
\begin{align}
    \int_u^{u+\omega_k} \zeta(z-z_i) \, dz &= \log\sigma(\omega_k+u-z_i) - \log\sigma(u-z_i) \nonumber\\
    &\hspace{-.25cm}\stackrel{\eqref{eq:sigmaPeriod}}{=} 
    2 \eta_k  \cdot
    (\omega_k+u-z_i) +i\pi \,,\label{eq:intZeta}\\
    \int_u^{u+\omega_k} \wp(z-z_i)\,dz &= -\zeta(\omega_k+u -z_i) + \zeta(u-z_i) \stackrel{\eqref{eq:ZetaPeriod}}{=}
    - 2 \eta_k \,,
    \label{eq:intWP}\\
    \int_u^{u+\omega_k} \zeta^{(n)}(z-z_i)\,dz &= 0\,. \quad \text{for} ~n>1\,,\label{eq:intHigherDer}
\end{align}

\noindent
Using these and once more $\sum_ic_{i,1} =0$, it follows that
\begin{align}\label{eq:C1}  
    \omega_k C = I_k + \sum_{i} 
    2\eta_k\,
    \left( c_{i,2} +  z_i c_{i,1} \right)\,.
\end{align}

\subsection{Ward identities and descendant fields on the torus}\label{sec:torus-correlators}

We will later only need one-point functions on the torus, but one can treat $n$-point functions without much extra effort, and we describe this more general case now. 

From here on we simplify notation slightly by no longer distinguishing between a point $p$ on the torus $T^2$ and a choice of representative for $p$ on $\mathbb{C}$.

\subsubsection{Ward identity for primary fields on the torus}

Here we review the effect of inserting a single copy of the stress tensor $T(z)$ in a correlator of Virasoro-primary fields on the torus. This has been derived in \cite{Eguchi:1986sb} (for general Riemann surfaces) from the Ansatz that an insertion of the stress tensor is the response of the unnormalised correlator to a point-deformation of the metric. We will take a different route here using complex analysis and the pole structure of correlators. The following can be regarded as a warm-up for the notationally more heavy case of torus correlators with arbitrary descendants. 

Let $L \in \mathbb{R}_{>0}$. 
We assume that the periods $\omega_1$, $\omega_2$ are of the form
\begin{equation}
  \omega_1 = |\omega_1| = L
  \qquad , \qquad  \mathrm{im}(\omega_2)>0 ~. \nonumber
\end{equation}

\noindent 
The local coordinates for field insertions $\phi(z)$ are implicitly taken to be the canonical ones for the complex plane (and hence we will defer the detailed discussion of the effect of local coordinates to Section~\ref{sec:bfields} below).

Fix a CFT $C$ and let $\phi_1,\dots,\phi_N$ be Virasoro-primary fields. 
Let us write 
\begin{equation}\label{eq:torus-correlator-unnormalised}
\big\langle \phi_1(z_1) \cdots \phi_N(z_N)
\big\rangle^\mathrm{un}_{\omega_1,\omega_2}
\end{equation}
for the amplitude of $C$ (the unnormalised correlator) on the torus with periods $\omega_1=L$ and $\omega_2$ and with field insertions $\phi_j(z_j)$. In particular, $\langle 1 \rangle^\mathrm{un}_{\omega_1,\omega_2}$ is just the torus partition function $Z(\omega_1,\omega_2)$ of $C$.
The correlator is then the normalised amplitude, 
\begin{equation}
\big\langle \phi_1(z_1) \cdots \phi_N(z_N) \big\rangle_{\omega_1,\omega_2} ~=~ \frac{\big\langle \phi_1(z_1) \cdots \phi_N(z_N) \big\rangle^\mathrm{un}_{\omega_1,\omega_2}}{\langle 1 \rangle^\mathrm{un}_{\omega_1,\omega_2}}~.
\end{equation}

\noindent
We want to express the amplitude
\begin{equation}
F(z) \equiv F(z,\{z_i\};\omega_1,\omega_2) := \big\langle T(z) \phi_1(z_1) \dots \phi_N(z_N)\big\rangle_{\omega_1,\omega_2}^\mathrm{un}
\end{equation}
on the torus with periods $\omega_1,\omega_2$ in terms of the amplitude
\begin{equation}
P \equiv P(\{z_i\};\omega_1,\omega_2) := \big\langle \phi_1(z_1) \dots \phi_N(z_N)\big\rangle_{\omega_1,\omega_2}^\mathrm{un}
\end{equation}
of just the primary fields. 

We know that $F(z)$ is periodic with respect to $\omega_1$, $\omega_2$, that its only poles are at $z_1,\dots,z_N$, and that these poles are of finite order. Thus $F(z)$ is an elliptic function. The pole structure of $F(z)$ is determined by the operator product expansion
\begin{equation}
    T(z)\,\phi_i(z_i)  ~=~ \frac{h_i \phi_i(z_i)}{(z-z_i)^2} ~+~ \frac{\partial_{z_i}\phi_i(z_i)}{z-z_i} ~+~ \mathcal{O}\!\left(\left(z-z_i\right)^0\right)\,.
\end{equation}
In terms of the notation introduced in \eqref{eq:f1}, this means
\begin{equation}\label{eq:corr-expansion-values}
    c_{i,1} = \partial_{z_i} P
    \quad , \quad
    c_{i,2} = h_i \, P
\end{equation}
for the coefficients of the poles appearing in $F$. 
By \eqref{eq:f2}, the elliptic function $F$ can be expressed in terms of Weierstrass functions as
\begin{align}
    F(z) &= C + \sum_{i=1}^N \Big( c_{i,1}\zeta(z-z_i) + c_{i,2} \wp(z-z_i) \Big)
    \nonumber \\
    &=C + \sum_{i=1}^N \Big( \zeta(z-z_i) \partial_{z_i} + \wp(z-z_i) h_i\Big) P \,.
    \label{eq:T-corr-with-constant}
\end{align}
As in \eqref{eq:C1} we can express the constant as a contour integral. We will integrate along $[0,L]$, assuming that none of the $z_i$ has imaginary part zero modulo $\Lambda$. We claim that
\begin{equation}\label{eq:intT2}
I_1 := \int_0^L F(z) \, dz = 2\pi i \,\frac{d}{d\omega_2} P~.
\end{equation}
To see this, we express the amplitude as a trace on the complex plane, namely\footnote{
    We do not treat Liouville factors in detail in this part of the series. But let us note that in this expression, the anomaly factor $|q|^{c/12}$ arises when changing the metric from the cylinder to the annulus.
    }
\begin{equation}
\big\langle\dots\big\rangle_\tau^\mathrm{un} = \Tr\left[\dots q^{L_0-\frac{c}{24}}\bar{q}^{\bar{L}_0-\frac{c}{24}}\right]\,,
\end{equation}
where $\tau$ is in the upper half plane and $q = e^{2 \pi i \tau}$. The trace is a sum over amplitudes on the complex plane with fields of an orthonormal basis inserted at $0$ and $\infty$. We use the conformal transformation $u = \exp(\frac{2\pi i}{L} z)$ from the torus to an annulus in the complex plane. Writing $\tilde\phi_i(u_i)$ for the transformed fields and $\tau = \omega_2/\omega_1$, we get
\begin{align}
I_1 &= \int_0^L dz \left\langle T(z) \phi_1(z_1) \cdots \right\rangle_{\omega_1,\omega_2}^\mathrm{un} = -\frac{4\pi^2}{L} \frac{1}{2\pi i}\oint du\left\langle\left( u\, T^\text{plane}(u) - \frac{c}{24 u} \right) \tilde \phi_1(u_1) \dots\right\rangle_{\tau}^\mathrm{un}
\nonumber \\
&
= -\frac{4\pi^2}{L} \left\langle\left(L_0-\frac{c}{24}\right)\tilde \phi_1(u_1) \dots\right\rangle_{\tau}^\mathrm{un}
= \frac{2\pi i}{L} \frac{d}{d\tau} \left\langle \tilde \phi_1(u_1) \dots\right\rangle_{\tau}^\mathrm{un}
\,,
\end{align}
and noting that $\frac{d}{d\tau} = L \frac{d}{d\omega_2}$ results in \eqref{eq:intT2}.
Inserting \eqref{eq:corr-expansion-values} and \eqref{eq:intT2} into \eqref{eq:C1} allows us to express the constant $C$ in \eqref{eq:T-corr-with-constant} as
\begin{align}
 C = \frac1L I_1 + \frac1L \sum_{i} 2\eta_1 
 \Big( h_i \, P(\{z_i\}) +  z_i \partial_{z_i} P(\{z_i\}) \Big)\,,
\end{align}
where $\eta_1 = \zeta(\omega_1/2) = \zeta(L/2)$ as in \eqref{eq:eta-i-def}.
Collecting all the terms gives the final Ward identity
\begin{align}
     &\big\langle T(z) \phi_1(z_1) \dots \phi_N(z_N)\big\rangle_{\omega_1,\omega_2}^\mathrm{un} 
     \nonumber
     \\
     &\quad =~ 
     \frac{2\pi i}{L} \frac{d}{d\omega_2} \big\langle\phi_1(z_1) \dots \phi_N(z_N)\big\rangle_{\omega_1,\omega_2}^\mathrm{un}
     \nonumber \\
     &\qquad + \sum_{i=1}^N \left( \zeta(z-z_i)+ \frac{2\eta_1}{L} 
     z_i \right) \partial_{z_i} \big\langle\phi_1(z_1) \dots \phi_N(z_N)\big\rangle_{\omega_1,\omega_2}^\mathrm{un}
     \nonumber \\
     &\qquad + \sum_{i=1}^N\left(\wp(z-z_i)+\frac{2\eta_1}{L}
     \right) h_i \big\langle\phi_1(z_1) \dots \phi_N(z_N)\big\rangle_{\omega_1,\omega_2}^\mathrm{un}\,.
\end{align}

\noindent
For $\omega_1 = L = 1$, this is \cite[Eq.\,(28)]{Eguchi:1986sb} (rewritten for unnormalised correlators).

\subsubsection{Generic descendant correlators on the torus} \label{sec:torusDes}

Torus correlation functions of descendant fields can be computed with the recursive relations of \cite{Zhu:1996,Gaberdiel:2008pr,Gaberdiel:2012yb}. In these references, the relations were derived from the perspective of vertex operators and mode expansions. Here we review the recursion from a complex-analytic point of view.

\medskip

Let $\Phi_i$, $i=1,\dots,N$ be fields of the CFT $C$ which are not necessary primary. Their OPE with the energy momentum tensor is given by 
\begin{equation}\label{eq:TOPE}
    T(z)\,\Phi_i(z_i) = \sum_{k=-M_i}^\infty
\frac{(L_{-k}\Phi_i)(z_i)}{(z-z_i)^{2-k}}\,,
\end{equation}
where $M_i = \text{lvl}(\Phi_i)$ is the level of the descendant field $\Phi_i$. 
The sum truncates at $-M_i$ because $L_n \Phi_i=0$ for $n > M_i$, and so correlators will only involve
finite order poles.
The correlators 
\begin{equation}
    \big\langle T(z) \Phi_1(z_1) \dots \Phi_N(z_N)\big\rangle_{\omega_1,\omega_2}^\mathrm{un} 
\end{equation}
are elliptic functions in $z$. Now consider an elliptic function $g(z)$ whose poles are allowed to coincide with the insertion points. Write the Laurant expansion of $g$ around $z_i$ as
\begin{equation}\label{eq:elliptic-g-factor}
    g(z) = \sum_{m=0}^\infty g_{-n_i+m}^{(i)} (z-z_i)^{-n_i +m}\,.
\end{equation}

\noindent 
The product
\begin{equation}
    F(z) := g(z) \, \big\langle T(z) \Phi_1(z_1) \dots \Phi_N(z_N)\big\rangle_{\omega_1,\omega_2}^\mathrm{un}
\end{equation}
is still an elliptic function. Its Laurent series around $z_i$ is given in terms of \eqref{eq:TOPE} and \eqref{eq:elliptic-g-factor} by
\begin{equation}
   F(z) = \sum_{k=-M_i}^\infty \sum_{m=0}^\infty g_{-n_i+m}^{(i)}\,
    \big\langle(L_{-k}\Phi_i)(z_i)\dots\big\rangle_{\omega_1,\omega_2}^\mathrm{un} \,(z-z_i)^{m+k-n_i-2}\,.
\end{equation}
The residues can be extracted to be
\begin{equation}
    \Res_{z_i}F(z) = \sum_{m=0}^{M_i+n_i+1} g_{-n_i+m}^{(i)} 
    \left\langle (L_{m-n_i-1}\Phi_i)(z_i)\dots\right\rangle_{\omega_1,\omega_2}^\mathrm{un}\,.
\end{equation}
Now assume that $g$ has no poles other than possibly the insertion points. Then, from the second Liouville theorem, it follows that
\begin{equation}\label{eq:master1}
0 ~=~ \sum_{i=1}^N \Res_{z_i}F(z) \\
~=~ \sum_{i=1}^N \sum_{m=0}^{M_i+n_i+1} g_{-n_i+m}^{(i)} \left\langle (L_{m-n_i-1}\Phi_i)(z_i)\dots\right\rangle_{\omega_1,\omega_2}^\mathrm{un}\,.    
\end{equation}

\noindent 
The latter equation relates the correlation function of different descendants with each other. These relations obviously depend on the choice of $g$. 

Let us now make the first choice, namely, 
\begin{equation}
    g(z) = \wp^{(n)}(z-z_i) = \frac{d^n \wp(z-z_i)}{dz^n}\,,\quad n\ge 0\,. 
\end{equation}

\noindent
It has only a single pole at $z=z_i$ of order $n+2$ with Laurent series that follows from \eqref{eq:wp1},
\begin{equation}
    g(z) = \frac{(-1)^{n}(n+1)!}{(z-z_i)^{2+n}} + \sum_{k=0}^\infty a_k^{(n)} (z-z_i)^{2k+\epsilon}\,,
\end{equation}
where $\epsilon = 1$ for odd $n$ and $\epsilon =0$ for even $n$, and $a_k^{(n)} = \frac{(2k+\epsilon+n)!}{(2k+\epsilon)!} a^{(0)}_{k+\frac{n+\epsilon}{2}}$. At any of the other insertion points $z_j\neq z_i$ we get an ordinary Taylor expansion
\begin{equation}
    g(z) = \sum_{k=0}^\infty \frac{\wp^{(n+k)}(z_j-z_i)}{k!} (z-z_j)^k\,.
\end{equation}

\noindent
Using these expansions in \eqref{eq:master1} we can write, for $n \ge 0$,
\begin{align}
    \frac{(n+1)!}{(-1)^{n+1}}\big\langle (L_{-3-n}\Phi_i)(z_i)\dots\big\rangle_{\omega_1,\omega_2}^\mathrm{un} =&\phantom{+} \sum_{k=0}^{\tfrac{M_i+1-\epsilon}{2}} a_k^{(n)} \big\langle (L_{2k+\epsilon-1}\Phi_i)(z_i)\dots\big\rangle_{\omega_1,\omega_2}^\mathrm{un}
     \label{eq:TC1}\\
&+ \sum_{j\neq i} \sum_{k=0}^{M_j+1} \frac{\wp^{(n+k)}(z_j-z_i)}{k!} \big\langle (L_{-1+k}\Phi_j(z_j)\dots\big\rangle_{\omega_1,\omega_2}^\mathrm{un}\,.\nonumber
\end{align}

\noindent 
The total level of each term on the r.h.s. is lower than the one on the left. Hence, we can use it recursively to express a correlator of descendants with $L_{-3-n}$, $n\ge0$\,, in terms of correlators of lower level that only consist of descendants that are built from $L_{-2}$ and $L_{-1}$\,.

To compute a recursive algorithm for descendants with $L_{-2}$'s appearing we need to take a slightly different approach. Because $L_{-2}$ appears in the constant part of the OPE \eqref{eq:TOPE}, to extract it we need to multiply by an elliptic function with a simple pole. However, there is no elliptic function with a single simple pole, so there has to be at least one other simple pole. A choice for such a function is
\begin{equation}
    g(z) = \zeta(z-z_0)-\zeta(z-z_i)\,,
\end{equation}
where $z_0$ is not one of the insertion points. Its Laurent expansion around $z_i$ is given by 
\begin{equation}
    g(z) = -\frac{1}{z-z_i} + \sum_{n=0}\left(\frac{\zeta^{(n)}(z_i-z_0)}{n!} - \zeta_n\right) (z-z_i)^n\,,
\end{equation}
where $\zeta_n$ are the Laurent coefficients of $\zeta(z)$ around $z=0$\,. 
\begin{equation}
    \zeta(z) = \frac1z + \sum_{n=0}^\infty \zeta_n z^n ~.
\end{equation}
They are given by $\zeta_{2k}=0$ and $\zeta_{2k+1}\equiv a_k^{\small(-1)}=-\frac{a_k^{(0)}}{2k+1}$ for $k\in\mathbb{N}$, see \eqref{eq:wp2} and \eqref{eq:zeta-def}.
At the other insertion points $z_j$, $j\neq i$, we get
\begin{equation}
    g(z) =  \sum_{n=0}\frac{1}{n!}\left(\zeta^{(n)}(z_j-z_0) - \zeta^{(n)}(z_j-z_i)\right) (z-z_i)^n\,.
\end{equation}

\noindent
Now using again \eqref{eq:master1}, but with the extra point $z_0$ and $T(z) = (L_{-2}\one)(z)$, we can write
\begin{align}
    \big\langle (L_{-2}\Phi_i)(z_i)\dots\big\rangle_{\omega_1,\omega_2}^\mathrm{un} = ~& \big\langle T(z_0) \dots\big\rangle_{\omega_1,\omega_2}^\mathrm{un} \nonumber\\
    &+ \sum_{n=0}^\infty \left(\frac{1}{n!}\zeta^{(n)}\!(z_i-z_0) - \zeta_n\right) 
    \big\langle (L_{-1+n}\Phi_i)(z_i)\dots\big\rangle_{\omega_1,\omega_2}^\mathrm{un}
   \\
& + \sum_{j\neq i} \sum_{n=0}^\infty \frac{\zeta^{(n)}\!(z_j-z_0) - \zeta^{(n)}\!(z_j-z_i)}{n!} 
\big\langle (L_{-1+n}\Phi_j)(z_j)\dots\big\rangle_{\omega_1,\omega_2}^\mathrm{un}\,.\nonumber 
\end{align}

\noindent
To get rid of the supposedly $z_0$ dependence on the r.h.s. of this equation we integrate $z_0$ over the $\omega_1$ cycle. For the integrals over the Weierstrass functions we use \eqref{eq:intZeta} -- \eqref{eq:intHigherDer} and for the integral over the $T$ insertion we can use \eqref{eq:intT2}, i.e. 
$\int_{z_0}^{z_0+L} dz \left\langle T(z) \dots \right\rangle_{\omega_1,\omega_2}^\mathrm{un} = 2\pi i \frac{d}{d\omega_2} \left\langle\dots\right\rangle_{\omega_1,\omega_2}^\mathrm{un}$. 
Finally, this allows us to write
\begin{align}\label{eq:TC2}
    \big\langle (L_{-2}\Phi_i)(z_i)\dots\big\rangle_{\omega_1,\omega_2}^\mathrm{un} = &\phantom{+} \frac{2\pi  i}{L} \frac{d}{d\omega_2} \big\langle\dots\big\rangle_{\omega_1,\omega_2}^\mathrm{un} 
\nonumber \\
& - \sum_{n=0}^\infty \zeta_n  \big\langle (L_{-1+n}\Phi_i)(z_i)\dots\big\rangle_{\omega_1,\omega_2}^\mathrm{un} 
\nonumber \\
& - \sum_{j\neq i} \sum_{n=0}^\infty\frac{1}{n!} \zeta^{(n)}(z_j-z_i) \big\langle (L_{-1+n}\Phi_j)(z_j)\dots\big\rangle_{\omega_1,\omega_2}^\mathrm{un}
 \\
& + \sum_j \frac{2 \eta_1}{L} \big( h_{\Phi_j} + 
z_j \partial_{z_j}\big) \big\langle \Phi_1(z_1) \dots \Phi_N(z_N)\big\rangle_{\omega_1,\omega_2}^\mathrm{un}\,. \nonumber
\end{align}

\noindent
Using this equation together with \eqref{eq:TC1} recursively and the simple identity $(L_{-1}\Phi_i)(z_i) = \partial_{z_i}\Phi_i(z_i)$ allows one to write any correlator of descendant fields on the torus in term of a differential operator acting on the respective primary correlator on the torus. In the recursion it will generically occur that $\omega_2$ derivatives of the Weierstrass functions and constants need to be evaluated, due to the first term on the r.h.s. of \eqref{eq:TC2}. The needed derivatives are given in Appendix~\ref{app:Weierstr}.

\subsubsection{Recursion formula for one-point functions on the torus}\label{sec:recusion-one-point}

For the applications in this paper, we only need to evaluate torus one-point functions. The above recursion simplifies in this case, and we state this simplified version here for later reference.

Due to translation invariance, $\left\langle(L_{-1}\Phi)(z)\right\rangle^{\mathrm{un}}_{\omega_1,\omega_2} = \partial_z \left\langle\Phi(z)\right\rangle^{\mathrm{un}}_{\omega_1,\omega_2} \equiv0$.
Thus,
any one-point function is independent of the insertion point $z$, 
and we will no longer write out the $z$-dependence to get more compact expressions.
However, it still depends on the periods of the torus.
For $\left\langle L_{-m}\Phi\right\rangle^{\mathrm{un}}_{\omega_1,\omega_2}$ with $m\ge 2$ we define $\epsilon=1$ for even $m$ and $\epsilon=0$ for odd $m$. Then
\eqref{eq:TC1} and \eqref{eq:TC2} result in
\begin{align}
    \big\langle L_{-m}\Phi\big\rangle^{\mathrm{un}}_{\omega_1,\omega_2} &= \frac{(-1)^\epsilon}{(m-2)!} \sum\limits_{k=0}^{\frac{M_\Phi+1-\epsilon}{2}} a_k^{(m-3)} \big\langle L_{-1+2k+\epsilon}\Phi\big\rangle^{\mathrm{un}}_{\omega_1,\omega_2} \nonumber\\
    & ~ + \delta_{m,2} \left[ \frac{2\pi i}{L} \frac{d}{d\omega_2} \big\langle \Phi\big\rangle^{\mathrm{un}}_{\omega_1,\omega_2} + 
    \frac{2\eta_1}{L} 
    h_{\Phi}\big\langle \Phi\big\rangle^{\mathrm{un}}_{\omega_1,\omega_2} \right]\,.
\end{align}

\noindent 
Note that due to the expansions of $\wp(z)$ in only even powers of $z$, the parity of the total holomorphic level is the same in each term. As a consequence, all descendants with odd total holomorphic level have to vanish. That is, for $\phi$ primary and $m_1,\dots,m_N \ge 0$,
\begin{equation}
    \big\langle L_{-m_1} \cdots L_{-m_N} \phi\big\rangle^{\mathrm{un}}_{\omega_1,\omega_2}
    = 0
    \quad \text{if $m_1+\cdots+m_N$ odd .}
\end{equation}

\noindent 
Anti-holomorphic descendants can be computed from the complex conjugated version of the above recursive formula, i.e. 
\begin{align}
    \big\langle \bar{L}_{-m}\Phi\big\rangle^{\mathrm{un}}_{\omega_1,\omega_2} &= \frac{(-1)^\epsilon}{(m-2)!} \sum\limits_{k=0}^{\frac{M_\Phi+1-\epsilon}{2}} \bar{a}_k^{(m-3)} \big\langle \bar{L}_{-1+2k+\epsilon}\Phi\big\rangle^{\mathrm{un}}_{\omega_1,\omega_2} \nonumber\\
    & ~ + \delta_{m,2} \left[ -\frac{2\pi i}{L} \frac{d}{d\bar{\omega}_2} \big\langle \Phi\big\rangle^{\mathrm{un}}_{\omega_1,\omega_2} + 
    \frac{2\bar\eta_1}{L} 
    h_{\Phi}\big\langle \Phi\big\rangle^{\mathrm{un}}_{\omega_1,\omega_2} \right]\,,
\end{align}
where the coefficients $\bar{a}^{(n)}_{k}$ are the expansion coefficients of the respective Weierstrass functions with complex conjugate period $\bar{\omega}_i = \left(\omega_i\right)^\star$ and are given by the complex conjugate of ${a}^{(n)}_{k}$\,. 

Let us give a few explicit examples for a primary field $\phi$ of conformal weight $(h_\phi,\overline h_{\phi})$:
\begin{align}
     \big\langle L_{-2}\phi\big\rangle^{\mathrm{un}}_{\omega_1,\omega_2} &= \frac{1}{L}\left(2\pi i \partial_{\omega_2} 
     + 2\eta_1 h_\phi\right) \big\langle\phi\big\rangle^{\mathrm{un}}_{\omega_1,\omega_2}\,,\label{eq:L-2_1pt}
     \nonumber \\
    \big\langle L_{-2}\bar{L}_{-2}\phi\big\rangle^{\mathrm{un}}_{\omega_1,\omega_2} &= \frac{1}{L^2}\left(4\pi^2 \partial_{\omega_2}
    \partial_{\bar{\omega}_2}  + 4\pi i\left( \bar{\eta}_1 \bar{h}_\phi\partial_{\omega_2}  
    -  \eta_1 h_\phi\partial_{\bar{\omega}_2}\right) + 4\bar{\eta}_1\eta_1 h_\phi \bar{h}_\phi\right) \big\langle\phi\big\rangle^{\mathrm{un}}_{\omega_1,\omega_2}\,,
    \nonumber \\
    \big\langle L_{-4}\phi\big\rangle^{\mathrm{un}}_{\omega_1,\omega_2} &= \frac{g_2 h_\phi}{20} \big\langle\phi\big\rangle^{\mathrm{un}}_{\omega_1,\omega_2}  ~, 
\end{align}
where $g_2$ is the Weierstrass invariant (see Appendix \ref{app:Weierstr}).

\subsubsection{Primary correlation functions from null states}\label{sec:modDiffEq}

In a rational CFT the torus correlation functions of primary fields are solutions to modular differential equations that come from null vectors.
In the state space, null vectors have been identified with zero, so that for each primary state $\phi$ one obtains a polynomial $Q$ of Virasoro-modes that acts as zero on that state. Hence, also the corresponding primary field $\phi(z)$ becomes zero after acting with $Q$, i.e.\ $(Q\phi)(z)=0$. 

On the one hand, inserting $(Q\phi)(z)$ in a correlator gives zero. On the other hand, the above recursive algorithm expresses it as a modular differential operator on the correlator of the primary field $\langle \phi(z) \cdots \rangle_{\omega_1,\omega_2}^\mathrm{un}$.

\subsubsection*{Example: Torus one-point function of the energy field in the Ising CFT}\label{sec:1ptEpsilon}

The Ising CFT is the simplest unitary minimal model. Its field content, modular $S$-matrix and bulk structure constants are given in Section \ref{sec:IsingIntro}. It, in particular, includes the energy field $\epsilon$, a primary field with conformal weights $(\frac12,\frac12)$, with a holomorphic and anti-holomorphic null-state at level two. 

The holomorphic null state is 
\begin{equation}
    \xi=\left(L_{-2} - \frac{3}{4} L_{-1}^2\right) \epsilon\,.
\end{equation}

\noindent 
Due to translation invariance on the torus, we can write
\begin{equation}
    0 = \left\langle \xi \right\rangle_{\omega_1,\omega_2}^\mathrm{un} = \left\langle L_{-2}\epsilon \right\rangle_{\omega_1,\omega_2}^\mathrm{un} \stackrel{\eqref{eq:L-2_1pt}}{=} \left(\frac{ \eta_1}{\omega_1} + \frac{2 \pi i }{\omega_1} \frac{d}{d\omega_2}\right) \left\langle \epsilon \right\rangle_{\omega_1,\omega_2}^\mathrm{un}\,. \label{eq:epsDGL}
\end{equation}

\noindent
Here we note the identity of the Dedekind-$\eta$ function (see e.g. \cite[Eq.\,(5.6)]{Kilford:2008})
\begin{equation}
    \frac{\eta'(\tau)}{\eta(\tau)} = \frac{i\pi}{12} E_2(\tau) \equiv \frac{i}{2\pi} \zeta\left(\frac{1}{2};\Lambda_\tau\right) = \frac{i}{2\pi} \eta_1(\Lambda_\tau)
\end{equation}
where $E_2(\tau)$ is an Eisenstein series\footnote{In contrast to the Eisenstein series' $E_{2k}$ for $k>1$ which are holomorphic forms, $E_2$ is a so-called \textit{almost holomorphic modular form} of weight 2 and level 1.} and $\Lambda_\tau$ is the lattice with periods $\omega_1 = 1$ and $\omega_2=\tau$\,. Using the general scaling behaviour $\zeta(az,a\Lambda) = a^{-1} \zeta(z,\Lambda)$ and $\frac{d}{d\tau} = \omega_1 \frac{d}{d\omega_2}$ one obtains above differential equation.
Hence 
\begin{equation}
     \left\langle \epsilon \right\rangle_{\omega_1,\omega_2}^\mathrm{un} = C_1 \, \eta\!\left(\frac{\omega_2}{\omega_1}\right)\,,
\end{equation}
is the solution to \eqref{eq:epsDGL} with an integration constant $C_1$ that is independent of $\omega_2$ but can still depend on $\omega_1$, $\overline\omega_1$, $\overline\omega_2$.
If we also consider the anti-holomorphic null state at level two we obtain 
\begin{equation}\label{eq:eps-1pt-aux}
     \left\langle \epsilon \right\rangle_{\omega_1,\omega_2}^\mathrm{un} = C_2 \, \eta\!\left(\frac{\omega_2}{\omega_1}\right) \eta\!\left(-\frac{\bar{\omega}_2}{\bar{\omega}_1}\right) \,,
\end{equation}
where the integration constant $C_2$ can still depend on $\omega_1$ and $\bar{\omega}_1$. 
To determine $C_2$, first note the behaviour under rescaling $(\omega_1,\omega_2) \mapsto (\lambda \omega_1,\lambda \omega_2)$ 
for some $\lambda \in \mathbb{C}^\times$:
\begin{align}
\big\langle \epsilon \big\rangle_{\omega_1,\omega_2}^\mathrm{un} = \sqrt{\lambda \bar{\lambda}} \, \big\langle \epsilon \big\rangle_{\lambda\omega_1,\lambda\omega_2}^\mathrm{un} ~,
\end{align}
where the prefactor arises from the transformation of $\epsilon$ which has weight $(\frac12,\frac12)$. This results in $C_2 = C_3 |\omega_1|^{-1}$ for some $C_3 \in \mathbb{C}$. Finally, to fix $C_3$ we look at the limit where only the leading primary contributes to the torus amplitude (expressed as a trace). On the one hand, the asymptotic behaviour of $\langle \epsilon \rangle_{\omega_1,\omega_2}^\mathrm{un}$ obtained from \eqref{eq:eps-1pt-aux} is
\begin{equation}
    \big\langle \epsilon   \big\rangle^\mathrm{un}_{\omega_1=2\pi,\frac{\omega_2}{i}=\beta\gg 1} 
    \approx \frac{C_3}{2\pi} \,e^{-\frac{\beta}{12}}\,.
\end{equation}

\noindent 
On the other hand, in this limit only the $\sigma$-channel contributes, resulting in the asymptotic behaviour
\begin{equation}
    \big\langle \epsilon
    \big\rangle^\mathrm{un}_{\omega_1=2\pi,\frac{\omega_2}{i}=\beta\gg 1} 
    \approx 
    C_{\epsilon\sigma}^{~\sigma} \, e^{-\beta\left(\Delta_\sigma-\frac{1}{24}\right)}\,.
\end{equation}
Since $\Delta_\sigma = \frac18$, the $\beta$-dependence matches.
The OPE coefficients of the Ising CFT are given in \eqref{eq:IsingOPEbulk}:
$C_{\epsilon\sigma}^{~\sigma} = \frac12$, and so $C_3= \pi$. Altogether we now have
\begin{equation}\label{eq:eps-1pt-final}
     \left\langle \epsilon \right\rangle_{\omega_1,\omega_2}^\mathrm{un} = \frac{\pi}{|\omega_1|}\, \eta\!\left(\frac{\omega_2}{\omega_1}\right) \eta\!\left(-\frac{\bar{\omega}_2}{\bar{\omega}_1}\right) \,,
\end{equation}
in agreement with \cite[Eqn.\,(3.13)]{DiFrancesco:1987ez}.

\section{The torus with a single hole: closed channel}\label{sec:torus-closed}

In this section, we give the closed channel approximation scheme for the amplitude of the torus with one hole. We first briefly review the description of boundary states of a diagonal rational CFT at arbitrary radius $R$. Via the state-field correspondence, the boundary can be replaced by a torus one-point correlator which in turn can be evaluated order by order in $R$ via the recursion obtained in Section~\ref{sec:TorusDescCorr}.

\subsection{Boundary states in diagonal rational CFTs}\label{sec:BSdiag}

Let $C$ be a diagonal rational CFT. 
The elementary boundary conditions for $C$ were described in \cite{Cardy:1989ir}, and we state the result to fix our notation. 

The closed channel state space of $C$ is given by $\mathcal{H} = \bigoplus_{i \in I} M_i \otimes \overline M_i$, where $I$ indexes the distinct irreducible representations $M_i$ of the chiral algebra.
In this section, we use the common notation in terms of ket-vectors $\ket{\cdot}$ when we talk about states.  
So let $\ket{s(i)_n}$ be a homogeneous orthonormal basis of the representation $M_i$, indexed by $n$. By homogeneous we mean that each $\ket{s(i)_n}$ is an $L_0$-eigenvector. The \textit{Ishibashi state} $\ket{i}\!\rangle$, $i \in I$, is given by \cite{Ishibashi:1988kg}
\begin{equation}
\ket{i}\!\rangle = \sum_{n} \ket{s(i)_n} \otimes \overline{\ket{s(i)_n}} ~.
\end{equation}
They are elements of $M_i \otimes \overline M_i$, or, to be precise, in the algebraic completion of that space, but we will ignore this point. The Ishibashi states satisfy the conformal invariance condition for a circular boundary of radius $1$ on the complex plane:
\begin{equation}
L_m \, \ket{i}\!\rangle = \overline L_{-m} \, \ket{i}\!\rangle ~.
\end{equation}

\subsubsection*{Example: Ishibashi states for the Virasoro algebra}\label{sec:ExVirIshibashi}

If the chiral algebra is just the Virasoro algebra, we also write $\ket{h}\!\rangle$ for the Ishibashi state of the irreducible representation of lowest conformal weight $h$, and we abbreviate $\ket{h} \equiv \ket{h} \otimes \ket{h} \in M_h \otimes \overline M_h$. For the vaccum module $M_0$, up to total level 4 the Ishibashi-state is
\begin{equation}
\ket{0}\!\rangle 
~=~ 
\ket{0} ~+~ \frac{2}{c}\,L_{-2} \overline L_{-2} \ket{0} ~+~ \dots ~.
\label{eq:vacIshi}
\end{equation}
Here, terms of total level 6 or higher are indicated by ``$\dots$''.
If $h \neq 0$ and if there is no null state at level $2$ in $M_h$, then up to total level 4 the Ishibashi state is given by
\begin{align}
\ket{h}\!\rangle 
~&=~ 
\ket{h} 
+ 
\frac{1}{2h}\,L_{-1} \overline L_{-1} \ket{h} 
+ 
\frac{1}{4 h(1+2h)} L_{-1}^2\overline{L}_{-1}^2 \ket{h} 
+ X\overline{X}  \ket{h} 
+ \dots \nonumber \\
&=~ \ket{h} 
+
\frac{2+4h}{c+2(c-5)h+16h^2}\,L_{-2} \overline L_{-2} \ket{h} 
+
\text{(der.)} + \dots\,.\label{eq:primIshi}
\end{align}

\noindent 
In the third line, ``der.'' stands for terms up to level 4 which start with $L_{-1}$ or $\overline L_{-1}$ and which will vanish in the application below. The mode combination $X \overline X$ is given by
\begin{equation}
X \overline{X} = \frac{2+4h}{(c+2(c-5) h +16h^2)} \left(L_{-2} - \frac{3}{2+4h} L_{-1}^2\right)\left(\overline L_{-2} - \frac{3}{2+4h} \overline L_{-1}^2\right)
\end{equation}

\noindent
If $M_h$ does have a level 2 null state, then this term is absent in $\ket{h}\!\rangle$ and up to total level 4 we have
\begin{equation}\label{eq:primIshi-2null}
\ket{h}\!\rangle = \ket{h} + \text{(der.)} + \dots
\qquad \text{(level 2 null state) .}
\end{equation}

\noindent 
The elementary conformal boundary conditions $|k\rangle$ of $C$ are labelled by irreducible representations $k\in I$ and are given as linear combinations of Ishibashi states:
\begin{equation}\label{eq:Cardy-boundary-state}
\ket{k} = \sum_i  \frac{S_{ki}}{\sqrt{S_{\one i}}} \ket{i}\!\rangle\,.
\end{equation}
Here, $S$ is the modular $S$-matrix that describes the transformation of characters, and the specific linear combinations above are determined by a consistency condition on the annulus (the Cardy condition), we refer to \cite{Cardy:1989ir} for details. All other conformal boundary conditions are non-negative integer linear combinations of these elementary ones. 

To be more specific, the boundary state \eqref{eq:Cardy-boundary-state} describes the state seen by fields on the complex plane with the open unit removed and the unit circle as boundary. The expression for a boundary state describing a hole of radius $R>0$ is given by
\begin{equation}\label{eq:Cardy-boundary-state-anyR}
    \ket{k;R} 
= R^{L_0+\overline L_0 - \frac c6}  \ket{k} \,.
\end{equation}
The operator $R^{L_0+\overline L_0}$ implements the rescaling from radius 1 to radius $R$. The factor $R^{-\frac c6}$ is a Liouville-factor which arises from the extrinsic curvature of the boundary \cite{Cardy:1988tk}.\footnote{
	As another example, for a unit disc the curvature is opposite and the Liouville factor there is $R^{\frac c6}$. In this paper, however, the Liouville factors will not play a role as we will consider ratios of amplitudes, and these factors cancel.}

Let us write $h(i)_k$ for the $L_0$-weight of the state $\ket{s(i)_k}$ of the ON-basis of $M_i$. Substituting the expression of Ishibashi states, the boundary state at radius $R$ reads
\begin{equation}\label{eq:Cardy-boundary-state-anyR-explicit}
\ket{k;R} = R^{- \frac c6} \sum_{i \in I}
 \frac{S_{ki}}{\sqrt{S_{\one i}}}
 \sum_{n} 
R^{2h(i)_{n}}  \ket{s(i)_{n}} \otimes \overline{\ket{s(i)_{n}}}
  \,.
\end{equation}

\noindent
As an aside, this shows that the normalisation factor $\delta_0$ in the cloaking boundary condition \eqref{eq:cloaking-def-cloaking-bc} needs to be set to $\delta_0 = (S_{\one\one})^{3/2}$ in order for its boundary state to produce the expansion \eqref{eq:bnd-state-leading-contrib}: 
$\ket{\gamma(\delta);R} = R^{- \frac c6} ( |0\rangle + \cdots )$.

\subsection{Torus with a hole via expansion of the boundary state}\label{sec:TorusBoundaryExpansion}

To compute the amplitude of a torus with periods $\omega_1,\omega_2$ and with a hole of radius $R$ labelled by the conformal boundary condition $k \in I$, we use the state field correspondence. Namely, we insert the field $V_{\ket{k;R}}(z)$ corresponding to the state \eqref{eq:Cardy-boundary-state-anyR-explicit} at $z=0$ and use the recursion from Section~\ref{sec:recusion-one-point} to compute the resulting amplitude. 

Strictly speaking, $V_{\ket{k;R}}(z)$ is not a field as it contains contributions of arbitrarily high conformal weight.
To be correct, and to be able to do the computation in practice, we include an energy-cutoff $\Delta_\mathrm{max}$ and only sum terms up to combined left/right conformal weight $\Delta_\mathrm{max}$:
\begin{equation}\label{eq:Cardy-boundary-state-V-anyR-explicit}
V_{\ket{k;R},\Delta_\mathrm{max}}(z)
 = R^{- \frac c6} \sum_{i \in I}
\frac{S_{ki}}{\sqrt{S_{\one i}}}
\underset{2h(i)_{n} \le\Delta_\mathrm{max}}{\sum_{n}} 
R^{2h(i)_{n}}  V_{\ket{s(i)_{n}} \otimes \overline{\ket{s(i)_{n}}}}(z)
\,.
\end{equation}

\noindent
Altogether, the amplitude of the torus with one hole and boundary condition $k$ is given by
\begin{align}
Z_k( \omega_1,\omega_2;R ) &= \lim_{\Delta_\mathrm{max} \to \infty} \left\langle V_{\ket{k;R},\Delta_\mathrm{max}}(0)\right\rangle_{\omega_1,\omega_2}^\mathrm{un}\nonumber \\
&=  R^{-\frac{c}{6}}\sum_i \sum_{n} \frac{S_{ki}}{\sqrt{S_{\one i}}} R^{2 h(i)_{n}  } \left\langle V_{ \ket{s(i)_{n}} \otimes \overline{\ket{s(i)_{n}}}}(0)\right\rangle_{\omega_1,\omega_2}^\mathrm{un}
\nonumber \\
&=  R^{-\frac{c}{6}}\sum_{\Delta} C_\Delta R^\Delta  
~.
\label{eq:Zopen1}
\end{align}
Here we use the notation \eqref{eq:torus-correlator-unnormalised} for torus amplitudes. The coefficients $C_\Delta$ of the resulting (fractional) power series in $R$ can be computed order by order via the recursion relation in Section~\ref{sec:recusion-one-point} (recall the convention $\omega_1 = |\omega_1|=L$ used there).
The series is (expected to) converge for radius $R < |\omega_1|/2$, that is, up to the point where the boundary circle touches itself.

If we consider the ratio of two torus amplitudes, the Liouville-divergence cancels and the $R\to 0$ limit is finite. If we assume the CFT $C$ to be in addition unitary, so that the lowest conformal weight is $0$, the limit is
\begin{equation}\label{eq:leading-beh-closed}
\lim_{R\to0}\frac{Z_a(\omega_1,\omega_2;R)}{ Z_b(\omega_1,\omega_2;R)} = \frac{S_{a \one}}{S_{b \one}}\,.
\end{equation}

\subsubsection*{Example: Expansion of the torus amplitude up to fourth order}

For the vacuum Ishibashi state, the primary contribution is given by the torus partition function $\left\langle\one\right\rangle^{\mathrm{un}}_{\omega_1,\omega_2} = Z(\tau,\bar{\tau})$ with $\tau = \omega_2/\omega_1$. From \eqref{eq:vacIshi} and \eqref{eq:L-2_1pt} with $h_\one =\bar{h}_\one =0$ we read off
\begin{align}
    Z_{\ket{0}\!\rangle}(\omega_1,\omega_2;R) = R^{-\frac{c}{6}} \left( Z(\tau,\bar{\tau}) + \frac{2}{c} \frac{4\pi^2}{L^4} \partial_\tau\partial_{\bar{\tau}}Z(\tau,\bar{\tau}) R^{4}+ \mathcal{O}(R^8)\right)\,, \label{eq:ZIshi0}
\end{align}
where we used $\omega_1\partial_{\omega_2} = \partial_\tau$ and $\omega_1 = L$\,.

Next, consider the Ishibashi state $\ket{h}\!\rangle$ with $h \neq 0$ such that $M_h$ has no null state at level 2 as in \eqref{eq:primIshi}, and denote the corresponding primary field by $\phi_h$. Since summands of the form $L_{-1}(\dots)$ and $\overline L_{-1}(\dots)$ contribute zero to the torus amplitude, from \eqref{eq:L-2_1pt} we get
\begin{align}
    Z_{\ket{h}\!\rangle}(\omega_1,\omega_2;R) = R^{-\frac{c}{6} +2 h}\! \left(\left\langle\phi_h\right\rangle^{\mathrm{un}}_{\omega_1,\omega_2} + \tfrac{2+4h}{c+2(c-5)h+16h^2} \!\left(\mathcal{D}\left\langle\phi_h\right\rangle^{\mathrm{un}}_{\omega_1,\omega_2}\right) R^{4} + \mathcal{O}\!\left(R^8\right) \right) \label{eq:ZIship}
\end{align}
with
\begin{equation}
    \mathcal{D}\left\langle\phi_h\right\rangle^{\mathrm{un}}_{\omega_1,\omega_2} = \frac{1}{L^2}\left(4\pi^2 \partial_{\omega_2}\partial_{\bar{\omega}_2}  + 4\pi i h \left( \bar{\eta}_1 \partial_{\omega_2}  -  \eta_1 \partial_{\bar{\omega}_2}\right) + 4\bar{\eta}_1\eta_1 h^2\right) \left\langle\phi_h\right\rangle^{\mathrm{un}}_{\omega_1,\omega_2}\,.
\end{equation}

\noindent 
Note that in case $M_h$ has a null state at level 2, as e.g.\ in the Ising CFT, by \eqref{eq:primIshi-2null} there will be no contribution from total level 4: 
\begin{equation}\label{eq-1h-torus-level2null}
    Z_{\ket{h}\!\rangle}(\omega_1,\omega_2;R) = R^{-\frac{c}{6} +2 h} \left(\left\langle\phi_h\right\rangle^{\mathrm{un}}_{\omega_1,\omega_2}  + \mathcal{O}\!\left(R^8\right) \right)
    \qquad \text{(level 2 null state) .}
\end{equation}

\section{Correlation functions on the clipped triangle}\label{sec:openstring}

In this section, we derive the main player of the lattice model constructed from a CFT, the interaction vertex $T^b_R$. We start by describing the uniformisation map for the clipped triangle and use it to express $T^b_R$ as a disc correlator with deformed local coordinates at the field insertion points, see Figure~\ref{fig:glue-open-uniformise}. Then we give a recursive formula to evaluate correlators of descendant fields on the disc. 

\subsection{Uniformisation of the clipped triangle} \label{sec:uniform}

\begin{figure}[t]
	\centering
	\begin{tikzpicture}
	\node at (0,0) {\includegraphics[scale=.8]{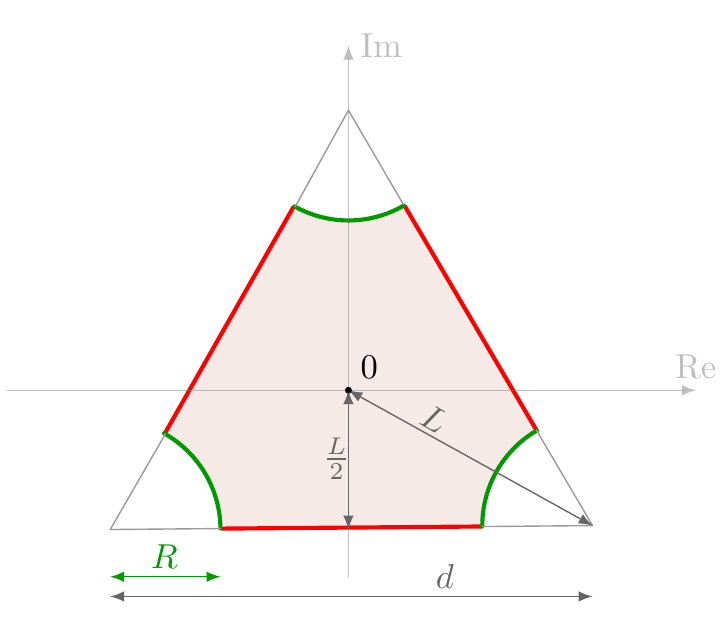}};
	\node at (0,-5.5) {\includegraphics{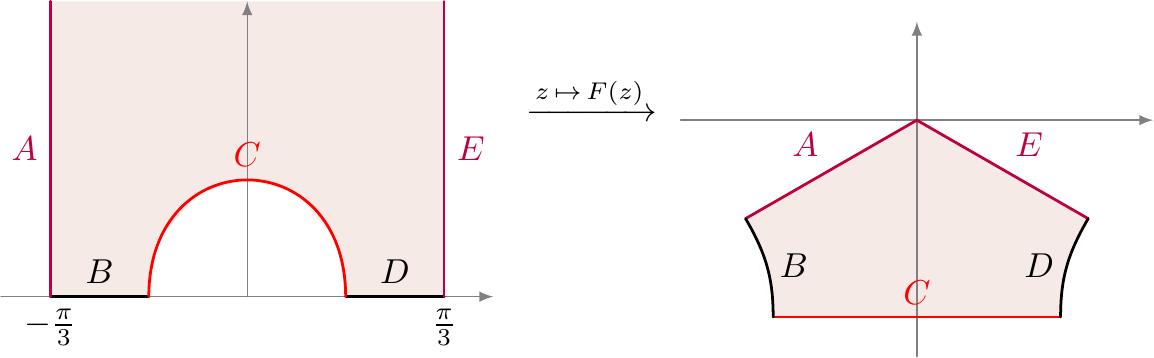}};
	\node at (-1.5,1) {a)};
	\node at (-5.9,-3.5) {b)};
	\end{tikzpicture}	
	\caption{a) Parametrisation of the clipped triangle and its position in the complex plane. b) The uniformising map $F$; the letters indicate which part of the domain is mapped to which part of the target.}
	\label{fig:clipped-triangle-geometry}
\end{figure}

The clipped triangle is parametrised by $L$, the distance from a corner of the equilateral triangle to its centre, and $R$, the radius of the clipping-circle at each corner. The sides of the unclipped triangle have length $d = \sqrt{3} L$. The clipped triangle is positioned in the complex plane as shown in Figure~\ref{fig:clipped-triangle-geometry}\,(a). It will be convenient to express $L,d,R$ in terms of an auxiliary parameter $t \in \mathbb{R}_{>0}$:
\begin{align}
    L(t) &= 2 \sqrt{\pi} \, 
    \frac{\cosh(\pi t)}{\cosh^2(\pi t)-\frac{3}{4}} \,
    \frac{\Gamma\!\left(\frac{7}{6}\right)}{\Gamma\!\left(\frac{5}{6}-i t\right)\Gamma\!\left(\frac{5}{6}+i t\right)}\,,
\nonumber
    \\
    d(t) &= \sqrt{3} \,L(t)\,,
\nonumber
    \\
    R(t) &= \frac{\sqrt{3}}{2} \, \frac{1}{\cosh(\pi t)} L(t)\,. 
\label{eq:L(t)etc}
\end{align}
see Appendix~\ref{sec:obtainF} for details on this and the following expressions. 
For $t\to 0$ one obtains the finite value 
$L(0) = 8\sqrt{\pi}\Gamma(\frac{7}{6}) / \Gamma(\frac{5}{6})^2 \approx 10.32$, 
and for $t\to\infty$ one finds 
\begin{equation}
L(t) = \frac{2\Gamma(\frac{7}{6})}{\sqrt{\pi} \, t^{2/3}} + \mathcal{O}\!\left(\frac{1}{t^{7/3}}\right) \,.
\end{equation}
 More important is the ratio $R/d = \frac{1}{2\cosh{\pi t}}$, which satisfies
\begin{align}
\frac{R(t)}{d(t)}  &\to 0 ~~ \text{ for } t \to \infty
&& \text{(small hole limit)}~,
\nonumber
\\ 
\frac{R(t)}{d(t)}  &\to \frac{1}2 ~~ \text{ for } t \to 0
&& \text{(touching hole limit)}~.
\end{align}

\noindent 
Any other value for the ratio, $0<\frac{R}{d}<\frac{1}{2}$, is realised for
\begin{equation}
    t = \frac{\cosh^{-1}\!\left(\frac{d}{2R}\right)}{\pi}\,.
\end{equation}

\medskip
\noindent 
Next, we give the uniformisation map $F$ for one segment of the clipped triangle as shown in Figure~\ref{fig:clipped-triangle-geometry}\,(b).
Let
\begin{equation}\label{eq:S-domain-F}
    S = \big\{ \, z \in \mathbb{C} \,\big|\, \mathrm{Im}(z) \ge 0 , -\tfrac{\pi}3 \le \mathrm{Re}(z) \le \tfrac{\pi}3 \, \big\}
\end{equation}
be a half-infinite strip. Then $F : S \setminus \{0\} \to \mathbb{C}$ is defined as
\begin{equation}\label{eq:F}
    F(z) ~=~ \frac{1}{i}\left(\frac1i \sin\big(\tfrac{3z}{2}\big)\right)^{-\frac{2}{3}}
    ~
    \frac{_2F_1\left(\frac{5}{12}+\frac{it}{2},\frac{5}{12}-\frac{it}{2},\frac{4}{3};
    \left(\sin\frac{3z}{2}\right)^{-2}
    \right)}{_2F_1\left(\frac{1}{12}+\frac{it}{2},\frac{1}{12}-\frac{it}{2},\frac{2}{3};
    \left(\sin\frac{3z}{2}\right)^{-2}
    \right)} ~.
\end{equation}
We explain in Appendix~\ref{sec:obtainF} how this function was obtained. The relatively simple expression in terms of $t$ is the reason to use $t$ as the fundamental parameter rather than any of $L,d,R$.
For $z = ir$ with $r>0$, the argument of the fractional power $(~)^{-\frac23}$ is real and positive, which is the reason to include the factor of $i$ in this way. For $r$ large enough, also $|\sin\left(\frac{3z}{2}\right)|>1$, so that the power series expressions for the hypergeometric functions converge. This selects the relevant branch of $F$, and the values on all of $S \setminus \{0\}$ are understood as analytic continuation from there.
The function $F$ is not one-to-one on all of $S \setminus \{0\}$, but it is a bijection between the shaded regions indicated in Figure~\ref{fig:clipped-triangle-geometry}\,(b).\footnote{
\label{fn:we-do-not-know}
We note that we cannot actually prove this bijectivity, but we have verified the properties of $F$ numerically. More supporting evidence will be given in Appendix~\ref{sec:obtainF}.}
The red curve in the domain $S$ forming part of the boundary of the shaded region is the preimage under $F$ of the horizontal state boundary of the clipped triangle (or rather the first preimage one encounters when moving in from $i\infty$). We do not know a closed expression for this curve, but we will also not need one.    
Note that as $t$ varies from 0 to $\infty$ the ratio $R/d$ takes all possible values from $\frac12$ to $0$. However, the hole radius $R(t)$ and hole distance $d(t)$ are fixed for specific values of $t$. Hence, if we want to choose some particular values for $d_0$ and $R_0$ we first need to do a global rescaling with $d(t)/d_0$.\footnote{
Due to the curved boundary, this will
introduce a Liouville factor in the partition function as in \eqref{eq:Cardy-boundary-state-anyR}. These factors will drop out in the ratios we consider.
}
\medskip

\begin{figure}[t]
	\centering
	\includegraphics[scale=1]{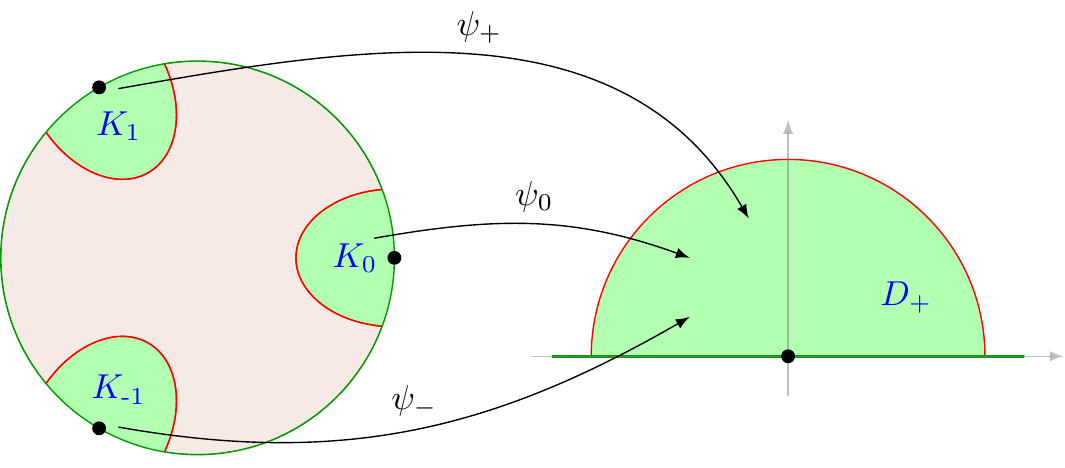}	
	\caption{Closed unit disc $D$ and the regions $K_n$, $n = 0, \pm1$ mapped by $\psi_n$ to the unit half-disc $D_+$. }
	\label{fig:maps-psi_n-from-disc}
\end{figure}

We will not work with three translated copies of the strip $S$ to compute correlators for the clipped triangle, but with the closed unit disc $D$.  Let $\widetilde S$ be the half-infinite strip given by
\begin{equation}
	\widetilde S = \big\{ \, z \in \mathbb{C} \,\big|\, \mathrm{Im}(z) \ge 0 , -\pi < \mathrm{Re}(z) \le \pi \, \big\} ~.
\end{equation}

\noindent
Then
\begin{equation}
	H : D \to \widetilde S
	\quad , \quad
	u \mapsto \frac1i \log(u)
\end{equation}
is a conformal bijection. If we restrict our attention to the wedge $D_\triangleleft$ consisting of points $u \in D$ with $-\frac{\pi}3 \le \mathrm{arg}(u) \le \frac{\pi}3$, then $H$ gives a bijection between $D_\triangleleft$ and $S$ as in \eqref{eq:S-domain-F}.
Let $K_0 \subset D$ be the preimage of the area in $S$ enclosed by the real axis and the curve $C$
as in Figure~\ref{fig:clipped-triangle-geometry}\,(b), see Figure~\ref{fig:maps-psi_n-from-disc}. 
\begin{figure}[t]
	\centering
	\begin{tikzpicture}[>=latex]
	\node (A) at (0,0) {\includegraphics[width=.5\textwidth]{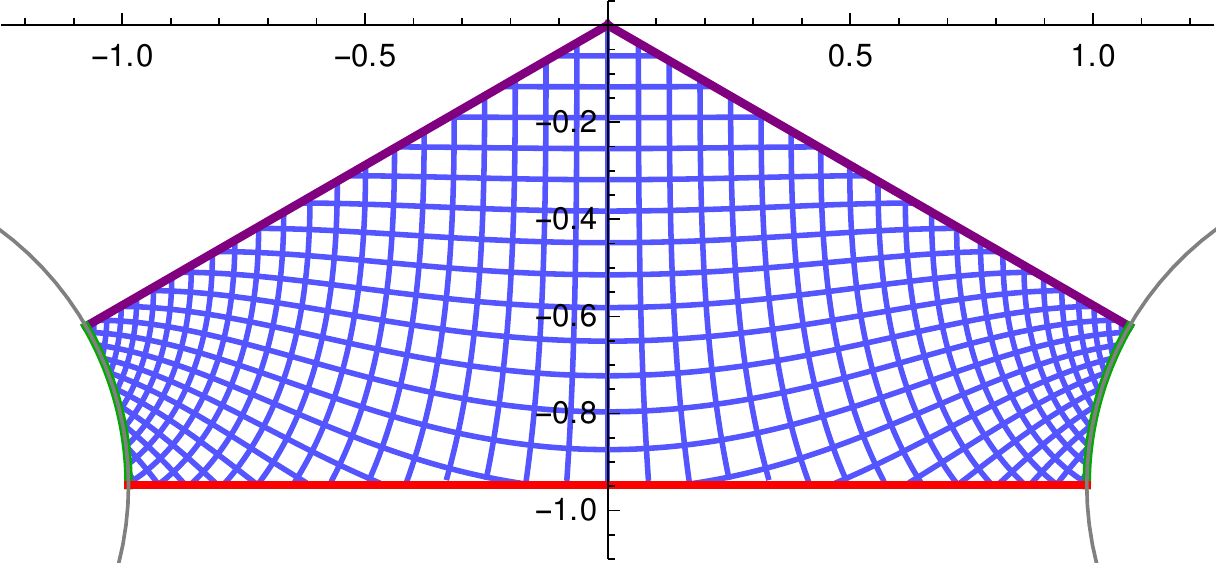}};
	\node (B) at (7,0) {\includegraphics[width=.3\textwidth]{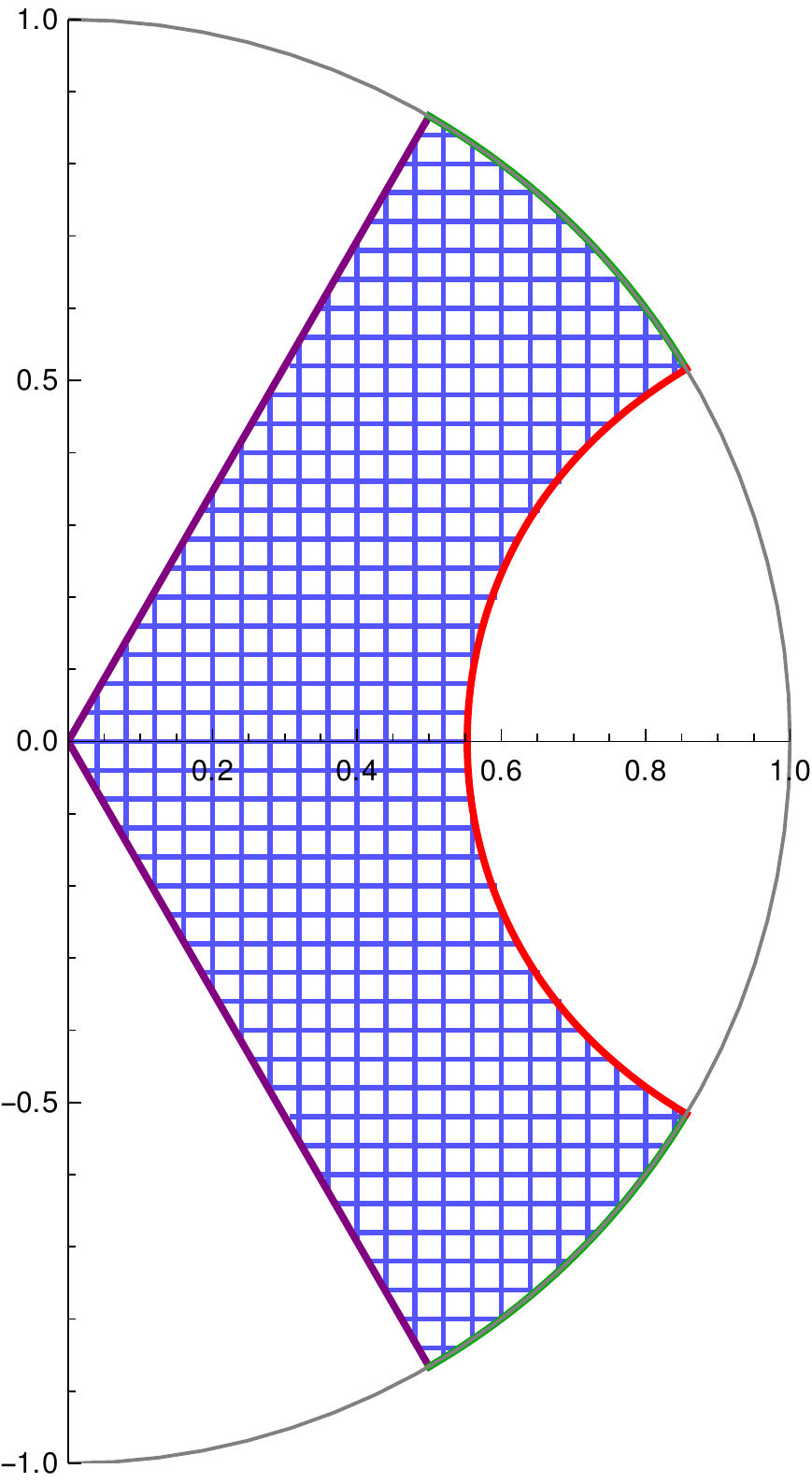}};
	\draw[->] (B) to[out=120,in=60] node [above,sloped]{$\tilde{F}(u)$} (A);
	\end{tikzpicture}
	\caption{Contour plot for the uniformisation $\tilde{F}$ between third of the unit circle to a third of the clipped triangle for $t=0.5$.}
	\label{fig:tildeF-plot}
\end{figure}

As was the case for the curve $C$, we do not have an explicit description for $K_0$, but is also not needed. For the other state boundaries of the clipped triangle, we obtain corresponding rotated copies of $K_0$:
\begin{equation}
	K_n := e^{2 \pi i n/3} K_0
\quad , \quad n \in \{ 0 , \pm1 \} ~.	
\end{equation}
We define the holomorphic map 
\begin{equation}
\tilde F : D \setminus \big( {\textstyle \bigcup_{i=n,\pm 1} K_n} \big) \to \mathbb{C}
~~,~~
\tilde F = F \circ H \,.
\end{equation}
It maps the red shaded area of the disc in Figure~\ref{fig:maps-psi_n-from-disc} to the clipped triangle. If we restrict $\tilde F$ to the wedge $D_\triangleleft$, the image is a third of the clipped triangle as shown in the contour plot in Figure~\ref{fig:tildeF-plot}.

\medskip

\begin{figure}[t]
	\centering	
	\includegraphics[scale=1.1]{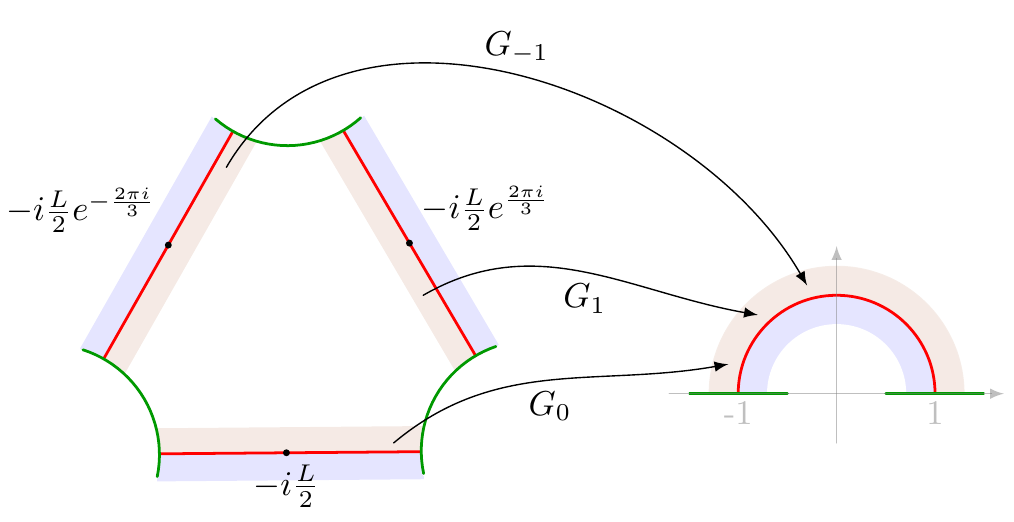}
	\caption{Conformal bijections between neighbourhoods of the three state boundaries and a neighbourhood of the unit half circle.}
	\label{fig:parametrise-state-bnd}
\end{figure}

Next, we turn to the parametrisation of the three state boundaries that we will use to insert a complete sum over intermediate states in Section~\ref{sec:intermediate_state_sum}. These are the line segments of length $d-2R$  with midpoints $-\frac{i}2 L e^{2 \pi i n /3}$, $n = 0, \pm 1$, cf.\ Figure~\ref{fig:parametrise-state-bnd}.
The holomorphic map
\begin{align}
	G_n(w) = i \exp\!\left(\frac{1}{2it} \log\frac{\frac{\sqrt{3}}{2} L \tanh\left(\pi t\right)+\frac{i}{2}L+ e^{-2\pi i n/3}w}{\frac{\sqrt{3}}{2} L \tanh\left(\pi t\right) -\frac{i}{2}L-e^{-2\pi i n/3}w}\right) 
\label{eq:Gn-def}
\end{align}
provides a bijection between a neighbourhood of the state boundary of the clipped triangle with midpoint $-\frac{i}2 L e^{2 \pi i n /3}$, and a neighbourhood of the unit half circle in the upper half plane, see Figure~\ref{fig:parametrise-state-bnd}.
The expression for $G_n$ is derived in Appendix~\ref{sec:G-and-psi}. Here we just note that the midpoint of the state boundary with index $n$ gets mapped to the point $i$ on the unit half-circle,
$G_n(-\frac{i}2 L e^{2 \pi i n /3}) = i$.

Write $\overline{\mathbb{H}} = \{ z \in \mathbb{C} \,|\, \mathrm{Im}(z)\ge 0 \}$ for the upper half plane together with the real line, and $D_+ = D \cap \overline{\mathbb{H}}$ for the upper half of the closed unit disc. We define the conformal maps
\begin{equation}\label{eq:psi_n-def}
	\psi_n : K_n \to D_+
	~~,~~
	\psi_n = G_n \circ F \circ H
	\quad , \quad n \in \{ 0 , \pm1 \} ~.	
\end{equation}
The composition is well-defined in a neighbourhood of the arc separating $K_n$ from the red-shaded inner part of the unit disc 
(Figure~\ref{fig:maps-psi_n-from-disc}), and from there $\psi_n$ is defined by analytic continuation on all of $K_n$.
Together with the abbreviations 
\begin{align}
    s &= \frac{1}{2i} \left( u^{\frac32} - u^{-\frac32} \right) \, ,
\label{eq:s(u)}    
    \\
    X &= \exp\!\left(\frac{1}{2i t} \log\!\left(\frac{\Gamma\!\left(\frac{1}{6}+i t\right)
    \Gamma\!\left(-it\right)}{\Gamma\!\left(\frac{1}{6}-i t\right)
    \Gamma\!\left(it\right)}\right)\right)
    \, ,
    \label{eq:X(t)} 
\end{align}
the map $\psi_0$ can be written as (see Appendix~\ref{sec:G-and-psi})
\begin{equation}\label{eq:psi}
    \psi_0(u) =  X\, \frac{s}{2} \,\exp\!\left(\frac{1}{2 i t} \log\frac{_2F_1\left(\frac{5}{12}+\frac{it}2,\frac{1}{12}+\frac{i t}2,1+i t; s^2\right)}{_2F_1\left(\frac{5}{12}-\frac{i t}2,\frac{1}{12}-\frac{i t}2,1-i t; s^2\right)}\right)
\end{equation}
and $\psi_n(u)$ is given accordingly by $\psi_n(u) = \psi_0(e^{-\frac{2 \pi i n}3} u)$. 
The map $\psi_0$ is regular at $u=1$ with $\psi_0(1) = 0$ and its power series expansion around $1$ is
\begin{align}
    \frac{\psi_0(u)}{i X} &=  -\frac{3}{4}\,(u-1)   +\frac{3}{8}\,(u-1)^2  -\frac{28t^2+43}{256\,(t^2+1)}(u-1)^3 
    \nonumber \\
    &\qquad + \frac{3\,(4t^2 -11)}{512\, (t^2+1)}(u-1)^4 + \mathcal{O}\!\left((u-1)^5\right)\,.
\end{align}
We see that $\psi_0$ is invertible in a neighbourhood of $1$.
Below we will need the inverse functions
\begin{equation}\label{eq:phi_n-def}
\phi_n := \psi_n^{-1} : D_+ \to K_n \,.
\end{equation}
The expansion of $\psi_0$ around $z=0$ is given by
\begin{equation}\label{eq:phi0}
    \phi_0(z) = 1 + \frac{4}{3} \frac{i z}{X} + \frac{8}{9}\left(\frac{i z}{X}\right)^2 + \frac{68 t^2+53}{81(t^2+1)} \left(\frac{i z}{X}\right)^3 +\frac{4\left(44t^2+29\right)}{243(t^2+1)} \left(\frac{i z}{X}\right)^4 + \mathcal{O}\!\left(z^5\right)\,.
\end{equation}

\noindent 
The maps to the other two insertion points at $e^{\pm\frac{2\pi i}{3}}$ are then given by
\begin{equation}
    \phi_n(z) = e^{\frac{2\pi i n}{3}} \phi_0(z) \,.
\end{equation}

\medskip

If we use the maps $G_n$ to glue the three upper-half discs to the three state boundaries of the clipped triangle as in Figure~\ref{fig:parametrise-state-bnd}, the resulting surface $\widetilde D$ has the topology of a closed disc. The various maps defined above provide a uniformisation of $\widetilde D$, that is, a biholomorphic 
bijection $D \to  \widetilde D$. On the blue shaded area of the disc $D$ it is given by $\tilde F$, see Figures~\ref{fig:tildeF-plot} and \ref{fig:maps-psi_n-from-disc}, and on $K_n$ it is given by $\psi_n$.

\medskip
For $t \to 0$ (touching hole limit), the factor $X$ in \eqref{eq:X(t)} diverges. Indeed, the limit of the $\log(\cdots)$-factor in the exponential is $- i \pi$ and we get
\begin{equation}\label{eq:X-t0-limit}
    \log X(t) = \frac{\pi}{2t} + \mathcal{O}(1) 
    \qquad
    \text{for $t \to 0$ .}
\end{equation}

\noindent
On the other hand,
the local coordinate transformations have a well-defined limit for $t \to \infty$ (the small hole limit). Namely by comparing series expansions around $z=0$, one finds 
\begin{equation}\label{eq:logFF-limit}
    \lim_{t \to \infty} 
    \frac{1}{2 i t} \log\!\left( \frac{_2F_1\left(\frac{5}{12}+\frac{it}2,\frac{1}{12}+\frac{i t}2,1+i t; z\right)}{_2F_1\left(\frac{5}{12}-\frac{i t}2,\frac{1}{12}-\frac{i t}2,1-i t; z\right)}\right)
    =
    - \log\!\left(\frac{1+\sqrt{1-z}}2\right)\,.
\end{equation}
Furthermore, $\lim_{t \to \infty} X(t) = 1$ and so
\begin{equation}\label{eq:psiphi-limit}
    \lim_{t \to \infty} \psi_0(u) 
    = i\, \frac{1-u^{\frac32}}{1+u^{\frac32}}
\quad , \quad    
    \lim_{t \to \infty} \phi_0(z) =  \left(\frac{1+iz}{1-iz}\right)^{\frac23}
    ~.
\end{equation}

\subsection{Boundary fields with local coordinates}\label{sec:bfields}

When working with general surfaces that are not embedded in the complex plane in some canonical way, it is not enough to specify the insertion point of a field, but one needs to give a small coordinate neighbourhood. 
Changing the local coordinates can then be traded for acting with Virasoro modes on the field inserted at that point.
In this section, we review this formalism in the case of boundary fields as that is what we will need for the sum over intermediate states. More details on local coordinates for field insertions can be found in \cite{Gaberdiel:1994fs} and e.g.\ in \cite[Sec.\,6.3.1]{Huang:2002mx}, \cite[Sec.\,5.4]{Frenkel:2004jn} or \cite[Sec.\,5.1]{Fuchs:2004xi}. See also \cite{Brehm:2020zri} for a summary and an application with a Mathematica implementation. 

\subsubsection{Local coordinates and transformations}\label{sec:local-coord-bndfield}

Let $\Sigma$ be a surface with boundary $\partial\Sigma$. 
In an approach which goes back to \cite{Cardy:1984bb}, the correlators on $\Sigma$ are expressed in terms of conformal blocks on the doubled surface $\widehat\Sigma$, see e.g.\ \cite{Blau:1987pn} or \cite[Sec.\,6.1]{Fuchs:2004xi} for a more detailed discussion. The conformal double $\widehat\Sigma$ is a complex manifold of (real) dimension two without boundary, together with an anti-holomorphic involution $\iota$. The surface on which the correlator is defined is recovered as the quotient
\begin{equation}
	\Sigma = \widehat\Sigma / \langle \iota \rangle~,
\end{equation}
where $p \in \widehat\Sigma$ is identified with $\iota(p)$. The boundary of $\Sigma$ consists precisely of the fixed points of $\iota$. For example, if $\Sigma = D$ is the unit disc in the complex plane, $\widehat\Sigma$ can be taken to be the Riemann sphere $\widehat\Sigma = \mathbb{C} \cup \{\infty\}$ with $\iota(z)=1/\overline{z}$. The fixed points of $\iota$ form precisely the unit circle. The orientation of $\Sigma$ gives a preferred embedding of $\Sigma$ into $\widehat\Sigma$, and we will often just think of the surface as a sub-manifold (with boundary) of the double. In our example we have $D \subset \mathbb{C} \cup \{\infty\}$.

Let $b$ be a conformal boundary condition and $\mathcal{H}_{bb}$ the space of boundary fields on a $b$-labelled boundary component. A boundary field insertion is a triple
\begin{equation}
	\psi(p;\varphi) ~,
\end{equation}
where $\psi \in \mathcal{H}_{bb}$ is the field being inserted, $p \in \partial\Sigma$ is a point on the boundary of $\Sigma$, and $\varphi : U_0 \to \widehat\Sigma$ is an injective holomorphic map from a neighbourhood $U_0 \subset \mathbb{C}$ of $0$ to $\widehat\Sigma$, such that $\varphi(0)=p$.
In this sense, giving $p$ is redundant, but it is useful to include the actual insertion point in the notation.
For example, take $\Sigma = \overline{\mathbb{H}}$ the closed upper half plane and $\widehat\Sigma = \mathbb{C}$ with $\iota(z)=\overline z$. In the usual notation $\psi(s)$, $s\in \mathbb{R}$ for boundary fields, there is an implied canonical local coordinate given by $\varphi(z) = s+z$.

If one includes the transformation of local coordinates, correlators are actually \textit{invariant} under conformal transformations, rather than just covariant. Namely, let $\Sigma$ and $\Sigma'$ be surfaces and $f : \Sigma \to \Sigma'$ be an (orientation preserving) conformal bijection. In more detail, this means that there is a biholomorphic map $\widehat f : \widehat \Sigma \to \widehat \Sigma'$ such that $\widehat f \circ \iota = \iota' \circ \widehat f$. Then we have the equality of correlators
\begin{equation}\label{eq:transformation-identity-for-corr}
\big\langle 
\psi_1(p_1;\varphi_1) \cdots
\psi_n(p_n;\varphi_n) \big\rangle_{\Sigma}
~=~
\big\langle 
\psi_1\big(f(p_1);\widehat f \circ \varphi_1\big) \cdots
\psi_n\big(f(p_n);\widehat f \circ \varphi_n\big) \big\rangle_{\Sigma'}
\end{equation}
Note that the fields $\psi_1,\dots,\psi_n \in \mathcal{H}_{bb}$ being inserted have not changed, just the insertion points and local coordinates. We will see in a moment why that is not at odds with factors like $f'(p_1)^{h_{\psi_1}}$, etc., which one may have expected. 

The equality in \eqref{eq:transformation-identity-for-corr} only holds for normalised correlators. For unnormalised correlators, there is an additional Liouville factor which we will not treat here.

\medskip

Next, we describe how deforming the local coordinates can be exchanged for acting with Virasoro modes on the field insertion. Here we follow \cite[Sec.\,5.4]{Frenkel:2004jn} and \cite[Sec.\,5.1]{Fuchs:2004xi}. Let $U_0, \tilde U_0$ be neighbourhoods of $0 \in \mathbb{C}$.    
Consider a boundary field $\psi(p;\varphi)$ with $\varphi : U_0 \to \widehat\Sigma$. Let $g : \tilde U_0 \to U_0$ be an injective holomorphic map with $g(0)=0$. Then also $\tilde\varphi :=  \varphi \circ g : \tilde U_0 \to \widehat\Sigma$ is a local coordinate for a field insertion at $p \in \partial\Sigma$. We will now describe a field $\tilde\psi \in \mathcal{H}_{bb}$ such that the identity 
\begin{equation}\label{eq:change-coord-for-Vir}
	\tilde\psi(p;\varphi) ~=~ \psi(p,\tilde\varphi)
\end{equation}
holds in all correlators on $\Sigma$.

As a first step, one needs to solve the equation
\begin{equation}\label{eq:vi-defining-relation}
    v_0 \,\exp\!\left(\sum_{j=1}^\infty v_j \,t^{j+1}\,\partial_t\right)t = g(t)
\end{equation}
for the coefficients $v_j$ order by order in $t$. For example, expanding $g$ as $g(z) = a_1 z + a_2 z^2 + a_3 z^3 + \dots$ one finds
\begin{equation}\label{eq:v0v1v2}
    v_0 = a_1 
    ~,~~
    v_1 = \frac{a_2}{a_1}
    ~,~~
    v_2 = \frac{a_3}{a_1} -\Big( \frac{a_2}{a_1} \Big)^2 
    ~,~
    \dots ~.
\end{equation}
In terms of the $v_j$ one then defines the operator
\begin{equation}\label{eq:Gamma}
    \Gamma_g := 
    v_0^{L_0} \,\exp\!\left(\sum_{j=1}^\infty v_j \,L_j\right)
    =
    \exp\!\left(\sum_{j=1}^\infty \frac{v_j}{v_0^{\,j}}\,L_j\right)\,v_0^{L_0} \,,
\end{equation}
where the second expression is more convenient for our application.
On any given vector of $\mathcal{H}_{bb}$, all but finitely many terms in the infinite sum and in the expansion of the exponential will act as zero.
The transformed field $\tilde\psi$ is given by
\begin{equation}\label{eq:FieldTrafo}
	\tilde\psi ~:=~ \Gamma_g \psi ~.
\end{equation}

\noindent 
Some simple examples for transformed fields and correlators that illustrate the above rule are given in Appendix \ref{app:TrafoExamples}. 

\subsubsection{Local coordinates for the clipped triangle}

Finally, we apply the local coordinate formalism to the clipped triangle in its uniformised form of the unit disc $D$ with three field insertions as in Figure~\ref{fig:maps-psi_n-from-disc}. Recall the maps $\phi_n : D_+ \to K_n$ from \eqref{eq:phi_n-def}.
For conformal boundary condition $b$,
fields $\alpha_0,\alpha_1,\alpha_{-1} \in \mathcal{H}_{bb}$, and hole radius $R = R(t)$, the interaction vertex of the lattice model (or rather its normalised ratio) is given by\footnote{Note the change in the order of the states in the interaction vertex (anti-clockwise, see Figure~\ref{fig:intro-cloaking-vertex}) and on the disc (clockwise). This change of ordering is made such that we have the familiar ordering of fields after transforming to the upper-half plane, from larger to smaller values on the real line. We stress that this is just a matter of notation. Since the insertion points are given, the order in which we write fields in a correlator is immaterial.
\label{fn:order-change}
}
\begin{equation}\label{eq:T-def}
\frac{T_R^b(\alpha_0,\alpha_1,\alpha_{-1})}{T_R^b(\one,\one,\one)}
~=~ 
\Big\langle 
\,\alpha_1\big(e^{\frac{2 \pi i}3};\phi_1\big)
\,\,\alpha_0\big(1;\phi_0\big)
\,\,\alpha_{-1}\big(e^{-\frac{2 \pi i}3};\phi_{-1}\big)
\,\Big\rangle_{\!D_b} ~.
\end{equation}
Here we implicitly used the state-field correspondence, taking the same vector space $\mathcal{H}_{bb}$ to describe states and fields, and expressed the vertex in terms of field labels.

To compute the correlator on the right hand side, we need to change the local coordinates to standard ones. The standard coordinate $\sigma_s$ at the point $e^{2 \pi i s}$ on the unit disc we will use is given by a simple rotation which makes the image of the real axis tangent to the boundary,
\begin{equation}\label{eq:disc-standard-ciird}
\sigma_s(z) := e^{2 \pi i s} (1+ i z) ~.
\end{equation}
Then
\begin{equation}\label{eq:phi_n-coord-deform}
\phi_n(z) = \sigma_{n/3} \circ g_n(z) 
\quad , \quad
\text{with}
~~
g_n(z) = \frac1i \Big( e^{-\frac{2\pi i n}3} \phi_n(z)-1 \Big)
= \frac1i \big( \phi_0(z)-1 \big)~.
\end{equation}

\noindent 
From \eqref{eq:phi0} the first few expansion coefficients can be read off to be
\begin{equation}
    g_n(z) = \frac{4}{3X} z + \frac{8 i}{9 X^2}z^2 - \frac{68 t^2+53}{81 X^3(t^2+1)} z^3 -\frac{4i\left(44t^2+29\right)}{243 X^4(t^2+1)} z^4 + \mathcal{O}\left(z^5\right)\,,
\end{equation}
with $X$ given in \eqref{eq:X(t)}. The first few coefficients of the resulting vector in \eqref{eq:vi-defining-relation} are
\begin{align}
    v_0 &= \frac{4}{3X}\,, &v_1 &= \frac{2 i}{3 X}\,,
    \nonumber\\
    v_2 &= -\frac{5}{108 X^2} \frac{1+4t^2}{1+t^2}\,,& v_3 &= \frac{5 i}{324 X^3} \frac{1+4t^2}{1+t^2}\,.
\end{align}
Up to level $-3$ the coordinate change operator in \eqref{eq:Gamma} is given by\footnote{
  We observed that up to at least level 13, once one expands the exponential, apart from $L_1$ no odd modes of the Virasoro algebra appear in the ordering we use (neither individually nor as a factor in a product of $L_n$-modes). In the expression here this manifests itself in the absence of the $L_3$-term. This requires non-trivial cancellations for which so far we have no conceptual explanation. 
}
\begin{equation}\label{eq:Gamma_h}
\begin{split}
    \Gamma &= 
    \exp\!\left(\frac{v_1}{v_0} L_1 + \frac{v_2}{v_0^{\,2}} L_2 + \frac{v_3}{v_0^{\,3}} L_3 +\cdots\right) \left(\frac{4}{3X}\right)^{L_0} 
    \\
    &= \Big(1 + \frac{i}{2} L_1 - \frac{5}{192} \frac{1+4t^2}{1+t^2} L_2 - \frac{1}{8} L_1^2
    \\   
    &\qquad\qquad\qquad
    -\frac{5 i}{384}\frac{1+4t^2}{1+t^2} L_1 L_2 -\frac{i}{48} L_1^3 + \cdots\Big)
    \left(\frac{4}{3X}\right)^{L_0}
\,.
\end{split}
\end{equation}
The action of $\Gamma$ on a primary state $|h \rangle$ and its first few descendants is:
\begin{align}
    \Gamma |h \rangle &= \left(\frac{4}{3X}\right)^{\!h} |h \rangle
    \nonumber\\
    \Gamma L_{-1} |h \rangle &= \left(\frac{4}{3X}\right)^{\!h+1} \Big( L_{-1} |h \rangle + i h |h \rangle \Big)
    \nonumber\\
    \Gamma (L_{-1})^2 |h \rangle &= \left(\frac{4}{3X}\right)^{\!h+2} \Big( (L_{-1})^2 |h \rangle+ i(2h+1)L_{-1} |h \rangle - \frac{h (21 + 36 t^2 + 32 h (1 + t^2))}{32 (1 + t^2)} |h \rangle \Big)
    \nonumber\\
    \Gamma L_{-2} |h \rangle &= \left(\frac{4}{3X}\right)^{\!h+2} \Big( L_{-2} |h \rangle+ \tfrac32 i L_{-1} |h \rangle - \frac{5 c (1 + 4 t^2) + 8 h (41 + 56 t^2)}{384 (1 + t^2)} |h \rangle \Big)
\label{eq:Gamma-action-one-two-levels}
\end{align}

\noindent 
Altogether, in terms of the standard local coordinates $\sigma_s$ on the disc, the ratio of interaction vertices reads
\begin{equation}\label{eq:T-def-standard coord}
\frac{T_R^b(\alpha_0,\alpha_1,\alpha_{-1})}{T_R^b(\one,\one,\one)}
~=~ 
\Big\langle 
\,\tilde\alpha_1\big(e^{\frac{2 \pi i}3};\sigma_{1/3}\big)
\,\,\tilde\alpha_0\big(1;\sigma_0\big)
\,\,\tilde\alpha_{-1}\big(e^{-\frac{2 \pi i}3};\sigma_{-1/3}\big)
\,\Big\rangle_{\!D_b} ~,
\end{equation}
with the transformed fields given by
\begin{equation}\label{eq:tilde-alpha}
\tilde\alpha_n = \Gamma \alpha_n\,.
\end{equation}

\subsection{Correlation functions on the disc}\label{sec:disc-recursion-and-vertex}

In this section, we give a recursive formula to compute the disc correlators \eqref{eq:T-def-standard coord} which allows one to reduce correlators of descendant fields to those of the corresponding primaries. We proceed in three steps. First, we give the correlators of two and three boundary fields on the upper half plane in terms of structure constants as the reference case. Then we transform this result to the disc and, finally, we show how to move Virasoro modes between the insertion points on the disc.

\subsubsection{Correlators of primary fields on the disc}\label{sec:primary-corr-disc}

The boundary structure constants describe the contribution of primary fields in the operator product expansion of primary boundary fields. For a basis $\{ \psi_i \}$ of primary fields in $\mathcal{H}_{bb}$ and for $x>y$ we have
\begin{equation}
    \psi_i(x) \psi_j(y) ~=~ \sum_{k} c_{ij}^{(b)\,k} (x-y)^{h_k-h_i-h_j} \psi_k(y) ~+~ \text{(descendant fields)} ~.
\end{equation}
Here it is understood that for all fields one uses the standard local coordinates $\varphi_s(z) = s+z$ around an insertion point $s \in \mathbb{R}$. The normalised two and three point correlators on the closed upper half plane $\overline{\mathbb{H}}_b$ with boundary condition $b$ are given by, for $r>s>t$,
\begin{align}
    \big\langle \psi_i(r) \psi_j(s) \big\rangle_{\overline{\mathbb{H}}_b} &~=~ c^{(b) \, \one}_{ij}
    \, (r-s)^{-h_i-h_j}  ~,
    \nonumber \\
    \big\langle \psi_i(r) \psi_j(s) \psi_k(t) \big\rangle_{\overline{\mathbb{H}}_b} &~=~ c^{(b)}_{ijk} \,
    (r-s)^{h_k-h_i-h_j} 
    (r-t)^{h_j-h_i-h_k} 
    (s-t)^{h_i-h_j-h_k} ~,
\label{eq:UHP-2pt3pt}
\end{align}
where
\begin{equation}
    c^{(b)}_{ijk} := \sum_l c^{(b) \, l}_{jk} c^{(b) \, \one}_{il}  ~.
\end{equation}
Since $\psi_i,\psi_j$ are primary, the two-point correlator can be nonzero only if $h_i=h_j$.

Next, we use this result to compute boundary two- and three-point correlators on the disc with local coordinates $\sigma_s$ from \eqref{eq:disc-standard-ciird} at the insertion points. To move from the disc to the upper half plane we use the Möbius transformation
\begin{equation}\label{eq:Moebius-to-UHP}
f : D \setminus \{-1\} \to \overline{\mathbb{H}}
\quad , \quad
u \mapsto 2i \,\frac{1-u}{1+u} ~.
\end{equation}
Let $\frac12>r>s>t>-\frac12$.
For the two-point correlator one finds
\begin{align}
	&\big\langle \psi_i(e^{2\pi i r};\sigma_r) \psi_j(e^{2\pi i s};\sigma_s) \big\rangle_{D_b}
	\nonumber\\
	&=
	\big\langle \psi_i(f(e^{2\pi i r}); f \circ \sigma_r) \psi_j(f(e^{2\pi i s});f \circ \sigma_s) \big\rangle_{\overline{\mathbb{H}}_b}
	\nonumber\\
	&= (\cos(\pi r))^{-2h_i}(\cos(\pi s))^{-2h_j} 
	\big\langle \psi_i(2\tan(\pi s)) \psi_j(2\tan(\pi r)) \big\rangle_{\overline{\mathbb{H}}_b}
	\nonumber\\
	&= c^{(b) \, \one}_{ij} \big(2\sin \pi(r-s)\big)^{-2 h_i} ~,
\end{align}
where in the last step we used that $h_i=h_j$ if $c^{(b) \, \one}_{ij} \neq 0$.
Analogously for the three-point correlator one computes
\begin{align}
	&\big\langle \psi_i(e^{2\pi i r};\sigma_r) \, 
	\psi_j(e^{2\pi i s};\sigma_s) \,
	\psi_k(e^{2\pi i t};\sigma_t) 
	\big\rangle_{D_b}
	\nonumber\\
	&= c^{(b)}_{ijk}  
	    \big(2\sin \pi(r-s)\big)^{h_k-h_i-h_j} 
    \big(2\sin \pi(r-t)\big)^{h_j-h_i-h_k} 
    \big(2\sin \pi(s-t)\big)^{h_i-h_j-h_k} ~.
\label{eq:disc-3pt-arbitrary}
\end{align}

\subsubsection{Ward identities for descendant fields on the disc}

We start by treating a subtlety in the action of Virasoro modes. Namely, for $\phi \in \mathcal{H}_{bb}$ (not necessarily primary), on the one hand one can consider the field $L_n \phi$ inserted at the point $p := e^{2\pi i s}$ with local coordinate $\sigma_s$, i.e.\ $(L_n \phi)(p;\sigma_s)$. On the other hand, one can consider the operator $\widehat{L}_n(p)$ given by contour integration of $T$ around $p$:
\begin{equation}
\widehat{L}_n(p) \phi(p;\sigma_s)
:=
\frac{1}{2 \pi i} \oint_{\gamma_p} (z-p)^{n+1} T(z;\varphi_z) \phi(p;\sigma_s) dz ~,
\end{equation}
where $\gamma_p$ is a small circular contour running anti-clockwise around $p$.
Note that the $z$-integration leaves the unit disc $D$. Here we implicitly use that $\mathbb{C} \cup \infty$ is the conformal double of $D$, and insertions of $T$ outside of $D$ are to be understood as insertions of $\bar T$ inside $D$. We will use this identification frequently in the computations below, and we will only work with $T$, not $\bar T$.

The above two descendants of $\phi$ are related by
\begin{equation}\label{eq:Ln-Ln-relation}
\widehat{L}_n(p) \phi(p;\sigma_s) = \big( i e^{2 \pi i s} \big)^n (L_n \phi)(p;\sigma_s) ~,
\end{equation}
and we will briefly sketch how to see this. The field $L_n \phi$ is given by a contour integral with respect to the standard local coordinates on $\mathbb{C}$ as
\begin{equation}
(L_n \phi)(0;\varphi_0)
:=
\frac{1}{2 \pi i} \oint_{\gamma_0} u^{n+1} T(u;\varphi_u) \phi(0;\varphi_0) du ~.
\end{equation}
Transforming with $\sigma_s$ and setting $z = \sigma_s(u)$ gives $T(\sigma_s(u);\sigma_s \circ \varphi_u) = (i e^{2\pi i s})^2 T(z;\varphi_z)$. Furthermore, $u = (z-p)/(i e^{2 \pi i s})$ and so $du = (i e^{2 \pi i s})^{-1} dz$. Collecting factors produces \eqref{eq:Ln-Ln-relation}.

\medskip

We can now proceed very similarly to the torus case to obtain relations between correlators of descendant fields. Let $p_1,\dots, p_N$ be pairwise distinct points on the unit circle and $\phi_1,\dots,\phi_N \in \mathcal{H}_{bb}$. 
Consider a meromorphic function $\rho(z)$ that has singularities at most at $z\in \left\{p_1,\dots , p_N \right\}$. Let us make the particular choice
\begin{equation}
    \rho(z) = \prod_{i=1}^N (z-p_i)^{a_i}
\end{equation}
for $a_i\in\mathbb{Z}$. Contour deformation results in the following integral identity,
\begin{equation}\label{eq:disc-integral-identity}
    \sum_{i=1}^N \oint_{\gamma_{z_i}} \frac{dz}{2\pi i} \rho(z) \Big\langle T(z) \phi_i(p_i)  \prod_{j\neq i} \phi_j(p_j) 
    \,\Big\rangle_{\!D_b}
    = - \oint_{\gamma_\infty} \frac{dz}{2\pi i} \rho(z) 
    \,\Big\langle T(z) \prod_{j=1}^N \phi_j(p_j) \Big\rangle_{\!D_b}\,,  
\end{equation}

\noindent
Here the field insertions $\phi_i(p_i)$ are equipped with local coordinates $\sigma_{s_i}$, i.e.\ for $p_i = \sigma_{s_i}(0)$ we abbreviate $\phi_i(p_i) = \phi_i(p_i;\sigma_{s_i})$.
The r.h.s.\ of \eqref{eq:disc-integral-identity} vanishes for $\sum_{i=1}^N a_i \le2$. 

Denote by $\rho_i^{(n)}$ the coefficients of the Laurent expansion of $\rho(z)$ around $p_i$,
\begin{equation}\label{eq:rhoi}
    \rho(z) = \sum_{n=a_i}^\infty \rho_i^{(n)} \, (z-p_i)^n\,.
\end{equation}
for $z$ close enough to $p_i$.
Note that the expansion coefficients $\rho_i^{(n)}$ are some rational expressions that depend on all $p_j\neq p_i$ and $a_j$. 
Under the condition that $\sum a_i\leq 2$, we can now rewrite \eqref{eq:disc-integral-identity} as
\begin{equation}
    \sum_{i=1}^N \sum_{n=a_i}^\infty \rho_i^{(n)} 
    \left\langle \widehat{L}_{n-1}(p_i) \phi_i(p_i)
    \prod_{j\neq i} \phi_j(p_j)
    \right\rangle_{\!\!D_b} = 0~.
    \label{eq:WardIdNpt}
\end{equation}
Note that even if not written explicitly, the sums over $n$ do always terminate. 
Finally, one can use \eqref{eq:Ln-Ln-relation} to convert the $\widehat{L}_n$'s into $L_n$'s.

\subsubsection{Explicit recursion relation}\label{sec:disc-explicit-rec}

To compute the interaction vertex, we will need the disc correlator with insertion points at $p_n = e^{2 \pi i n/3}$, $n = 0,\pm 1$ and local coordinates $\sigma_{n/3}$. We abbreviate, for $\phi_0,\phi_\pm \in \mathcal{H}_{bb}$,\footnote{
    For the same reason as for the interaction vertex $T^b_R$ (cf.\ Footnote~\ref{fn:order-change}) we use a different order of arguments between the tensor $B_b$ and the disc correlator. The ordering convention for fields in the disc correlator is such that via the conformal transformation \eqref{eq:Moebius-to-UHP}
one obtains the standard order on the upper half plane in \eqref{eq:UHP-2pt3pt}, which is also the standard convention for the index ordering in OPE coefficients. This implies that there is a swap in the order of arguments between $B_b$ and the OPE coefficient in \eqref{eq:disc-3pt-fixed}.
}

\begin{equation}\label{eq:B_b-def}
B_b(\phi_0,\phi_+,\phi_-) := \big\langle 
    \phi_+(e^{2\pi i/3};\sigma_{1/3}) \, 
    \phi_0(1;\sigma_{0}) \, 
    \phi_-(e^{-2\pi i/3};\sigma_{-1/3}) 
	\big\rangle_{D_b}\,.
\end{equation}

\noindent
For the basis $\{\psi_i\}$ of primary fields introduced in Section~\ref{sec:primary-corr-disc},  we can read off from \eqref{eq:disc-3pt-arbitrary} that
\begin{equation} 
B_b(\psi_i,\psi_j,\psi_k) = c^{(b)}_{jik}  
    \big(\sqrt3\big)^{-h_i-h_j-h_k} \,.
	\label{eq:disc-3pt-fixed}
\end{equation}

\noindent 
The function $B_b$ is cyclically symmetric, and so it is enough to give a relation removing an $L_{-m}$ mode from $\phi_0$.
Let us abbreviate $\zeta = e^{2\pi i/3}$. We will apply \eqref{eq:WardIdNpt} to the function $\rho(z) = (z-1)^{-m+1}(z-\zeta)(z-\bar\zeta)$. For $m\ge 1$ this satisfies the asymptotic condition for $z \to \infty$. The first order poles at $\zeta$ and $\bar\zeta$ ensure that no $L_{-1}$ modes appear at these insertions. The Laurent expansion of $\rho(z)$ around $1$, $\zeta$, and $\bar\zeta$ is given by
\begin{align}
\rho(z) 
&\overset{z\approx 1}=
3 \, (z-1)^{-m+1}
+ 3 \, (z-1)^{-m+2}
+ (z-1)^{-m+3}
\nonumber \\
&\overset{z\approx \zeta}= i^m \, 3^{-\frac{m}2+1} \zeta^{m-1} \left[
(z-\zeta) + 
\sum_{k=1}^\infty \binom{1-m}{k-1} i^k \, 3^{-\frac{k}2} \zeta^{k}  \left( \tfrac{2-m-k}k - \bar\zeta \right) (z-\zeta)^{k+1} \right] \nonumber \\
&\overset{z\approx \bar\zeta}= i^{-m} \, 3^{-\frac{m}2+1} \bar\zeta^{m-1} \left[
(z-\bar\zeta) + 
\sum_{k=1}^\infty \binom{1-m}{k-1} i^{-k} \, 3^{-\frac{k}2} \bar\zeta^{k}  \left( \tfrac{2-m-k}k - \zeta \right) (z-\bar\zeta)^{k+1} \right] \,.
\end{align}
The resulting recursion relation is, for $m \ge 1$ and $\phi_0,\phi_\pm \in \mathcal{H}_{bb}$,
\begin{equation}\label{eq:recusive-relation}
 B_b(L_{-m} \phi_0,\phi_+,\phi_-)
 = 
B_b(A_0 \phi_0,\phi_+,\phi_-)
+
B_b(\phi_0,A_+ \phi_+,\phi_-)
+
B_b(\phi_0,\phi_+,A_- \phi_-)~,
\end{equation}
where
\begin{align}
A_0 &= -i L_{-m+1} + \tfrac13 L_{-m+2}~,
\nonumber \\
A_+ &= -(-1)^m \, 3^{-\frac{m}2} \, \zeta^{m-1} \left[
L_0 + 
\sum_{k=1}^\infty \binom{1-m}{k-1} (-1)^k \, 3^{-\frac{k}2} \, \bar\zeta^k \left( \frac{2-m-k}k - \bar\zeta \right) L_k \right]~,
\nonumber \\
A_- &= - 3^{-\frac{m}2} \, \bar\zeta^{m-1} \left[
L_0 + 
\sum_{k=1}^\infty \binom{1-m}{k-1}  3^{-\frac{k}2} \, \zeta^k \left( \frac{2-m-k}k - \zeta \right) L_k \right]~.
\end{align}

\noindent
Here we used the relation \eqref{eq:Ln-Ln-relation} to convert contour integrals in $\mathbb{C}$ to $L_n$-modes in the respective coordinates.

\subsection{Computing the interaction vertex}

Combining \eqref{eq:T-def-standard coord} with \eqref{eq:tilde-alpha} and \eqref{eq:B_b-def} we get the expression
\begin{equation}\label{eq:T-via_B}
\frac{T_R^b(\alpha_1,\alpha_2,\alpha_3)}{T_R^b(\one,\one,\one)}
~=~ B_b( \Gamma \alpha_1 ,  \Gamma \alpha_2 ,  \Gamma \alpha_3 ) ~,
\end{equation}
where $\alpha_1,\alpha_2,\alpha_3 \in \mathcal{H}_{bb}$ and $\Gamma$ is given in \eqref{eq:Gamma_h}.

Let us give a few concrete examples of evaluating \eqref{eq:T-via_B}. We start with the case where all $\alpha_i$ are taken from the vacuum module.
The vacuum, respectively the identity, does transform trivially, $\Gamma\one=\one$, and $B_p(\Gamma\one,\Gamma\one,\Gamma\one)=B_p(\one,\one,\one)=1$. The first excited state is $L_{-2}\ket{\one}$ which corresponds to $T = L_{-2}\one$. From \eqref{eq:Gamma-action-one-two-levels} with $h=0$ we read off the transformed state and obtain
\begin{align}
    \frac{T_R^b(T,\one,\one)}{T_R^b(\one,\one,\one)}
&= \frac{16}{9X}
\underbrace{B_b(  L_{-2}\one  ,  \one ,  \one )}_{=0} - \frac{5c}{216X^2}\frac{1+4t^2}{1+t^2} \underbrace{B_b(  \one,  \one ,  \one )}_{=1} 
\nonumber\\
&= - \frac{5c}{216X^2}\frac{1+4t^2}{1+t^2}\,. \label{eq:L-2vertex}
\end{align}

\noindent 
Our next example is $\alpha_1=T = \alpha_2$ and $\alpha_3=\one$\,. One obtains
\begin{align}
    \frac{T_R^b(T,T,\one)}{T_R^b(\one,\one,\one)} &= \frac{256}{81 X^4} B_b(  L_{-2}\one  ,  L_{-2}\one ,  \one ) + \frac{25 c^2}{46656 \,X^4} \left(\frac{1+4t^2}{1+t^2}\right)^2 
    \nonumber\\
    &= \frac{128 c}{729 X^4}  + \frac{25 c^2}{46656 \,X^4} \left(\frac{1+4t^2}{1+t^2}\right)^2\,.
\end{align}
In the first step we used $B_b(  \one,  \one ,  \one ) = 1$ and $B_b(  L_{-2}\one  ,  \one ,  \one ) = 0 = B_b(  \one ,L_{-2}\one  ,    \one )$\,. For the second step we compute the remaining correlator with the recursive algorithm: All terms in $A_0$ and $A_-$ vanish and in $A_+$ only $k=2$ survives, so that we obtain
\begin{align}
    B_b( L_{-2}\one  ,  L_{-2}\one ,  \one ) &= \left(- \frac{\zeta}{3}\right) \cdot \left(- \frac{\bar{\zeta}^2}{3}(-1-\bar{\zeta}) \right)  B_p(L_2L_{-2} \one,\one,\one) = \frac{c}{18}\,.
\end{align}

\noindent 
For completeness, we also give the result for $\alpha_i = T$, $i=1,2,3$, 
\begin{align}
    \frac{T_R^b(T,T,T)}{T_R^b(\one,\one,\one)} &= 
    \frac{4096\,c}{19683\, X^6} - \frac{80\,c^2(1+4t^2)}{6561 \,(1+t^2)\, X^6} - \frac{125\,c^3 (1+4t^2)^3}{10077696\, (1+t^2)^3 X^6}\,.
\end{align}

\noindent
As a final example we choose $\alpha_1=L_{-1}\phi_1, \alpha_2 =  \phi_2$ and $\alpha_3 = \phi_3$ for primary fields $\phi_i$, $i=1,2,3$. We obtain
\begin{align}
    \frac{T_R^b(L_{-1}\phi_1,\phi_2,\phi_3)}{T_R^b(\one,\one,\one)} &= 
    \left(\frac{4}{3X}\right)^{h_1+h_2+h_3 +1} 
    \Big(i h_1 B_b(\phi_1,\phi_2,\phi_3) + 
    \hspace{-1.2em}
    \underbrace{B_b(L_{-1}\phi_1,\phi_2,\phi_3)}_{= (-ih_1 + \frac{h_2 -h_3}{\sqrt{3}})B_b(\phi_1,\phi_2,\phi_3)}
    \hspace{-1.2em} 
    \Big)\nonumber\\
    &= c^{(b)}_{213} \, (h_2-h_3)\,
\left(\frac{4}{3\,\sqrt{3}\,X}\right)^{h_1+h_2+h_3 +1}\,,
\end{align}
where we used \eqref{eq:Gamma-action-one-two-levels}, \eqref{eq:disc-3pt-fixed} and \eqref{eq:recusive-relation}. 

The expressions for higher descendants become quite large and it is not illuminating to present them here. However, both the transformation as well as the recursive relation are straightforward to implement on a computer. 

\section{The torus with a single hole: open channel}\label{sec:open-channel-approx}

By ``open channel'' we mean that we cut the torus with a single hole into two uniform pieces along the three shortest straight lines between the boundary. Each piece is a clipped triangle as discussed in the previous section. 
In this section, we describe how to carry out the sum over intermediate open states for general surfaces, and we then illustrate the procedure by computing the first subleading correction for the torus with one hole.

\subsection{Sum over intermediate states in the open channel}\label{sec:intermediate_state_sum}

We would like to cut a surface $\Sigma$ along a simple curve $\gamma$ stretching between two boundary points and insert a complete sum over intermediate states without changing the value of the unnormalised correlator. We will use the state-field correspondence and glue half discs with a boundary field insertion at zero on the free boundary and sum over an appropriate basis of boundary fields.
Since the sum over intermediate states involves unnormalised correlators, we need to keep track of the metric $g$ on $\Sigma$, not just of its conformal class. As before, this dependence will cancel in the ratios we consider in the end.

\begin{figure}[t]
	\centering	
	\includegraphics[scale=1.15]{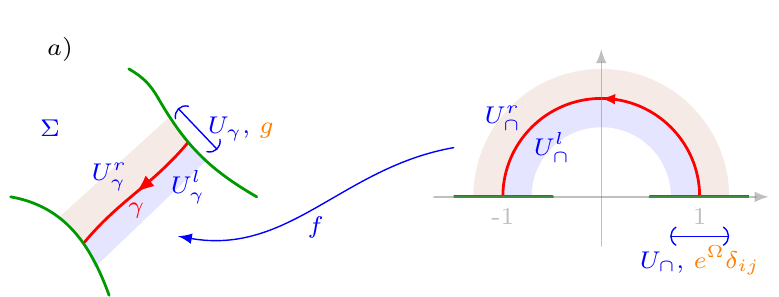}
	\includegraphics[scale=1.15]{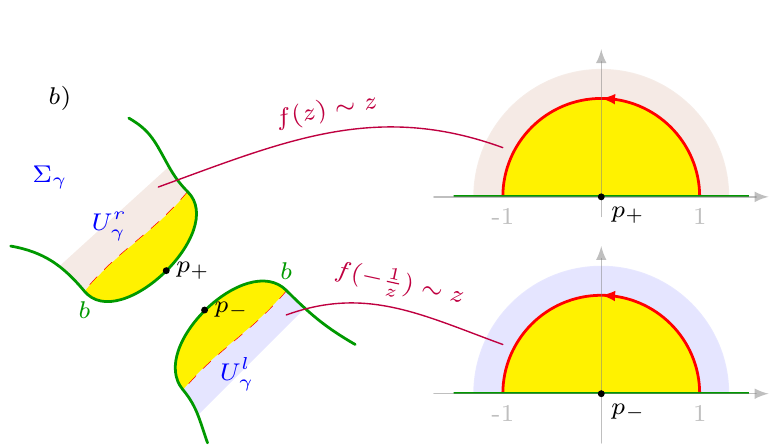}
	\caption{(a) The surface $\Sigma$ and a parametrisation of a neighbourhood $U_\gamma$ of the cutting curve $\gamma$. (b) The surface $\Sigma(\gamma)$ obtained by cutting $\Sigma$ along $\gamma$ and glueing in two half-discs where $z \in U^r_\cap$ is identified with $f(z)$, resp.\ $f(-1/z)$ on the surface $\Sigma \setminus \gamma$.}
	\label{fig:parametrise-cut-curve}
\end{figure}

\medskip
As a first step, one needs to parametrise a neighbourhood $U_\gamma$ of $\gamma$ by giving a biholomorphic map $f: U_\cap \to U_\gamma$ of a neighbourhood $U_\cap \subset  \overline{\mathbb{H}}$ of the unit half-circle to $U_\gamma$, see Figure~\ref{fig:parametrise-cut-curve}\,(a). 
We require that $f$ preserves also the curve-orientation, where we take the unit half-circle to be oriented anticlockwise (from $1$ to $-1$ in $\overline{\mathbb{H}}$). 
The inverses of the maps $G_n$ in Figure~\ref{fig:parametrise-state-bnd} are examples of such maps $f$. 
Removing $\gamma$ from $U_\gamma$ results in two open sets: $U^\ell_\gamma$ to the left of $\gamma$ (looking along the curve) and $U^r_\gamma$ to the right of $\gamma$. Analogously, $U^\ell_\cap = U_\cap \cap \{ |z|<1 \}$ and $U^r_\cap = U_\cap \cap \{ |z|>1 \}$. By restriction, $f$ gives bijections $U_\cap^{\ell/r} \to U_\gamma^{\ell/r}$. 
Since $f$ is conformal, the pullback metric $f^*g$ on $U_\cap$ is of the form
\begin{equation}
    (f^*g)(z)_{ij} ~=~ e^{\Omega(z)} \delta_{ij}
    ~,
\end{equation}
where $\Omega : U_\cap \to \mathbb{R}$ is smooth.

As before, let $D_+ = D \cap \overline{\mathbb{H}}$ be the closed unit half disc. 
Pick a smooth extension $\Omega_\ell :D_+ \to \mathbb{R}$ of $\Omega$ from $U^\ell_\cap$ to $D_+$.
We need an analogous extension for the metric on $U_\cap^r$. To this end we map $\{ |z| \ge 1 \}$ to $D_+$ via $j : z \mapsto -1/z$ and extend the pullback metric $(f \circ j)^*g$ to $D_+$, resulting in a function $\Omega_r : D_+ \to \mathbb{R}$. 
Explicitly we have $\Omega_r(z) = \Omega(-1/z)-4\log|z|$ 
for suitable $z$ with $|z|<1$. 
We write
\begin{equation}\label{eq:Omega_lr}
    D_+(\Omega_{\ell/r})
\end{equation}
for the two half discs with their respective metrics.

We cut $\Sigma$ along $\gamma$, that is, we consider $\Sigma \setminus \gamma$, and we glue in two copies of $D_+$, one to each side of the cut. For the copy joined to $U_\gamma^r$ we identify points of $U_\cap^r$ (seen as a subset of $D_+ \cup U_\cap$) and $U_\gamma^r$ via $z \sim f(z)$. 
For the copy joined to $U_\gamma^\ell$ we first map $D_+$ to $\{ |z| \ge 1 \}$ via $j$ and the use $f$ to identify points of $U_\cap^\ell$ and $U_\gamma^\ell$, that is, $z \sim f(-1/z)$. This is illustrated in Figure~\ref{fig:parametrise-cut-curve}\,(b). We denote the cut surface with half discs glued to the cuts by $\Sigma(\gamma)$. By construction, $f|_{U_\cap^{\ell/r}}$ are isometries and the resulting metric on $\Sigma(\gamma)$ is again smooth.
On each copy of $D_+$ we place a field insertion at $0$ with standard local coordinates, and we denote the two resulting insertion points in $\Sigma(\gamma)$ by $p^\pm$.

\medskip

Suppose the boundary condition on either end of the curve $\gamma$ is $b$ (which need not be elementary), and $\mathcal{H}_{bb}$ is the space of boundary fields on $b$. Our next task is to find the correct bilinear pairing on $\mathcal{H}_{bb}$ so that for an orthonormal basis $\kappa_j$, $j \in \mathbb{N}$ of $\mathcal{H}_{bb}$ with respect to this pairing, we have the identity
\begin{equation}\label{eq:sum-intermediate}
\big\langle \text{(some fields)} \big\rangle^\mathrm{un}_\Sigma
~=~
\sum_{j=1}^\infty
\big\langle \text{(some fields)} \kappa_j(p_+) \kappa_j(p_-) \big\rangle^\mathrm{un}_{\Sigma(\gamma)}~.
\end{equation}
Here, ``some fields'' refers to field insertions  that are the same on both sides of the equation. 
Note that we use the notation $\langle \cdots \rangle^\mathrm{un}_{\Sigma}$ also for non-connected $\Sigma$. In particular, $\Sigma(\gamma)$ can be non-connected, even if $\Sigma$ was connected. 

The pairing  on $\mathcal{H}_{bb}$ is uniquely fixed by applying the above procedure to the unit disc with $\gamma$ a straight line from $i$ to $-i$. Consider the bijections $\rho_{\pm1} : \overline{\mathbb{H}} \to D \setminus \{\mp1\}$ given by
\begin{equation}\label{eq:disc-coordinates-for-pairing}
\rho_{\pm 1}(z) = \pm \frac{i-z}{i+z} \,,
\end{equation}
which satisfy $\rho_{\pm 1}(0) = \pm 1$.
We equip $D$ with the two field insertions $\alpha(-1;\rho_{-1})$ and $\beta(1;\rho_{+1})$, for $\alpha,\beta \in \mathcal{H}_{bb}$.
Since $\rho_{+1}$ maps the unit half circle to the curve $\gamma$, preserving orientation, we can choose $f = \rho_{+1}$ for the parametrisation of a neighbourhood of $\gamma \subset D$. 
Let $g$ be the metric on $D$ such that $\rho_{\pm 1} |_{D_+} : D_+(\Omega_{\ell/r}) \to D$ become isometries (note that $f(-1/z) = \rho_{-1}(z)$). Explicitly, $g(u)_{ij} = e^{\Omega_g(u)} \delta_{ij}$ with $\Omega_g(u) = \Omega_+(\rho_+^{-1}(u))+\log \frac{4}{|1+u|^4}$.
With this choice of metric, the surface $D(\gamma)$ consists of two copies of $D$, both again with metric $g$. Altogether, \eqref{eq:sum-intermediate} turns into
\begin{align}
&\big\langle
    \alpha(-1;\rho_{-1}) \beta(1;\rho_{+1}) 
\big\rangle^\mathrm{un}_{D_b(g)}
~=~
\sum_{i=1}^\infty
\big\langle \alpha(-1;\rho_{-1}) \beta(1;\rho_{+1}) \kappa_j(p_+) \kappa_j(p_-) \big\rangle^\mathrm{un}_{D_b(g;\gamma)}
\nonumber\\
~&=~
\sum_{i=1}^\infty
\big\langle
\alpha(-1;\rho_{-1}) \kappa_j(1;\rho_{+1}) 
\big\rangle^\mathrm{un}_{D_b(g)}
\,
\big\langle
\kappa_j(-1;\rho_{-1}) \beta(1;\rho_{+1})
\big\rangle^\mathrm{un}_{D_b(g)} ~,
\label{eq:disc-check-1}
\end{align}
where all discs $D$ carry metric $g$ and boundary condition $b$.
This shows that the pairing $\langle - , - \rangle_b : \mathcal{H}_{bb} \times \mathcal{H}_{bb} \to \mathbb{C}$ with respect to which the $\kappa_j$, $j \in \mathbb{N}$ have to be orthonormal is
\begin{equation}\label{eq:pairing}
	\big( \alpha , \beta \big)_b
	:= \big\langle \alpha(-1;\rho_{-1}) \beta(1;\rho_1) \big\rangle^\mathrm{un}_{D_b(g)} \,.
\end{equation}
A standard computation using contour integrals of the stress tensor shows that
\begin{equation}\label{eq:pairing2}
	\big( \alpha , L_m \beta \big)_b
	~=~
	(-1)^m\big( L_{-m} \alpha , \beta \big)_b
	\qquad , ~~ m \in \mathbb{Z} \,,
\end{equation}
where the sign $(-1)^m$ comes from the factor $z^{m+1}$ in the integrand and the change of integration direction. 
This allows one to reduce the pairing of two arbitrary vectors to that of Virasoro primaries. Let $\psi,\xi \in \mathcal{H}_{bb}$ be Virasoro-primary and recall the primary two-point correlator on the upper half plane from \eqref{eq:UHP-2pt3pt},
\begin{equation}
    \big\langle \psi(x) \xi(0) \big\rangle_{\overline{\mathbb{H}}_b} ~=~ c^{(b) \, \one}_{\psi\xi} x^{-h_\psi-h_\xi}  ~.
\end{equation}
For $\vartheta \in \mathbb{R}$ consider the local coordinate $\rho_{e^{i\vartheta}}(z) = e^{i \vartheta} \rho_1(z)$, with $\rho_1$ as in \eqref{eq:disc-coordinates-for-pairing}. For $\vartheta = \pi$ this is consistent with the definition of $\rho_{-1}$ in \eqref{eq:disc-coordinates-for-pairing}.
The normalised two-point correlator on the disc is given by, for $\vartheta \in (0,\pi)$,
\begin{align}
	\big\langle \psi(e^{i\vartheta};\rho_{e^{i\vartheta}}) \xi(1;\rho_1) \big\rangle_{D_b}
	&=
	\big\langle \psi(\rho_1^{-1}(e^{i\vartheta});\rho_1^{-1} \circ \rho_{e^{i\vartheta}}) \xi(0;\varphi_0) \big\rangle_{\overline{\mathbb{H}}_b}
	\nonumber\\
	&= (\cos\tfrac\vartheta2)^{-2 h_\psi}
	\big\langle \psi(\tan\tfrac\vartheta2)  
	\xi(0) \big\rangle_{\overline{\mathbb{H}}_b}
	\nonumber\\
	&= c^{(b) \, \one}_{\psi\xi} (\sin\tfrac\vartheta2)^{-2 h_\psi} ~,
\end{align}
where $\varphi_0(z)=z$ is the standard local coordinate on $\mathbb{C}$. For $\vartheta \to \pi$, and including the factor $\langle \one \rangle^\mathrm{un}_{D_b(g)}=e^{A_D}\langle \one \rangle^\mathrm{un}_{D_b}$
to obtain the unnormalised correlator, altogether we obtain
\begin{equation}
	\big( \psi , \xi \big)_b ~=~ c^{(b) \, \one}_{\psi\xi} 
	\,
	e^{A_D}\langle \one \rangle^\mathrm{un}_{D_b} \,.
	\label{eq:pairing-primary-norm}
\end{equation}

\noindent
Here, $e^{A_D}$ is the Liouville factor for changing the metric to the flat standard metric on the unit disc. 
Note that choosing different extensions $\Omega_{\ell/r}$ of the metric in \eqref{eq:Omega_lr} affects the pairing \eqref{eq:pairing-primary-norm} only through $e^{A_D}$. 
This completes the definition of the pairing on $\mathcal{H}_{bb}$.

\subsubsection*{Orthonormal basis}

First, consider the standard bi-linear pairing $(\cdot,\cdot)_2$ with $(\cdot,L_m \cdot )_2  = (L_{-m} \cdot, \cdot )_2$ instead of \eqref{eq:pairing2}. A basis orthonormal w.r.t.\ $(\cdot,\cdot)_2$ can be constructed from real linear combinations of states at fixed descendant level in an irreducible lowest weight Virasoro-representation.
Given such a basis, a basis $\mathcal{B}$ orthonormal w.r.t.\ \eqref{eq:pairing} is then obtained by simply multiplying every basis vector $\beta$ with $i^n$, where $n = \mathrm{lvl}(\beta)$ is the descendant level of $\beta$. 

For example up to level $4$ a basis orthonormal w.r.t. \eqref{eq:pairing} in the vacuum representation is the union of 
\begin{align}
    \mathcal{B}_{\one,0} &= \{\one'\}\,, & \mathcal{B}_{\one,2} &= \left\{\sqrt{\frac{2}{c}} L_{-2}\one' \right\}\,,
    \nonumber \\
    \mathcal{B}_{\one,3} &= \left\{\frac{i}{\sqrt{2c}} L_{-3}\one' \right\}\,,& \mathcal{B}_{\one,4} &= \left\{\frac{L_{-4}\one}{\sqrt{5c}}  , \frac{6 L_{-4} \one' -10 L_{-2}^2 \one'}{\sqrt{10c (22+5c)}}\right\}\,,
\label{eq:BasisOne}
\end{align}
where 
\begin{equation}\label{eq:normalised-vac}
    \one' =  \left(e^{A_D}\langle \one \rangle^\mathrm{un}_{D_b}\right)^{-\frac{1}{2}} \one
\end{equation}
ensures the correct normalisation of the identity when inserted in \eqref{eq:pairing-primary-norm}.

In the open state space $\mathcal{H}_{bb}$ there may be several primaries $\psi_{i}$, $i=1,\dots,N$ of a given $L_0$-weight $h$. Generically, the two-point structure constants $c_{i,j}^{(b)\one}$ are non-zero also for $i\neq j$. To produce an orhonormal basis of $\mathcal{H}_{bb}$ one first has to give primaries $\phi_p$, $p=1,\dots,N$ which are orthonormal for the matrix $M_{ij} = e^{A_D} \langle \one \rangle^\mathrm{un}_{D_b} c_{i,j}^{(b)\one}$.
Then up to level two, the orthonormal descendant of each $\phi_p$ are
\begin{align}
    \mathcal{B}_{p,0} &= \{\phi_p\}\,, \qquad  \mathcal{B}_{p,1} = \left\{\frac{i}{\sqrt{2h_p}} L_{-1}\phi_p\right\}\,,
    \nonumber \\
    \mathcal{B}_{p,2} &= \left\{\frac{L_{-2}\phi_p}{\sqrt{\frac{c}{2}+4h_p}},\frac{6h_p L_{-2} \phi_p -\frac{c+8h_p}{2} L_{-1}^2 \phi_p}{\sqrt{h_p(c+8h_p)(c+2h_p+2h_p(8h_p-5))}}\right\}\,.
    \label{eq:BasisP}
\end{align}

\noindent 
Note that in case of a null state at level 2 the second state in $\mathcal{B}_{p,2}$ is absent. 

\subsection{Computing the torus amplitude}\label{sec:OpenChannelTorus}

\begin{figure}[t]
    \centering
    \includegraphics[scale=1]{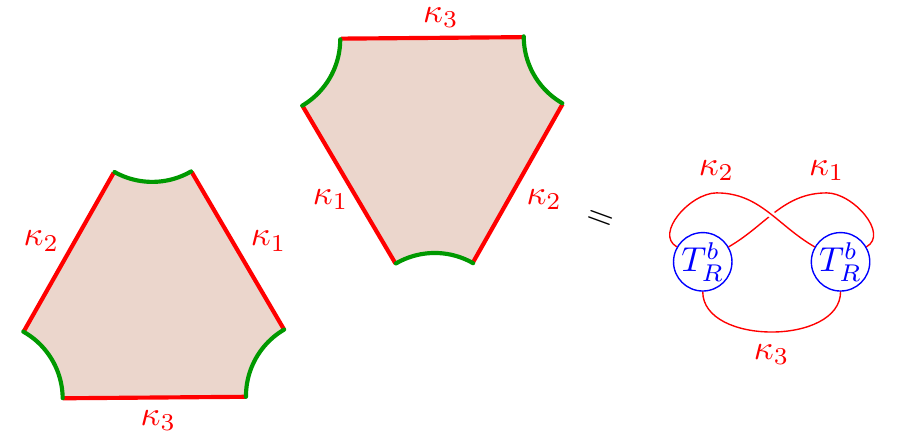}
    \caption{Visualisation of the contraction scheme for the one-hole partition function.}
    \label{fig:contraction}
\end{figure}

We now have all the necessary tools at hand for the open channel computation of the ratio of two partition functions for the torus with one hole.

Before passing to ratios to cancel the Liouville factors, let us start by expressing the partition function itself as a sum over open states.
Consider a torus with periods $\omega_1 = d$ and $\omega_2 = e^{\frac{i\pi}{3}} d$, and with a single hole of radius $R$ 
with boundary condition $b$. 
After the choice of an orthonormal basis $\mathcal{B}_b$ of boundary fields as in Section~\ref{sec:intermediate_state_sum}, it is given by
\begin{align}
    Z_b(R) &= \sum_{\kappa_1,\kappa_2,\kappa_3\in\mathcal{B}_b} \left( T_R^b(\kappa_1,\kappa_2,\kappa_3)\right)^2 \nonumber\\
    &= \Big( T_R^b(\one,\one,\one)\Big)^2 \sum_{\kappa_1,\kappa_2,\kappa_3\in\mathcal{B}_b} \Big( B_b(\Gamma\kappa_1,\Gamma\kappa_2,\Gamma\kappa_3)\Big)^2\,, \label{eq:OpenChannelSum}
\end{align}
where we used \eqref{eq:T-via_B} to express the correlators on the clipped triangle $T_R^b(\alpha_1,\alpha_2,\alpha_3)$ in terms of normalized correlators $B_b$ on the disc. When we regard the correlators as three tensors, then the triple sum over intermediate open channel states can be regarded as a tensor contraction as indicated in Figure \ref{fig:contraction}. Note that the contraction is such that the fields appear in the same order. The transformation $\Gamma$ was discussed in Section \ref{sec:bfields} and is given explicitly for the first few levels in \eqref{eq:Gamma_h}. 
A recursive algorithm to compute the disc correlators of any descendant is stated in Section \ref{sec:disc-explicit-rec}. 
The vacuum amplitude $T_R^b(\one,\one,\one)$, which appears in the overall factor, is given by the product of the unnormalised disc correlator $\left\langle\one\right\rangle_{D_b}^\mathrm{un}$ and a Liouville factor $e^{A_T}$ arising from transforming the flat metric from the clipped triangle to the disc,
\begin{equation}\label{eq:T111}
    T_R^b(\one,\one,\one) = e^{A_T} \left\langle\one\right\rangle_{D_b}^\mathrm{un} ~.
\end{equation}

\noindent
The unnormalised correlator $\left\langle\one\right\rangle_{D_b}^\mathrm{un}$ is the partition function on the (flat) unit disc with boundary $b$. 
It is given by the overlap of the boundary state $\ket{b}$ with the vacuum, 
\begin{equation}\label{eq:disc-unnorm}
    \left\langle\one\right\rangle_{D_b}^\mathrm{un} = 
    \langle b \ket{0}
    \overset{(*)}= \frac{S_{b\one}}{\sqrt{S_{\one\one}}}
    \,,
\end{equation}
where $\langle\cdot\vert\cdot\rangle$ is the standard pairing on the space of closed channel states. 
In $(*)$ we substituted the expression \eqref{eq:Cardy-boundary-state} for diagonal rational CFTs.

The Liouville factor $e^{A_T}$ will be treated in a future part of this series. 
In the following, just two observations are needed. Firstly, it is independent of the choice of conformal boundary condition, and secondly, it cancels altogether in the ratio 
\begin{equation}\label{eq:ZaZb-ratio}
    \frac{Z_a(R)}{Z_b(R)}
\end{equation}
for two conformal boundary conditions $a,b$. We now compute the leading and first subleading contribution to this ratio. The results of numerical computations of higher order contributions in the Ising example will be described in Section~\ref{sec:Ising-compare} below.

\subsubsection{Leading contribution}

In the limit $R\to d/2$, which is equivalent to $t\to0$, the vacuum amplitude dominates the sum in \eqref{eq:OpenChannelSum}. 
This can be seen from the explicit form of $\Gamma$ in \eqref{eq:Gamma_h}, where a state of $L_0$-weight $h$ contributes with a factor of $X(t)^{-h}$. 
In \eqref{eq:X-t0-limit} we saw that for $t \to 0$, the function $X(t)$ behaves as $X(t) \sim \exp\big(\frac{\pi}{2t}\big)$, 
and thus any state with non-zero conformal dimension is heavily suppressed in this limit. 
Therefore, only the (properly normalised) vacuum vector $\one'$ from \eqref{eq:normalised-vac} contributes to the limit. 
The sum \eqref{eq:OpenChannelSum} can thus be approximated as
\begin{equation}\label{eq:open-channel-vac-contrib}
    Z_b(R) ~\approx~  T_R^b(\one,\one,\one)^2  B_1(\one',\one',\one')^2
     \,=\, \left( e^{A_T} \left\langle\one\right\rangle_{D_b}^\mathrm{un} \right)^{2} \left(e^{A_D}\langle \one \rangle^\mathrm{un}_{D_b}\right)^{-3}
     =\, \frac{\mathcal{N}}{\left\langle\one\right\rangle_{D_b}^\mathrm{un}}
     \,,
\end{equation}
where in the second step we substituted \eqref{eq:normalised-vac} and \eqref{eq:T111}, and $\mathcal{N} = e^{2A_T -3A_D}$ is an overall normalization.
In the ratio \eqref{eq:ZaZb-ratio} the (divergent) Liouville factor cancels and we obtain
\begin{equation}\label{eq:R1/2Result}
    \lim_{R\to\frac{d}{2}}  \frac{Z_a(R)}{Z_b(R)} 
    = \frac{\left\langle\one\right\rangle_{D_b}^\mathrm{un}}{\left\langle\one\right\rangle_{D_a}^\mathrm{un}}
    \overset{(*)}= \frac{S_{b\one}}{S_{a\one}}
    \,,
\end{equation}
where $(*)$ holds for elementary boundary condition in diagonal rational theories, see \eqref{eq:disc-unnorm}.
Note that this is precisely the inverse of the $R\to 0$ limit in \eqref{eq:leading-beh-closed}.

\subsubsection{First order corrections}

For smaller radii $R$ we can use the explicit expressions given in the last sections to compute corrections to \eqref{eq:open-channel-vac-contrib}. 
The results depend on the spectrum of boundary fields $\mathcal{H}_{bb}$ 
and in particular on the conformal dimensions of its primaries. However, a universal contribution to $Z_b$ will be the one from the vacuum representation whose first correction will originate from the insertion of the normalised energy momentum tensor $T' = (e^{A_D} \langle \one \rangle^\mathrm{un}_{D_b})^{-1/2}\, \sqrt{2/c}\,L_{-2}\one$ (see \eqref{eq:BasisOne}) for any one of the states $\alpha_i$. Note that in diagonal rational unitary theories we can choose the Cardy boundary condition labelled by $\one$ whose spectrum only consists of the vacuum representation.
In that case, from \eqref{eq:L-2vertex} one finds the leading correction to \eqref{eq:open-channel-vac-contrib} to be
\begin{align}
    Z_\one(R) &~\approx~ T_R^\one(\one,\one,\one)^2 \Big(B_\one(\one',\one',\one')^2 + 3 B_\one(\Gamma T',\one',\one')^2\Big) \nonumber \\
   &\qquad =~\frac{\mathcal{N}}{\left\langle\one\right\rangle_{D_\one}^\mathrm{un}}
   \left(1 +  \frac{25 c}{7776 \, X(t)^4} \left(\frac{1+4t^2}{1+t^2}\right)^{\!2} \right) \,.
\end{align}

\noindent
In the more general case, when there will be contributions from non-vacuum primaries and their descendants, the first order correction comes from the lightest primary $\xi$ for any two of the states $\alpha_i$ and is more dominant than the contribution from $T$ as $R\to\frac{1}{2}$ if $2h_\xi<2 = h_T$. Let us assume this is the case for the boundary condition $b$. 
We take $\xi$ to have OPE-coefficient $c_{\xi\xi}^{(b)\,\one}=1$ and set $\xi' = (e^{A_D} \langle \one \rangle^\mathrm{un}_{D_b})^{-1/2}\, \xi$ so that it is normalised with respect to the pairing \eqref{eq:pairing-primary-norm}. Then to first subleading order, \eqref{eq:OpenChannelSum} becomes
\begin{align}
    Z_b(R) &~\approx~ T_R^b(\one,\one,\one)^2 \Big(B_b(\one',\one',\one')^2 + 3 B_b(\Gamma \xi',\Gamma \xi',\one')^2\Big)
    \nonumber \\
   &\qquad =~\frac{\mathcal{N}}{\left\langle\one\right\rangle_{D_b}^\mathrm{un}}\left(1 + 3 \left(\frac{16}{27 X(t)^2}\right)^{2h_\xi} \right)\,.\label{eq:openLeading}
\end{align}

\section{Comparison of both approximations in the Ising CFT}\label{sec:Ising-compare}

In this section, we compare the closed and open channel approximation schemes for the partition function of the one-holed torus developed in Sections~\ref{sec:torus-closed} and \ref{sec:open-channel-approx}, respectively.
The example we study is the Ising CFT. We begin with reviewing the relevant defining data. Then we give the numerical results for the closed and open approximation and find good agreement for intermediate values of $R$.

\subsection{The Ising CFT}\label{sec:IsingIntro}

\subsubsection*{Chiral data}

The Ising CFT is the simplest non-trivial unitary conformal field theory. It has central charge $c=\tfrac{1}{2}$. There are three irreducible unitary lowest weight representations of the Virasoro algebra at that central charge:
\begin{center}
{\def\arraystretch{1.4}
		\begin{tabular}{ccl}
		label & $h$ & character 
		\\ \hline
		$\one$  & 0 & $\chi_\one = \chi_{1,1}^{3,4}$
		\\
		$\sigma$  & $\frac1{16}$ & $\chi_\sigma = \chi_{1,2}^{3,4}$
		\\
		$\epsilon$  & $\frac12$ & $\chi_\epsilon = \chi_{1,3}^{3,4}$
	\end{tabular}
}
\end{center}
Here, $h$ denotes the lowest $L_0$-weight of the representation and the characters of the minimal model $(x,y)$ are given by \cite[Eq.\,(8.17)]{DiFrancesco:1997nk}
\begin{equation}
	\chi^{x,y}_{r,s}(\tau) = \frac{1}{\eta(\tau)} \sum_{n\in\mathbb{Z}} \Big(
	q^{(2 x y n + r y - s x)^2 / (4 x y)} 
	- q^{(2 x y n + r y  + s x)^2/(4 x y)} \Big)\,,
\end{equation}
where $q = e^{2 \pi i \tau}$. The modular $S$-matrix is
\begin{equation} \label{eq:Ising-S}
	(S_{ij}) = \frac{1}{2} 
	\begin{pmatrix}
		1&\sqrt{2}& 1\\ 
		\sqrt{2} & 0 & -\sqrt{2} \\
		1 & -\sqrt{2} & 1 
	\end{pmatrix}\,,\quad \text{with}~ i,j\in \{\one,\sigma,\epsilon\}\,.
\end{equation}
The fusion rules of the chiral Ising CFT are
\begin{equation}
	\epsilon \times \epsilon = \one
	~,~~
	\epsilon \times \sigma = \sigma
~,~~
	\sigma \times \sigma = \one + \epsilon ~.
\end{equation}
These fix the fusion rule coefficients $N_{ab}^{~c}$ via $a \times b = \sum_c N_{ab}^{~c} \, c$. For example, $N_{\sigma\sigma}^{~\epsilon}=1$ and $N_{\sigma\sigma}^{~\sigma}=0$. 

Denote by $M_a$ the Virasoro representation labelled by $a \in \{\one,\sigma,\epsilon\}$, and whose character $\chi_a = \mathrm{tr}_{M_a} q^{L_0 - c/24}$ is as in the above table. By definition, the coefficients $N_{ab}^{~c}$ give the dimension of the space of conformal 3-point blocks on the sphere with insertions of $M_a$, $M_b$ and $M_c$ (unitary representations of the Virasoro algebra are self-dual).

\subsubsection*{Bulk theory}

The bulk state space of the Ising CFT is
\begin{equation}
	\mathcal{H} 
	\,=\, 
	M_\one \otimes \overline M_\one 
	\,\oplus\,
	M_\sigma \otimes \overline M_\sigma
	\,\oplus\,
	M_\epsilon \otimes \overline M_\epsilon
	~.
\end{equation}
Correspondingly, the partition function is given by
\begin{equation}\label{eq:IsingPatition}
	Z(\tau,\bar{\tau}) 
	= \chi_\one(\tau) \chi_\one(-\bar{\tau}) + \chi_{\sigma}(\tau) \chi_{\sigma}(-\bar{\tau}) + \chi_{\epsilon}(\tau) \chi_{\epsilon}(-\bar{\tau})~.
\end{equation}
We will denote the three primary bulk fields in $\mathcal{H}$ by $\one$ (identity field), $\sigma$ (spin field) and $\epsilon$ (energy field). The non-zero OPE-coefficients for these fields are (see e.g. \cite[Sec.\,12.3.3]{DiFrancesco:1997nk})
\begin{align}
	C_{11}^1 &= 1\,, &&C_{\epsilon\epsilon}^1 = C_{\epsilon 1}^\epsilon = C_{1\epsilon}^\epsilon = 1\,,
	\nonumber\\
	C_{\sigma\sigma}^\epsilon &= C_{\sigma\epsilon}^\sigma = C_{\epsilon\sigma}^\sigma = \frac{1}{2}\,, &&C_{\sigma\sigma}^1 = C_{\sigma 1}^\sigma = C_{1\sigma}^\sigma = 1\,.\label{eq:IsingOPEbulk}
\end{align}

\noindent
Since $C_{a\sigma}^{~a}=0$ for all $a$, the torus one-point functions of the primary field $\sigma$ vanishes. The torus one-point function of $\epsilon$ 
was already considered in Section~\ref{sec:modDiffEq}. Altogether: 
\begin{equation}\label{eq:IsingEps}
	\left\langle \one \right\rangle^\mathrm{un}_{\omega_1,\omega_2} = Z\!\left(\frac{\omega_2}{\omega_1},-\frac{\bar{\omega}_2}{\bar{\omega}_1}\right) 
~,~~
	\left\langle \sigma \right\rangle^\mathrm{un}_{\omega_1,\omega_2} = 0 
~,~~
	\left\langle \epsilon \right\rangle^\mathrm{un}_{\omega_1,\omega_2} = \frac{\pi}{|\omega_1|} \eta\!\left(\frac{\omega_2}{\omega_1}\right) \eta\!\left(-\frac{\bar{\omega}_2}{\bar{\omega}_1}\right) \,.
\end{equation}

\subsubsection*{Boundary theory}

The Ising CFT has three elementary conformal boundary conditions, labelled 
by $\one$ (fixed spin $+$), $\epsilon$ (fixed spin $-$), and $\sigma$ (free). Their boundary states are
\begin{align}
    \ket{\one} &= \frac{1}{\sqrt{2}} \ket{\one}\!\rangle +  \frac{1}{\sqrt{2}} \ket{\epsilon}\!\rangle + 2^{-\frac{1}{4}} \ket{\sigma}\!\rangle\,,
    \nonumber\\
    \ket{\epsilon} &= \frac{1}{\sqrt{2}} \ket{\one}\!\rangle +  \frac{1}{\sqrt{2}} \ket{\epsilon}\!\rangle - 2^{-\frac{1}{4}} \ket{\sigma}\!\rangle\,,
    \nonumber\\
    \ket{\sigma} &=  \ket{\one}\!\rangle -  \ket{\epsilon}\!\rangle\,.
\label{eq:Ising-bnd-state}
\end{align}

\noindent 
The state space $\mathcal{H}_{ab}$ of open channel states between boundaries $a$ and $b$ is given by $\mathcal{H}_{ab} = \bigoplus_{c}  N_{ab}^{~c}\, M_c$, see \cite{Cardy:1989ir}. We will only use
\begin{equation}\label{eq:Ising-openstates}
    \mathcal{H}_{\one\one} = M_{\one}
    \quad , \quad
    \mathcal{H}_{\sigma\sigma} = M_{\one} + M_{\epsilon} ~.
\end{equation}
Correspondingly, the only boundary OPE coefficients we need are already fixed by the normalisation of boundary fields: $c_{\one\one}^{(\one)\,\one} = 1$ and $c_{aa}^{(\sigma)\,\one} = 1$ for $a = \one, \epsilon$.

\subsection{Comparison of both channels}\label{sec:explicit_results_Ising}

We will now compare the approximate results in the open and closed channel in the Ising CFT. 
The approximation is necessary because the theoretical methods that we explained ultimately lead to a point where we need to solve recursive algorithms. This becomes harder and harder the higher the conformal dimensions (of descendants) that are involved. Therefore, in both channels, we introduce cutoffs.
However, the two channels have complementary regimes in which the approximation can be trusted most. The closed channel approximation becomes exact for $R\to 0$, and the open channel approximation for $R\to \frac{d}{2}$.

Ultimately we will be interested in the open channel approximation because it is defined in terms of the interaction vertex of the lattice model we introduced in Section~\ref{sec:lattice-model-def}, or rather in terms of its truncated version discussed in Section~\ref{sec:truncated-sym} (but for elementary boundary conditions, not for the cloaking boundary condition).
Comparing approximate results for the ratio between the one-hole partition functions in the open and closed channel will give an indication to which extend the truncated lattice model still provides a good approximation to the exact CFT result.

\medskip

In what follows we will compute the ratio of the one-hole partition functions with boundary conditions $\sigma$ and $\one$. The torus is chosen to have periods $\omega_1 = d = 1$ and $\omega_2 = e^{\frac{i\pi}{3}}$, and hence $\tau= e^{\frac{i\pi}{3}}$.  

\subsubsection{Closed channel}

We start by discussing the first few orders in the approximation of the one-hole partition function explicitly. The general formula to compute it is \eqref{eq:Zopen1}. There will be no contribution from the $\sigma$-Ishibashi state because we saw in \eqref{eq:IsingEps} that the corresponding primary one-point function vanishes. 
The first two orders for the $\one$-Ishibashi state can be read off from \eqref{eq:ZIshi0} with $c=\frac{1}{2}$, and those for the $\epsilon$-Ishibashi state from \eqref{eq-1h-torus-level2null} with $h=\frac12$.
The two one-hole partition functions with elementary boundary conditions are then simply $Z_\sigma(R) = Z_{\ket{\one}\!\rangle}(R) - Z_{\ket{\epsilon}\!\rangle}(R)$ and $Z_\one(R) = \frac{1}{\sqrt{2}} Z_{\ket{\one}\!\rangle}(R) +\frac{1}{\sqrt{2}} Z_{\ket{\epsilon}\!\rangle}(R)$\,.
Altogether, up to total descendant level $4$ we get 
\begin{align}
    \frac{Z_{\sigma}(R)}{Z_{\one}(R)} &=  \frac{Z(\tau,\bar{\tau})  - \pi\, \left|\eta\left(\tau\right)\right|^2  R + 16 \pi^2 \partial_\tau\partial_{\bar\tau}Z(\tau,\bar{\tau})  R^4 +\dots }{\frac{1}{\sqrt{2}}\left(Z(\tau,\bar{\tau})  + \pi\,  \left|\eta\left(\tau\right)\right|^2  R + 16 \pi^2 \partial_\tau\partial_{\bar\tau}Z(\tau,\bar{\tau})  R^4 + \dots\right)} 
    \nonumber \\
    &= \sqrt{2} \Big(1 -  \frac{2 \pi\,  \left|\eta\left(\tau\right)\right|^2 R}{Z\left(\tau,\bar\tau\right)} + \frac{2 \pi^2 \left|\eta\left(\tau\right)\right|^4 R^2 }{Z(\tau,\bar\tau)^2} - \frac{2 \pi^3 \left|\eta\left(\tau\right)\right|^6 R^3 }{Z(\tau,\bar\tau)^3} 
    \nonumber \\
    &\qquad\qquad+ \frac{2 \pi^4 \left|\eta\left(\tau\right)\right|^8 R^4 }{Z(\tau,\bar\tau)^4}  + \dots\Big)\,.
\end{align}

\noindent
As a reminder, $\tau = e^{\frac{\pi i}{3}}$, $Z(\tau,\bar{\tau})$ is given by \eqref{eq:IsingPatition}, and $\eta$ is the  Dedekind $\eta$-function. 

Higher order contributions can be computed explicitly using the recursive algorithm discussed in Section \ref{sec:recusion-one-point}. 
For the cutoffs we will use below, the number of summands in the Ishibashi state $\ket{i}\!\rangle$ is as follows, in terms of the cutoff $\Delta_\mathrm{max}$: 

\begin{center}
    \begin{tabular}{c|c|c|c|c}
        $\Delta_\mathrm{max}$ & 4 & 12 & 20 & 28 \\
        \hline
        $\ket{\one}\!\rangle$ & 2 & 10 & 30 & 77 \\
        $\ket{\epsilon}\!\rangle$ & 2 & 8 & 26 & 69
    \end{tabular}
\end{center}

While the algebraic expressions quickly become cumbersome,  
since all parameters are fixed we can still give numerical values. For example, the numerical expansion for the one-hole partition function with the $\one$-Ishibashi state up to level 31 is 
\begin{align}
    R^{\frac{1}{12}} Z_{\ket{\one}\!\rangle}(R) ~\approx~ &  1.89 + 16.4 \,R^4 + 3.57 \cdot 10^2 \,R^8 + 7.99 \cdot 10^2 \,R^{12} + 7.65 \cdot 10^4 \,R^{16}\nonumber \\
    &+ 1.27 \cdot 10^6 \,R^{20} + 1.40 \cdot 10^7 \,R^{24} + 3.39 \cdot 10^8 \,R^{28} + \mathcal{O}(R^{32}) 
    \label{eq:NresultOne}
\end{align}
and for the $\epsilon$-Ishibashi we obtain
\begin{equation}
    R^{-1+\frac{1}{12}} Z_{\ket{\epsilon}\!\rangle}(R) ~\approx~ 2.01 + 8.65\cdot 10^3\, R^{12} + 3.16\cdot 10^7 \, R^{24} +\mathcal{O}(R^{32})\,.
    \,. \label{eq:NresultEpsilon}
\end{equation}

\noindent 
As explained in Section \ref{sec:recusion-one-point}, the one-point functions of odd holomorphic or odd anti-holomorphic descendants vanish, which explains why only powers $R^{4k}$, $k=0,1,2,\dots$ appear in \eqref{eq:NresultOne}. In addition to the odd descendants, in \eqref{eq:NresultEpsilon} more terms vanish because of the modular differential equation from the null state. 

The numerical results allow us to estimate the
radius of convergence. A power series $\sum_{n} c_n (z-a)^n$
converges for $|z-a|<r$ with
\begin{equation}
    r^{-1} = \limsup\limits_{n\to\infty} \sqrt[n]{|c_n|} \,.
\end{equation}

\noindent
The approximate non-zero numerical values for $\sqrt[4k]{|c_{4k}|}$ for the $\one$-Ishibashi state for $k$ from $1$ to $7$ are 
$\{2.014,2.085,1.745,2.019,2.019,1.985,2.017\}$ and in case of the $\epsilon$-Ishibashi state we obtain $\sqrt[13]{c_{13}} \approx 2.008$ and $\sqrt[25]{c_{25}} \approx 1.995$\,. All of these numerical values do not show any strong deviation from the expected value $r=1/2$ (touching hole limit) for the radius of convergence and, hence, support the claim that our theoretical framework and the numerics are correct and give a good approximation to one-hole torus partition functions.  

\begin{figure}[t]
    \centering
    \includegraphics[width=.85\textwidth]{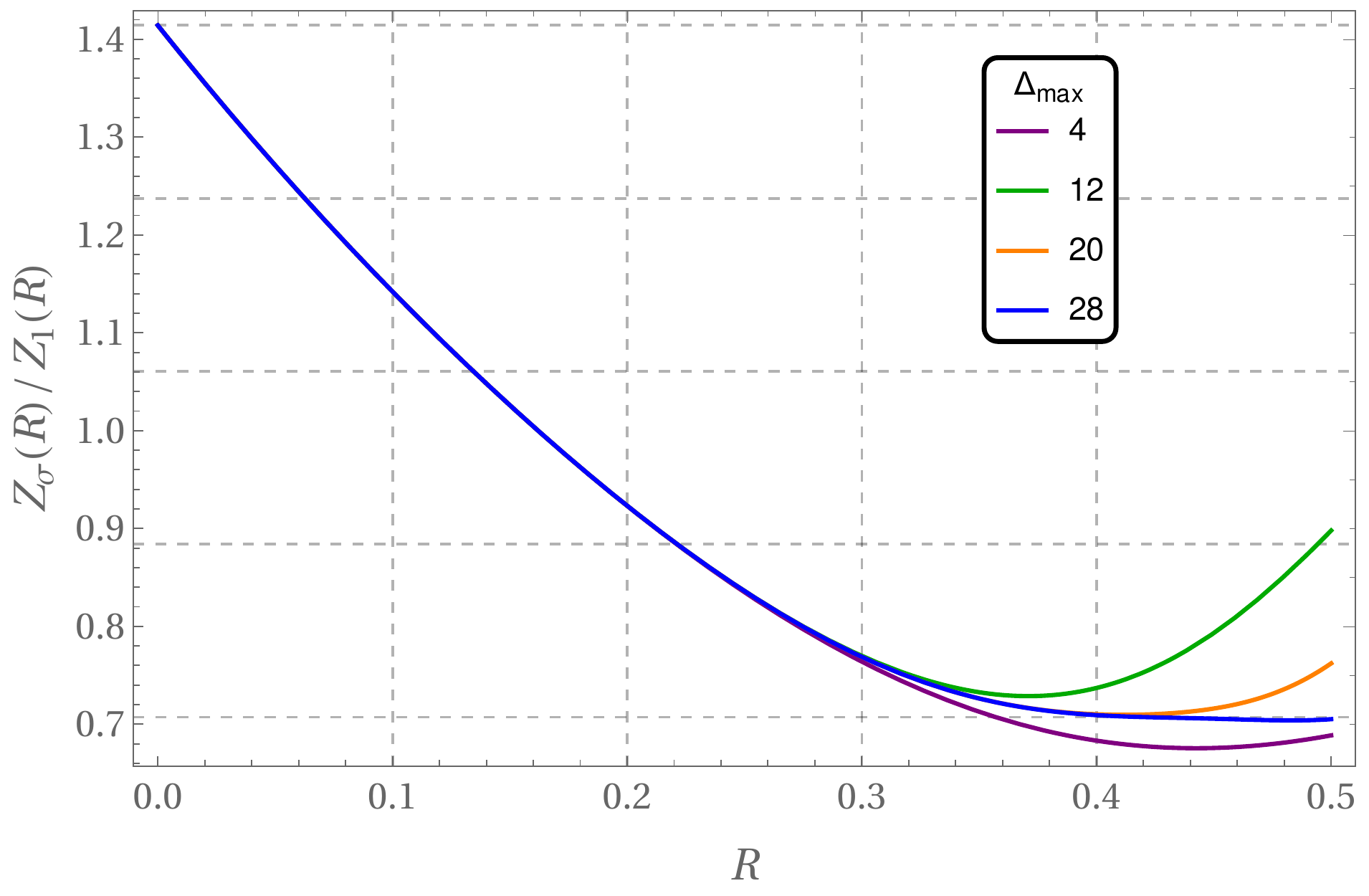}
    \caption{Results for the approximation in the closed channel for different values of the cutoff $\Delta_\mathrm{max}$. From the open channel we know the exact result as $R\to1/2$ to be $\frac{1}{\sqrt{2}}$. One needs to go up to $\Delta_\mathrm{max} =28$ to come close to this value. }
    \label{fig:cl_channel_Ising}
\end{figure}

We finally present plots of the approximation in Figure \ref{fig:cl_channel_Ising}. It shows our results for the ratio $Z_\sigma(R)/Z_{\one}(R)$ in the Ising CFT for different cutoffs $\Delta_\mathrm{max}$ in the $(L_0+\overline L_0)$-weight of the summands in the boundary state. 
The four plots show the behaviour expected from a convergent series.

\subsubsection{Open channel}

\begin{figure}[t]
    \centering
    \includegraphics[width=.85\textwidth]{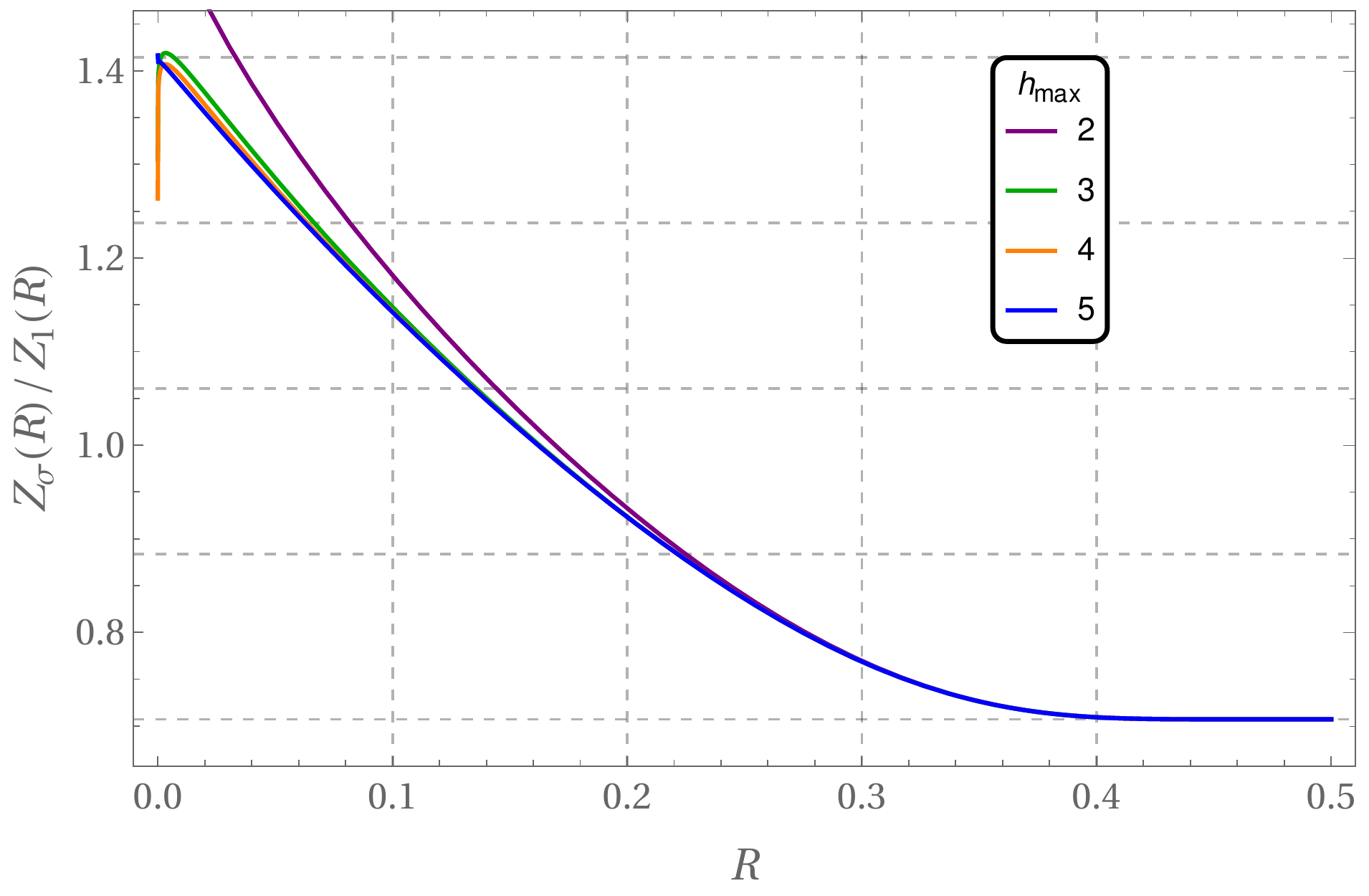}
    \caption{Results for the approximation in the open channel for different values of the cutoff $h_\mathrm{max}$. From the closed channel we know the exact result as $R\to0$ to be $\sqrt{2}$. 
    Already for small values of the cutoff we see very good agreement with this behaviour and the limit.   }
    \label{fig:op_channel_Ising}
\end{figure}

Now let us discuss the first few orders in an approximation of the one-hole partition function by introducing a cutoff in the open spectra, i.e. in \eqref{eq:OpenChannelSum} we only take basis vectors $\kappa_i$ 
up to some specific conformal weight $h_\mathrm{max}$\,. 
For example if we choose $h_\mathrm{max} = \frac12$ then only the vacuum and, in case of the $\sigma$ boundary, $\epsilon$ can appear in the sums, cf.\ \eqref{eq:Ising-openstates}. Hence, in this approximation $Z_\one(R)$ consist of the single term from the vacuum as in \eqref{eq:open-channel-vac-contrib},
\begin{equation}
    Z_\one(R) 
    \approx \mathcal{N} \, \sqrt{2} 
     \,,
\end{equation}
where we substituted $\left\langle\one\right\rangle_{D_\one}^\mathrm{un} = \sqrt{S_{\one\one}}$ from \eqref{eq:disc-unnorm} and the explicit value from \eqref{eq:Ising-S}.

The approximation of $Z_\sigma(R)$ also includes contributions from $\epsilon$ and is given by setting $h_\xi = \frac12$ 
and $\left\langle\one\right\rangle_{D_\sigma}^\mathrm{un} = S_{\sigma\one} /\sqrt{S_{\one\one}} = 1$
in \eqref{eq:openLeading},
\begin{equation}
    Z_\sigma(R) ~\approx~ \mathcal{N} \left(1 +  \frac{16}{9 X(t)^2} \, \right)\,.
\end{equation}

\noindent
Altogether, the ratio is given by
\begin{equation}
    \frac{Z_\sigma(R)}{Z_\one(R)} \approx \frac{1}{\sqrt{2}} \left(1 + \frac{16}{9\,X(t)^2}  \right)\,.
\end{equation}

\noindent 
Remember that $X(t)$ is given in \eqref{eq:X(t)} and $t = \frac1\pi \cosh^{-1}\!\left(\frac{1}{2R}\right)$\,.

If we instead choose $h_\mathrm{max} = 2$, then the basis vectors appearing in the sum are those of \eqref{eq:BasisOne} and \eqref{eq:BasisP}. The interaction vertex
$T_R^b(\kappa_1,\kappa_2,\kappa_3)$ is then a three-tensor with dimension $2\times2\times2$ in case of the $b=\one$ and with dimension $4\times4\times4$ in case of $b=\sigma$.
After contraction, the explicit expressions involve terms up to order $X(t)^{-12}$ and rational functions with polynomials in $t$ up to order 12: 
\begin{align}
    Z_{\one}(R) \approx& 
    \,\mathcal{N} \,\sqrt{2}\, 
    \Bigg(1 + 
    \frac{25 \left(4 t^2+1\right)^2}{15552 \left(t^2+1\right)^2 X^4}
    +\frac{\left(16784 t^4+32968 t^2+16409\right)^2}{725594112
   \left(t^2+1\right)^4 X^8}\nonumber\\
   &\hspace{4em}
   +\frac{\left(36
   t^2+41\right)^2 \left(22832 t^4+44824 t^2+22067\right)^2}{1253826625536
   \left(t^2+1\right)^6 X^{12}}\Bigg) 
   \nonumber \\
    Z_{\sigma}(R) \approx&\, 
        \,\mathcal{N}\,
    \Bigg(1
    +\frac{16}{9 X^2}
    -\frac{7792 t^4+16184 t^2+8167}{15552 \left(t^2+1\right)^2 X^4}
    +\frac{37264 t^4+77768 t^2+40729}{26244 \left(t^2+1\right)^2X^6}\nonumber\\
    &-\frac{16\left(65960176 t^6+258894064 t^4+380541786 t^2+248258959\right) t^2+970366351}{725594112\left(t^2+1\right)^4 X^8}\nonumber\\
    &+\frac{100 \left(36 t^2+41\right)^2}{531441 \left(t^2+1\right)^2 X^{10}}+\frac{\left(22832 t^4+44824 t^2+22067\right)^2 \left(36 t^2+41\right)^2}{1253826625536
   \left(t^2+1\right)^6 X^{12}}\Bigg)
\end{align}

\noindent
Note that higher cutoffs $h_\mathrm{max}$ can still change the coefficients of $X^{-n}$ for $n>4$ e.g. due to high energy excitation along a single open channel. 

We present the results for the ratio $Z_\sigma(R)/Z_\one(R)$ for different cutoffs in Figure~\ref{fig:op_channel_Ising}. Again the plots show the behaviour expected from a convergent series. 
The dimensions $D_{\one\one}$ and $D_{\sigma\sigma}$ of the open state spaces $\mathcal{H}_{\one\one}$ and $\mathcal{H}_{\sigma\sigma}$ after truncating at the values $h_\mathrm{max}$ used in Figure~\ref{fig:op_channel_Ising} are as follows:

%

\begin{center}
    \begin{tabular}{c|c|c|c|c}
        $h_\mathrm{max}$ & 2 & 3 & 4 & 5 \\
        \hline
        $D_{\one\one}$ & 2 & 3 & 5 & 7 \\
        $D_{\sigma\sigma}$ & 4 & 6 & 9  & 13 
    \end{tabular}
\end{center}

\noindent
Note that an increase in the dimension $D$ of the truncated open state space
increases the complexity of the computation much more than increasing the cutoff in the closed channel. This is because in the open channel we have to compute the $\mathcal{O}(D^3)$ amplitudes. Their computation needs several steps -- particularly the transformations and three-point correlators of descendant states on the disk -- which occupy computational resources. 
Still, we expect that this can be pushed to much higher levels with a more optimised computer code.

Introducing more holes eventually makes the open channel computation more efficient than the closed channel one, as the computation of torus $N$-point functions for increasing $N$ quickly becomes computationally expensive even for vacuum descendants.   

\subsubsection{Comparison}

The visualisation in Figure \ref{fig:cl_channel_Ising} and Figure \ref{fig:op_channel_Ising} already show large similarities, that become more apparent the larger the cutoffs are. It seems that the error in the approximation scales much better with the cutoff in the open channel. Already for small cutoffs $h_\mathrm{max}$ the results change very little, even close to $R=0$. In the closed channel we see quite some changes close to $R=\frac{1}{2}$ up to cutoffs of $\Delta_\mathrm{max} = 28$\,. 

\begin{figure}[t]
    \centering
    \includegraphics[width=.85\textwidth]{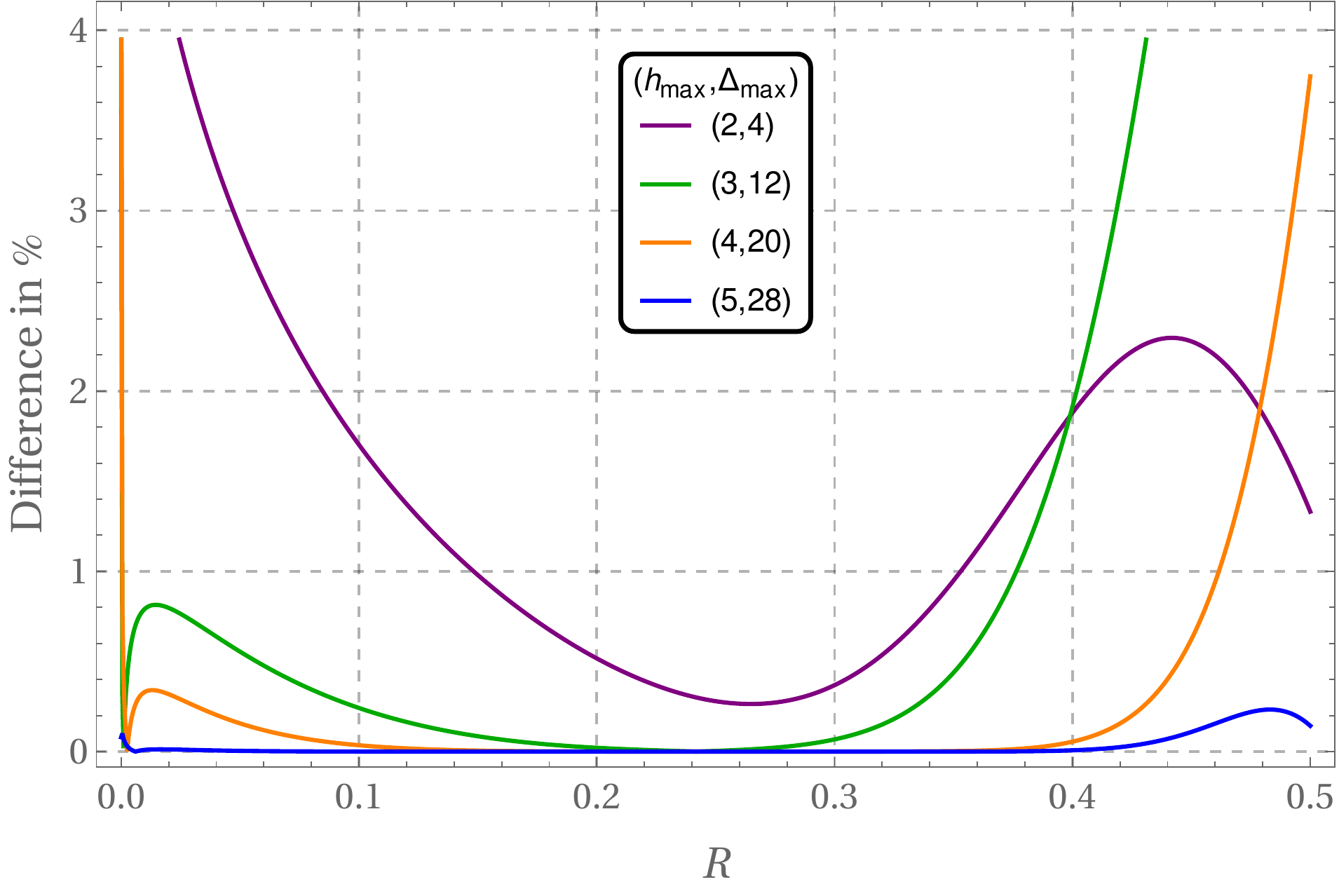}
    \caption{Difference in relation to the average (in $\%$) of the approximations for increasing cutoffs in the closed and open channel. We can in particular see that the approximation in the open channel is very good already for rather small values of the cutoff $h_\mathrm{max}$\,.}
    \label{fig:difference_Ising}
\end{figure}

A direct comparison of the two channels is given in Figure \ref{fig:difference_Ising}.
It is apparent that the difference shrinks with increasing cutoffs. Note again that the behaviour closer to $R=0$ correlates with the changes in $h_\mathrm{max}$ whereas the behaviour closer to $R=1/2$ does with the change in $\Delta_\mathrm{max}$\,.

The good results and the apparent fast convergence in the open channel are quite promising. It seems not illusory that a lattice model defined in terms of the vertices we constructed in this paper -- but for the cloaking boundary condition -- will give a good approximation to the real $N$-hole partition function of the original two-dimensional conformal field theory and that a sensible thermodynamic/continuum limit of that model with potentially fine-tuned parameters $R/d$ and $h_\mathrm{max}$ can reproduce the Ising CFT. 

We will investigate this in more detail in future parts of this series of papers.

\newpage
\appendix

\section{More on the Weierstrass functions}\label{app:Weierstr}

\subsubsection*{Invariants and expansion coefficient}

The first two coefficients $a_k^{\small(0)}$ of $\wp$ in \eqref{eq:wp2} are $a_1^{\small(0)} = \frac{g_2}{20}$ and $a_2^{\small(0)}=\frac{g_3}{28}$ \cite[Eq.\,18.5.2]{Abramowitz:1970} with the \textit{invariants} \cite[Eq.\,18.1.1]{Abramowitz:1970}
\begin{align}
    g_2 = 60 \sum_{w\in \Lambda\backslash\{0\}} \frac{1}{w^4}
    \quad , \quad
    g_3 = 140 \sum_{w\in \Lambda\backslash\{0\}} \frac{1}{w^6}\,.
\end{align}

\noindent
All higher coefficients follow from the recursive formula \cite[Eq.\,18.5.3]{Abramowitz:1970}
\begin{equation}
    a_k^{(0)} = \frac{3}{(2k+3)(k-2)}\sum\limits_{l=1}^{k-2} a_{l}^{(0)}a_{k-l-1}^{(0)}\,,
\end{equation}
and, hence, are polynomials in the invariants $g_2$ and $g_3$.

\subsubsection*{Derivatives of Weierstrass functions}

We can use the following identities for the derivatives of the Weierstrass functions
\begin{align}
    \frac{d \zeta(z)}{d\omega_2} = & -\frac{i}{2\pi}  \left(\frac{\omega_1}{2}\left(\wp'(z) + 2 \zeta(z)\wp(z) - \frac{g_2 z}{6}\right) + 2\eta_1 \left(\zeta(z)-z\wp(z)\right) \right)\,,
    \nonumber\\
    \frac{d \wp(z)}{d\omega_2} = & \frac{i}{\pi}  \left(\frac{\omega_1}{2}\left(2\wp(z)^2 + \zeta(z)\wp'(z) - \frac{g_2}{3}\right) - \eta_1 \left(2\wp(z)+z\wp'(z)\right) \right)\,,
    \nonumber\\
    \frac{d\wp'(z)}{d\omega_2} = & \frac{i}{2\pi}  \Big( \frac{\omega_1}{2} \left(6 \wp(z) \wp'(z) + 12 \zeta(z)\wp(z)^2 -g_2 \zeta(z)\right)\nonumber\\
    & \phantom{-\frac{i}{2\pi}  \Big(}- \eta_1\left(6\wp'(z) +12 z \wp(z)^2 -g_2 z\right)\Big)\,,
\end{align}
where $\eta_i = \zeta(\omega_i/2)$ as in \eqref{eq:eta-i-def} and $g_i$ are the Weierstrass invariants as before. This e.g. follows from \cite[Eq.\,18.6.19--24]{Abramowitz:1970} together with $g_2(\Lambda_\tau) = \frac{4\pi^2}{3}E_4$, $g_3(\Lambda_\tau)= \frac{8\pi^2}{27}E_6$ and $\eta_1(\Lambda_\tau) = \frac{\pi^2}{6} E_2$\,, and \cite{Eichler:1982}
\begin{align}\label{eq:EisensteinDer}
    \frac{1}{2\pi i} \partial_\tau E_4 = \frac{1}{3} \left(E_2E_4-E_6\right)\,,\quad \frac{1}{2\pi i} \partial_\tau E_6 \left(E_2E_6-E_4^2\right)\,,
\end{align}
where $E_n$ is the $n$th Eisenstein series and $\Lambda_\tau$ is the lattice with periods $\omega_1=1$ and $\omega_2=\tau$.\footnote{
    The
above expressions for the derivatives of the Weierstrass functions can also be found 
    e.g.\ in 
the online documentation of Wolfram on elliptic functions, \url{https://functions.wolfram.com/EllipticFunctions/}\,.} 

To compute the descendant correlator recursively we also define 
\begin{align}
    e_i &:= \wp(\omega_i/2)
    \nonumber\\
    f_i &:= \partial_z\wp(\omega_i/2)
\end{align}
and use the above identities to obtain
\begin{align}
    \frac{d\eta_1}{d\omega_2} =& -\frac{i}{2\pi}  \left(\frac{\omega_1}{2}\left(f_1 - \frac{g_2 \omega_1}{12}\right) + 2\eta_1^2   \right)\,, 
    \nonumber\\
    \frac{de_1}{d\omega_2} =& \frac{i}{\pi}\left(\frac{\omega_1}{2}\left(2e_1^2 - \frac{g_2}{3}\right) - 2\eta_1 e_1 \right)\,,
    \nonumber\\
    \frac{df_1}{d\omega_2} = & \frac{i}{2\pi}  \left( \frac{\omega_1}{2} \left(6 e_1 f_1 -g_2 \eta_1\right)-\eta_1\left(6f_1 -\frac{\omega_1 g_2 }{2}\right)\right)\,.
\end{align}

\noindent 
Finally, we also need the derivatives of the Weierstrass invariants which are given by
\begin{align}
    \frac{dg_2}{d\omega_2} = \frac{i}{2\pi} \big(6 g_3 \omega_1-8g_2\eta_1 \big)
    \quad , \quad
    \frac{dg_3}{d\omega_2} = \frac{i}{2\pi} \Big( \frac{g_2^2 \omega_1}{3}-12g_3\eta_1 \Big)\,,
\end{align}
which follow directly from \eqref{eq:EisensteinDer}.

\section{Uniformising maps}

\subsection{The map \texorpdfstring{$F$}{F} for the clipped triangle}\label{sec:obtainF}

\begin{figure}[t]
    \centering
    \includegraphics{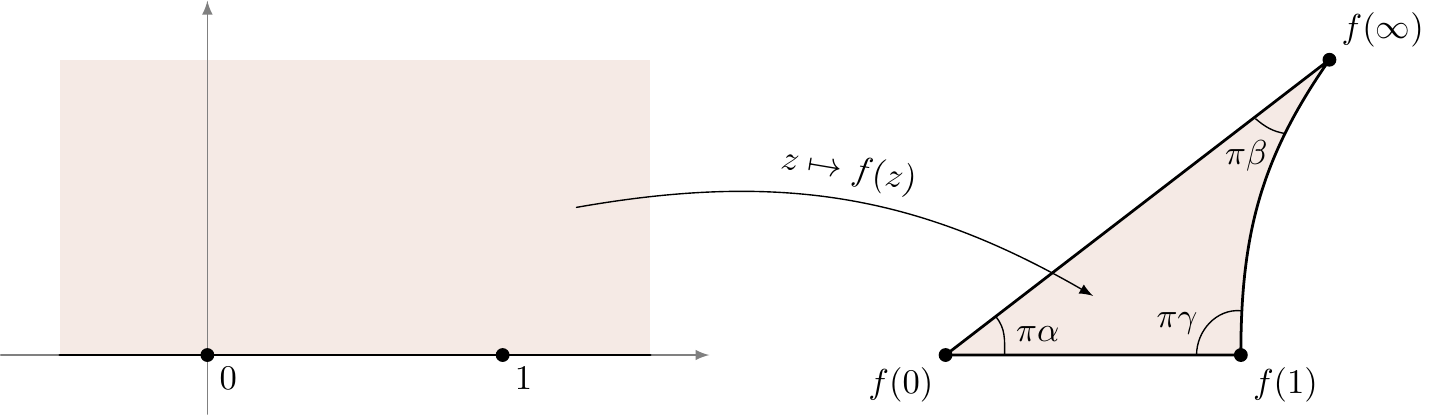}
    \caption{The function $f$ maps the upper half plane to the interior of a curvilinear triangle with straight lines between the point $f(0)$ and $f(1)$ and $f(0)$ and $f(\infty)\equiv\lim_{z\to\infty} f(z)$ and circle arc between the points $f(1)$ and $f(\infty)$.  
    }
    \label{fig:curvilinear}
\end{figure}

The conformal map $F$ is an adaptation of the conformal mapping of domains which are bounded by a finite number of circular arcs as discussed in \cite[Sec.\,V.7]{nehari1975conformal}. The map between the upper half plane and the interior of a curvilinear triangle with angles $\pi \alpha$, $\pi\beta$ and $\pi \gamma$ (see the right of Figure~\ref{fig:curvilinear}) is given by
\begin{equation}
    f(z) = \frac{y_2(z)}{y_1(z)}\,,
\end{equation}
where $y_1$ and $y_2$ are linear independent solutions of the hypergeometric equation
\begin{equation}
    z(1-z)y'' + (c-(a+b+1)z)y' - a b y = 0\,,
\end{equation}
with 
\begin{equation}
    a = \frac{1+\beta-\alpha-\gamma}{2}\,,\quad b = \frac{1-\alpha-\beta-\gamma}{2}\,,\quad c=1-\alpha\,.
\end{equation}

\noindent
Two independent solutions to the hypergeometric equation are \cite[9.153]{zwillinger2014table}
\begin{equation}
    \begin{split}
        y_1(z) &= \,_2F_1(a,b,c;z)\,\quad\text{and}\\
        y_2(z) &= z^{1-c} \,_2F_1(a-c+1,b-c+1,2-c;z)\,.      
        \,
    \end{split}
\end{equation}

\noindent 
For our purpose we fix $\alpha = \frac{1}{3}$ and $\gamma=\frac{1}{2}$, s.t. 
\begin{equation}\label{eq:f-map}
    f(z) = z^{\frac{1}{3}} \frac{_2F_1\left(\frac{5}{12}+\frac\beta2,\frac{5}{12}-\frac\beta2,\frac{4}{3};z\right)}{_2F_1\left(\frac{1}{12}+\frac\beta2,\frac{1}{12}-\frac\beta2,\frac{2}{3};z\right)}\,.
\end{equation}

\noindent 
It follows immediately that $f(z) = z^{\frac{1}{3}} + \mathcal{O}(z^{\frac{4}{3}})$ for an expansion around $z=0$. For the expansion around $z=1$ one can use \cite[Eq.\,9.131\,(2)]{zwillinger2014table}
\begin{equation}
\begin{split}
    _2F_1(a,b,c;1-z) &= \frac{\Gamma(c)\Gamma(c-a-b)}{\Gamma(c-a)\Gamma(c-b)}\,_2F_1(a,b,a+b-c+1;z) \\
    &~~+ z^{c-a-b} \frac{\Gamma(c)\Gamma(a+b-c)}{\Gamma(a)\Gamma(b)}\,_2F_1(c-a,c-b,c-a-b+1;z)
\end{split}
\end{equation}
and some identities for the $\Gamma$-function to obtain
\begin{equation}
\begin{split}
    f(z) &= \frac{2\cos(\pi\beta)-\sqrt{3}}{\pi^{\frac{3}{2}}} \Gamma\!\left(\tfrac{7}{6}\right) \Gamma\!\left(\tfrac{1}{6}-\beta\right)\Gamma\!\left(\tfrac{1}{6}+\beta\right) \\
    &~~+ i \frac{ 2^\frac{4}{3}\Gamma\!\left(\frac{1}{6}\right)}{\sqrt{3\pi}\left(2\cos(\pi\beta)+\sqrt{3}\right)} \frac{\Gamma\!\left(\frac{7}{12}-\frac{\beta}{2}\right)\Gamma\!\left(\frac{7}{12}+\frac{\beta}{2}\right)}{\Gamma\!\left(\frac{5}{12}-\frac{\beta}{2}\right)\Gamma\!\left(\frac{5}{12}+\frac{\beta}{2}\right)}  \sqrt{z-1}  +\mathcal{O}\left(z-1\right)\,.
\end{split}
\end{equation}

\noindent
The hypergeometric function has branch cut from $1$ to $+\infty$, and $\sqrt{z-1}$ has branch cut from $z=1$ to $-\infty$ and is positive for $z>1$. The above expansion is valid for $z$ in the upper half plane. 
Note that the qualitative behaviour of $f$ around $z=0,1$ does not change for $\beta \in \mathbb{R}$ and $\beta\in i \mathbb{R}$\, and, in particular, $f(1)\in \mathbb{R}^+$ and $f'(1) \in i \mathbb{R}^+$\, in both cases.

\medskip

We want to set $\beta\equiv it$ and claim that this will give the function we aim for. As a check, we will now verify that the line $z \in \mathbb{R}_{>1}$ is mapped onto a circle, and we compute centre and radius of that circle. To evaluate $f(z)$ for $z>1$ we use the hypergeometric identity \cite[Eq.\,9.132\,(2)]{zwillinger2014table}
\begin{align}
_2F_1(\alpha,\beta;\gamma;z) &= \frac{\Gamma\left(\gamma\right)\Gamma\left(\beta-\alpha\right)}{\Gamma\left(\beta\right)\Gamma\left(\gamma-\alpha\right)} (-z)^{-\alpha}\, _2F_1\left(\alpha,\alpha+1-\gamma;\alpha+1-\beta;\frac{1}{z}\right) 
\nonumber \\
&\quad + \frac{\Gamma\left(\gamma\right)\Gamma\left(\alpha-\beta\right)}{\Gamma\left(\alpha\right)\Gamma\left(\gamma-\beta\right)} (-z)^{-\beta}\, _2F_1\left(\beta,\beta+1-\gamma;\beta+1-\alpha;\frac{1}{z}\right)
\label{eq:hypergeo_id1/z}
\end{align}
for $\vert\text{arg}(z)\vert<\pi$ and $\alpha-\beta \notin \mathbb{Z}$\,. 
This allows us to rewrite $f$ in \eqref{eq:f-map} as
\begin{equation}\label{eq:f-transformed}
    f(z) = e^{\frac{\pi i}{3}} \, \frac{a(t) \,U(z) + b(t)\, V(z)}{c(t) \,U(z) +d(t)\, V(z)}\,,
\end{equation}
with
\begin{align}
   a(t) &= \frac{\Gamma\left(\frac{4}{3}\right)\Gamma\left(i t\right)}{\Gamma\left(\frac{5}{12}+ \frac{it}{2}\right)\Gamma\left(\frac{11}{12}+\frac{i t}{2}\right)} = b(-t)\,,\nonumber \\
   c(t) &=  \frac{\Gamma\left(\frac{2}{3}\right)\Gamma\left(i t\right)}{\Gamma\left(\frac{1}{12}+ \frac{i t}{2}\right)\Gamma\left(\frac{7}{12}+\frac{i t}{2}\right)} = d(-t)\,,
   \nonumber\\
    U(z) &= (-z)^{\frac{it}{2}} \,_2F_1 \left(\frac{5}{12}-\frac{it}{2},\frac{1}{12}-\frac{it}{2}\,;1-it\,;\frac{1}{z}\right) \nonumber \\
    V(z) &= (-z)^{\frac{-it}{2}} \,_2F_1 \left(\frac{5}{12}+\frac{it}{2},\frac{1}{12}+\frac{it}{2}\,;1+it\,;\frac{1}{z}\right)\,.\label{eq:abcduv-abbr}
\end{align}

\noindent
For $z>1$ and $t\in\mathbb{R}$, the relevant branch is $(-z)^{\pm it/2} = e^{\pm\pi t/2} e^{\pm\log(z) i t/2}$ and one gets $V=e^{-\pi t} \overline{U}$. 
Hence, the function $z \mapsto \frac{U(z)}{V(z)}$ is of the form 
$\frac{U(z)}{V(z)}= e^{\pi t} e^{i \theta(z)}$ with $\theta(z) \in \mathbb{R}$\,, s.t.\ the half line $\{z|z\in \mathbb{R},z>1\}$ is mapped onto the circle of radius $e^{\pi t}$ with its center around the origin. This is followed in \eqref{eq:f-transformed}
by the M\"obius transformation $w\mapsto \frac{a(t) w+b(t) }{c(t) w+d(t)}$.
M\"obius transformations map circles to circles,\footnote{Here circles include lines which are regarded as 'circles through infinity'.} and we now review from \cite[Sec.\,2.9--2.15]{priestley2003introduction} 
how to obtain the transformation of radius and centre. 
Any circle on the complex plane can be represented by the equation 
\begin{equation}\label{eq:circle-equation}
    \frac{|z-\mu|}{|z-\nu|}=\lambda\,,\quad \text{with $\mu \neq \nu$ and $\lambda>0$ .}
\end{equation}

\noindent 
The points $\mu$ and $\nu$ are called inverse points, and for $\lambda=1$ the latter equation gives a line. For example, the circle of radius $r$ around the origin can be represented by $\mu = \lambda r$, $\nu=r/\lambda$ for any $\lambda>0$ and $\lambda\neq1$. 
The center $q$ and the radius $r$ of the circle \eqref{eq:circle-equation} are given by
\begin{equation}
    q=\frac{\mu-\lambda^2\nu}{1-\lambda^2} 
    \quad\text{and}\quad 
    r=\left|\frac{\lambda(\mu-\nu)}{1-\lambda^2}\right|\,. 
    \label{eq:center+radius}
\end{equation}

\noindent
Under a M\"obius transformation 
\begin{equation}
    M(z) = \frac{a z +b}{c z + d}
\end{equation}
a circle of inverse points $\mu$ and $\nu$ is mapped onto a circle with inverse point $M(\mu)$ and $M(\nu)$. It is then the solution to
\begin{equation}
    \left|\frac{w-M(\mu)}{w-M(\nu)}\right| = \tilde{\lambda}=\lambda \left|\frac{\nu c +d}{\mu c+d}\right|\quad \mathrm{if}~ \mu c +d \neq 0 ~\mathrm{ and }~ \nu c +d \neq 0\,.
\end{equation}

\begin{figure}[t]
    \centering
    \includegraphics{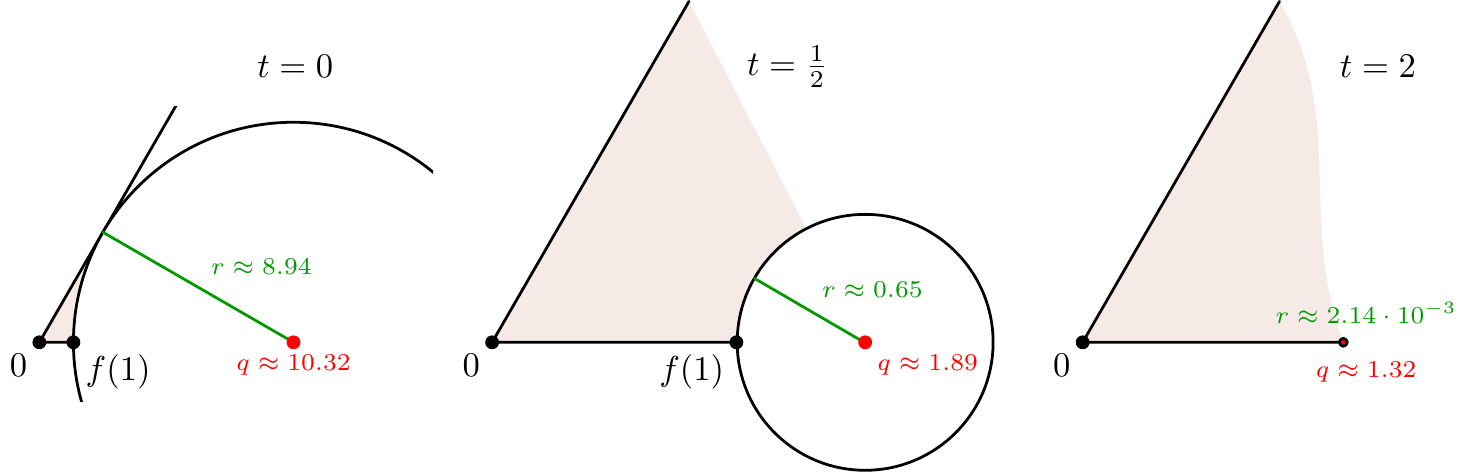}
    \caption{The image $f$ for $\alpha=\frac{1}{3},\gamma=\frac12$ and different values of the $\beta=it$.
}
    \label{fig:touching_limit}
\end{figure}

\noindent
Using these results one can compute that the circle with radius $r$ around the origin gets mapped to a circle with centre $q$ and radius $\tilde r$ given by
\begin{align}
    q &= \frac{M(\lambda r)- \tilde{\lambda}^2 M(r/\lambda)}{1-\tilde{\lambda}^2} = \frac{b \bar{d} - a \bar{c} r^2}{|d|^2 - |c|^2  r^2}
    \nonumber\\
    \tilde{r} &= \left| \frac{\tilde{\lambda} \big(M(\lambda r)-M(r/\lambda) \big)}{1-\tilde{\lambda}^2} \right| = \frac{|ad-bc|\,r}{\big|\,|d|^2-|c|^2 r^2\,\big| }\,.
\end{align}

\noindent 
We can apply these formulas to $f$: Using $\bar{d}(t) = d(-t) = c(t)$ and $\bar{b}(t) = b(-t) = a(t)$ it follows that $f$ maps $z>1$ to the circle around
\begin{align}
    q &= e^{\frac{\pi i}{3}}\frac{1}{1-e^{2\pi t}}\left(\frac{b(t)}{d(t)}-\frac{a(t)}{c(t)}e^{2\pi t}\right)
    \nonumber\\
    &= 2\sqrt{\pi} \frac{\cosh(\pi t)}{\cosh^2(\pi t)-\frac{3}{4}} \frac{\Gamma\left(\frac{7}{6}\right)}{\Gamma\left(\frac{5}{6}+i t\right)\Gamma\left(\frac{5}{6}-i t\right)} \equiv L(t)
\end{align}
with a radius 
\begin{equation}
    \tilde{r} = \frac{\sqrt{3\pi}}{\cosh^2(\pi t)-\frac{3}{4}} \frac{\Gamma\left(\frac{7}{6}\right)}{\Gamma\left(\frac{5}{6}- it\right)\Gamma\left(\frac{5}{6}+ it\right)} \equiv R(t)\,,
\end{equation}
where $L(t)$ and $R(t)$ are as in \eqref{eq:L(t)etc}.
Since $f(1)+\tilde{r} = q$ we see that the circle indeed touches the image of $0<z<1$ and we obtain the situation illustrated in Figure~\ref{fig:touching_limit}.

We also want to check the asymptotic values: For $t=0$ the center and the radius of the circle $f(z>1)$ is given by
\begin{equation}
    q = 8\sqrt{\pi}\frac{\Gamma\left(\frac{7}{6}\right)}{\Gamma\left(\frac{5}{6}\right)^2}
    \quad , \qquad
    \tilde{r} = 4\sqrt{3\pi}\frac{\Gamma\left(\frac{7}{6}\right)}{\Gamma\left(\frac{5}{6}\right)^2}\,.
\end{equation}

\noindent 
Then $\frac{\tilde{r}}{q} = \frac{\sqrt{3}}{2} = \sin\frac{\pi}{3}$, i.e. the circle touches the line $f(z<0)$ (see the left of Figure~\ref{fig:touching_limit}). 
For $t\to \infty$, $\tilde{r}\to 0$ and $q\to 0$. 
However the ratio also vanishes as 
\begin{equation}
    \left.\frac{\tilde{r}}{q}\right|_{t\gg 1} \sim \frac{\sqrt{3}}{2} e^{-\pi t}\,.
\end{equation}

\noindent
The above analysis indicates that, for $\beta=it$, $f$  maps (part of) the upper half plane to a sixth of the clipped triangle. The 'radius of the hole' and the 'distance to the center of the triangle' depend on $t$ and their ratio takes all values between $0$, the vanishing holes limit, and $\frac{\sqrt{3}}{2}$, the touching holes limit. 

\begin{figure}[t]
    \centering
    \includegraphics{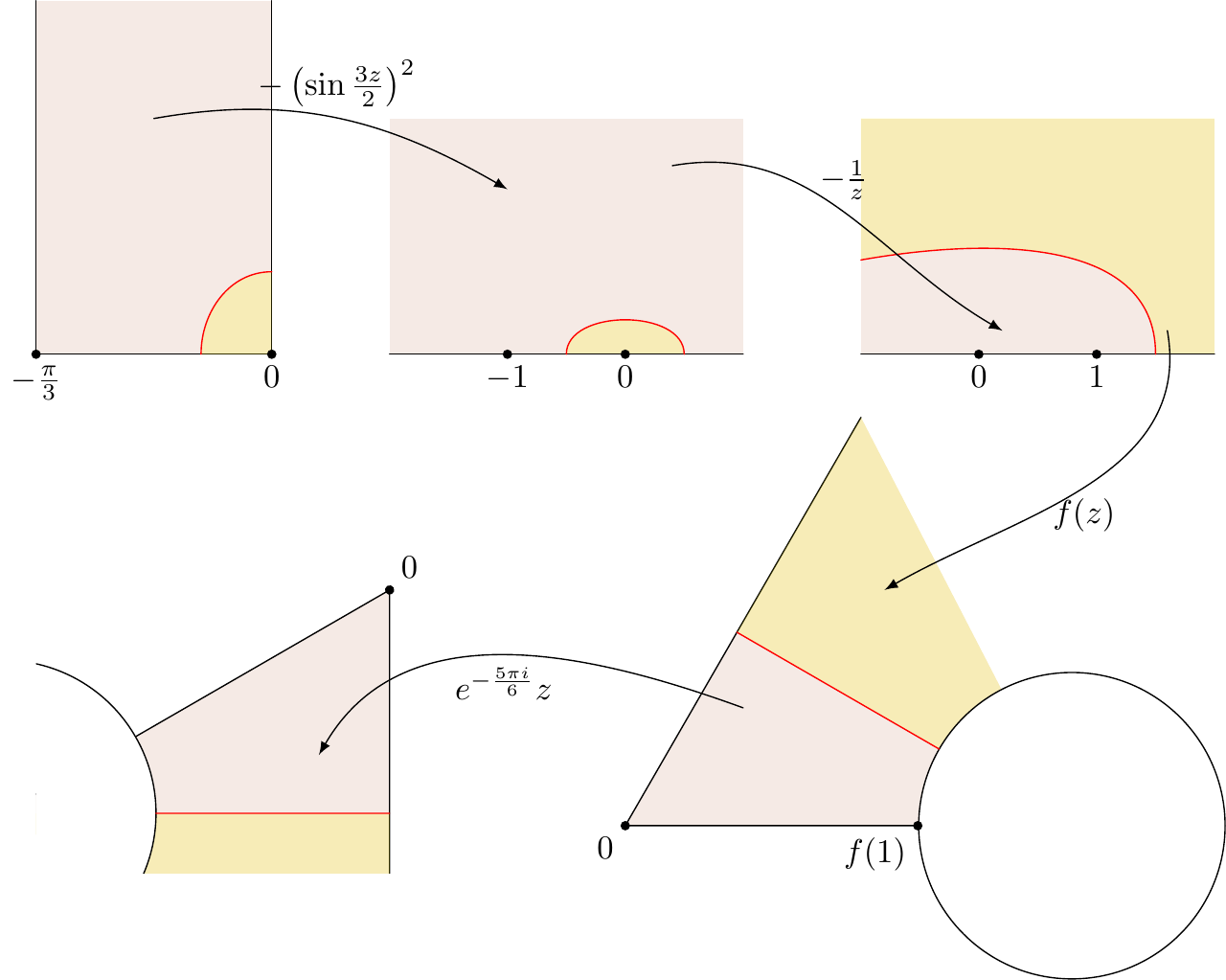}
    \caption{Visualisation of the steps for the map from the half strip with $\frac{\pi}{3}\le\mathrm{Re}(z)\le0$ to a sixth of sixth of the clipped triangle with a cut (red line) parallel to the real axis. Note that the exact form of the pre-image of the cut is not known, but also not needed. }
    \label{fig:decompose_maps}
\end{figure}

Now, we use the result that $z=\sin(w)$ maps the half-strip $\{|\mathrm{Re}(w)|< \frac{\pi}{2},\mathrm{Im}(w)>0\}$ onto the upper half plane $\mathrm{Im}(z)>0$ \cite[Sec. V.6]{nehari1975conformal}. 
Then, 
$\frac{1}{2} \sin\left(3 w +\frac{\pi}{2}\right)-\frac{1}{2} = - \left(\sin\frac{3w}{2}\right)^{2}$ 
maps the half-strip with $-\frac{\pi}{3}<\mathrm{Re}(w)<0$ onto the upper half plane with 
    $-\frac{\pi}{3}\mapsto -1$ 
and $0\mapsto 0$. We also use a rotation $e^{-\frac{5\pi i}{6}}$, such that state boundary, where we insert the sum over intermediate states, is parallel to the real axis. All of this is visualised in Figure~\ref{fig:decompose_maps}.

Finally, with all the above evidence and numerical checks (see e.g.\ Figure \ref{fig:tildeF-plot}), we claim that the function 
\begin{equation}
    F(z) = e^{-\frac{5\pi i}{6}} f\!\left(\left(\sin\tfrac{3z}{2}\right)^{-2}\right) \, ,
\end{equation}
as stated in \eqref{eq:F}, is the function we were looking for. 

\subsection{The maps \texorpdfstring{$G_0$}{G} and \texorpdfstring{$\psi_0$}{psi}}
\label{sec:G-and-psi}

\begin{figure}[t]
    \centering
    \includegraphics{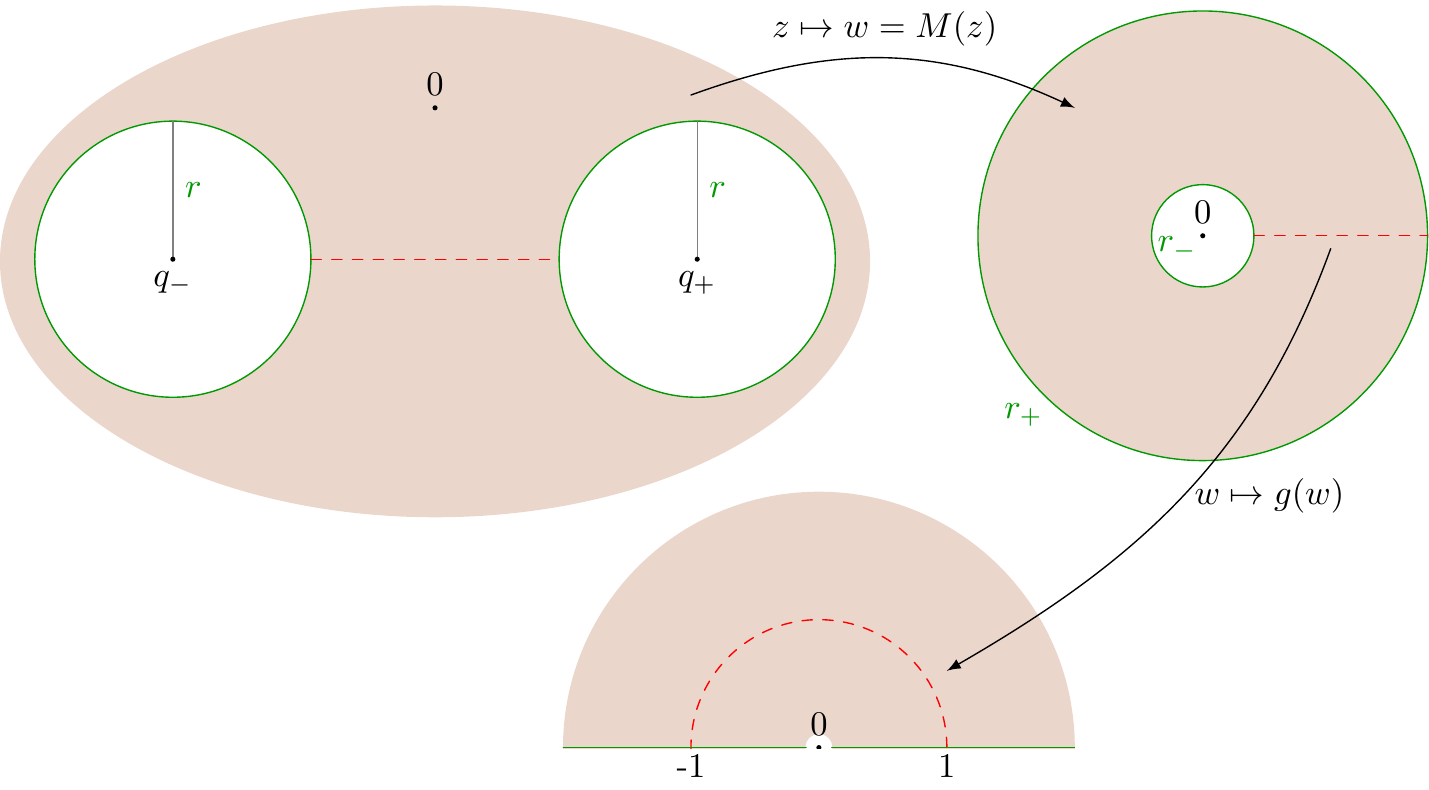}
    \caption{Visualisation of the steps to compute $G$ which allows to parametrize the cut (red dotted line) between two holes by the upper unit circle. 
    }
    \label{fig:G_derivation}
\end{figure}

Consider two circles on the complex plane with centers at $q_{\pm} = \frac{d}{2} (\pm 1 - \frac{i}{\sqrt{3}})$ and radius $r<\frac{d}{2}$ as in Figure~\ref{fig:G_derivation}.
They can be represented by the equations
\begin{align}
    \left|\frac{z-\mu_\pm}{z-\nu_\pm}\right| = \lambda\,,
\end{align}
for $\lambda>0$, $\lambda\neq1$\,, with $\mu_\pm = q_{\pm}+\lambda r$ and $\nu_\pm =q_{\pm}+\frac{r}{\lambda}$. The M\"obius transformation 
\begin{equation}\label{eq:Moebius2}
    M(z) = -\frac{z + \sqrt{\frac{d^2}{4} - r^2} + i \frac{d}{2\sqrt{3}}}{z - \sqrt{\frac{d^2}{4} - r^2} + i \frac{d}{2\sqrt{3}}}
\end{equation}
maps the two circles on circles around the origin with radii (using \eqref{eq:center+radius})
\begin{align}
    r_\pm = \frac{d\pm\sqrt{d^2-4r^2}}{2r}\,.
\end{align}

\noindent
We now use $\frac{\log\left(x\pm\sqrt{x^2-1}\right)}{\log\left(x+\sqrt{x^2-1}\right)}=\pm1,$ for $x>1$, and  $\cosh^{-1}(x) = \log\left(x+\sqrt{x^2-1}\right)$ to obtain
\begin{equation}
    i \exp\!\left(\frac{-i\pi \log(r_\pm e^{-i \varphi}) }{2\cosh^{-1}\left(\frac{d}{2r}\right)} \right)
    =  \pm e^{\frac{-\pi \varphi}{2\cosh^{-1}\left(\frac{d}{2r}\right)}}\,.
\end{equation}
This shows that 
\begin{equation}
w \,\longmapsto\, g(w)  = 
    i \exp\!\left(\frac{-i\pi \log w}{2\cosh^{-1}\left(\frac{d}{2r}\right)} \right)
\end{equation}
maps the two circles onto the two lines on the real axis with 
\begin{equation}
\exp\left(-\frac{\pi^2}{2\cosh^{-1}\left(\frac{d}{2r}\right)}\right) \le |x| \le  \exp\left(\frac{\pi^2}{2\cosh^{-1}\left(\frac{d}{2r}\right)}\right)
\,,
\end{equation}
see Figure~\ref{fig:G_derivation}. 

The line that connects the centers of the two original circles, which in particular includes the segment of the cut, also crosses the inverse points. First the M\"obius transformations maps it onto the real line and the cut is mapped onto $r_- < x < r_+$. Then $g$ maps this segment onto the upper unit half circle. 

We now use the parametrisation of the hole distance and radius as in \eqref{eq:L(t)etc}, s.t. 
\begin{align}
    d&=\sqrt{3}L(t)
    \,,& r = R(t) &= \frac{\sqrt{3}}{2 \cosh{\pi t}}L(t)\,, \quad \text{and hence} 
    \nonumber \\
    t &= \frac{\cosh^{-1}\!\left(\frac{d}{2r}\right)}{\pi}\,, & \frac{d^2}{4} - r^2 &= \frac{3}{4} L(t)^2 \left(\tanh{\pi t}\right)^2\,.
\end{align}

\noindent 
With this we get $g(w) = i\exp\left(\frac{1}{2it} \log w \right)$ and 
\begin{equation}
    M(z) = \frac{\frac{\sqrt{3}}{2} L(t) \tanh(\pi t) + i\frac{L(t)}{2} + z}{\frac{\sqrt{3}}{2} L(t) \tanh(\pi t) - i\frac{L(t)}{2} - z}\,.
\end{equation}

\noindent 
Hence $g(M(z))$ reproduces the map $G_0$ in \eqref{eq:Gn-def} which, by above considerations, obeys all the properties we were looking for. It maps the state boundary onto the unit circle, is well-defined in an open neighbourhood of the state boundary, and maps the boundary in that neighbourhood on the real axis. 

\medskip

To obtain the expression in \eqref{eq:psi} we use \eqref{eq:hypergeo_id1/z}. 
This allows us to rewrite $F$ in \eqref{eq:F} as
\begin{equation}
    F(z) = \frac{1}{i} \frac{a(t) \,\tilde{U}(z) + b(t) \,\tilde{V}(z)}{c(t)\, \tilde{U}(z) + d(t)\, \tilde{V}(z)}
\end{equation}
with $a,b,c,d$ as in \eqref{eq:abcduv-abbr}, and 
\begin{align}
   \tilde{U}(z) &= \left(\frac{\sin\frac{3z}{2}}{i}\right)^{-it} \,_2F_1 \left(\frac{5}{12}-\frac{it}{2},\frac{1}{12}-\frac{it}{2}\,;1-it\,;
       \left(\sin\frac{3z}{2}\right)^2
   \right)
   \nonumber \\
   \tilde{V}(z) &= \left(\frac{\sin\frac{3z}{2}}{i}\right)^{it} \,_2F_1 \left(\frac{5}{12}+\frac{it}{2},\frac{1}{12}+\frac{it}{2}\,;1+it\,;
       \left(\sin\frac{3z}{2}\right)^2
   \right) 
   \,.
\end{align}

\noindent 
With this, and using the abbreviation $\xi = \frac{\sqrt{3}}{2} L \tanh\left(\pi t\right)$ the composition of the two maps is given by
\begin{align}
    \psi_0(z) = G_0(F(z)) &= i \exp\left(\frac{1}{2it} \log\left(\frac{\xi+\frac{iL}{2} + \frac{1}{i} \frac{a(t) \,\tilde{U}(z) + b(t) \,\tilde{V}(z)}{c(t) \,\tilde{U}(z) + d(t) \,\tilde{V}(z)}}{\xi-\frac{iL}{2} - \frac{1}{i} \frac{a(t) \,\tilde{U}(z) + b(t) \,\tilde{V}(z)}{c(t) \,\tilde{U}(z) + d(t) \,\tilde{V}(z)}}\right)\right)
    \nonumber\\
    &= i \exp\left(\frac{1}{2it} \log\left(\frac{\left(\left(i\xi-\frac{L}{2}\right) c + a\right)\tilde{U}  + \left(\left(i\xi-\frac{L}{2}\right) d + b\right)\tilde{V}}{\left(\left(i\xi+\frac{L}{2}\right) c - a\right)\tilde{U}  + \left(\left(i\xi+\frac{L}{2}\right) d - b\right)\tilde{V}}\right)\right)\,.
\end{align}

\noindent 
Now, one can show (using Gamma-function and hyperbolic identities) that 
\begin{align}
    i\xi -\frac{L}{2} = -\frac{a}{c}\,, \quad \text{and} \quad i\xi + \frac{L}{2} = \frac{b}{d}
\end{align}
such that we get ($s(z) = \sin\frac{3z}{2}$)
\begin{align}
    \psi_0(z) &= i \exp\left(\frac{1}{2it} \log\left(\frac{d(t)}{c(t)}\frac{\tilde{V}(z)}{\tilde{U}(z)}\right)\right)\\
            &= i \exp\!\left(\frac{1}{2it} \log\!\left(e^{2it\log\frac{s(z)}{2i}} \frac{\Gamma\!\left(\frac{1}{6}\!+\!it\right)\Gamma\!\left(-it\right)}{\Gamma\!\left(\frac{1}{6}\!-\!it\right)\Gamma\!\left(it\right)}\,\frac{_2F_1\!\left(\frac{5}{12}\!+\!\frac{it}2,\frac{1}{12}\!+\!\frac{i t}2,1\!+\!i t; s(z)^2\right)}{_2F_1\!\left(\frac{5}{12}\!-\!\frac{i t}2,\frac{1}{12}\!-\!\frac{i t}2,1\!-\!i t; s(z)^2\right)}\right)\right)\,.\nonumber
\end{align}

\noindent 
Finally, using the map from the disc to the cylinder, $z(u) = \frac{\log(u)}{i}$, s.t. $s(z(u)) = s$ from \eqref{eq:s(u)} above expression is exactly the one in \eqref{eq:psi}.

\section{Example of transforming fields and correlators}\label{app:TrafoExamples}

In this appendix we give an example to illustrate how to work with local coordinates and their transformations as reviewed in Section~\ref{sec:local-coord-bndfield}.
    
Consider a Virasoro-primary $\psi$ of conformal weight $h$ and $\eta = L_{-1} \psi$. The transformed fields for a change of local coordinats $z \mapsto g(z)$ are (see \eqref{eq:v0v1v2} and \eqref{eq:Gamma})
\begin{align}
	\tilde\psi &= \Gamma_g \psi = (v_0)^{L_0} \, \psi = (g')^{h} \, \psi \,,
\nonumber\\	
	\tilde\eta &= \Gamma_g \eta = (v_0)^{L_0} \, (id + v_1 L_1) \, L_{-1}\psi 
	= (g')^{h+1} \,\eta + h \,(g')^{h-1} g'' \psi~,
\end{align}
where all derivatives of $g$ are taken at $z=0$.

Next, we work through the example of a two-point correlator on the upper half plane. In terms of the canonical local coordinates on the upper half plane, we have, for $x>y \in \mathbb{R}$,
\begin{align}
\big\langle  \psi(x) \psi(y) \big\rangle &= (x-y)^{-2h} ~,
\nonumber\\	
\big\langle  \eta(x) \psi(y) \big\rangle &= 
\frac{\partial}{\partial x}\big\langle  \psi(x) \psi(y) \big\rangle = -2h (x-y)^{-2h-1}~,
\label{eq:local-coord-example1}
\end{align}
where we assume that $\psi$ is normalised so that the leading coefficient is $1$ in the first correlator. Recall that here the canonical local coordinates $\varphi_s(z) = s+z$ at $s=x$ and $s=y$ are implied in the notation.

Now assume further that $x,y>0$ and apply the conformal bijection $f(z) = -1/z$. 
The canonical local coordinate $\varphi_s$ transforms as
\begin{equation}\label{eq:local-coord-example-aux1}
	\tilde\varphi_s(z) = f \circ \varphi_s(z) = \frac{-1}{s+z}
	= \frac{-1}{s} + \frac{1}{s} + \frac{-1}{s+z}
	= \varphi_{-1/s} \circ g_s(z)
\end{equation}
with $g_s(z) = \frac{1}{s} - \frac{1}{s+z}$. Then $g_s(0)=0$ as required and $g'_s(0) = s^{-2}$, $g''_s(0) = -2s^{-3}$.
Consequently
\begin{equation}\label{eq:local-coord-example-aux2}
	\tilde\psi_s = s^{-2h} \psi
\quad , \quad
	\tilde\eta_s = s^{-2h-2} \eta - 2 h s^{-2h-1} \psi~.
\end{equation}
Using the conformal transformation $f$, we can now express the correlator $\langle  \eta(x) \psi(y) \rangle$ in terms of correlators with insertions at $-\frac1x$ and $-\frac1y$:
\begin{align}
\big\langle  \eta(x) \psi(y) \big\rangle 
&= 
\big\langle  \eta(x;\varphi_x) \psi(y;\varphi_y) \big\rangle 
\nonumber\\	
&\overset{\eqref{eq:transformation-identity-for-corr}}{=}
\big\langle  \eta(f(x);f \circ \varphi_x) \, \psi(f(y);f \circ \varphi_y) \big\rangle 
\nonumber\\	
&\overset{\eqref{eq:local-coord-example-aux1}}{=} 
\big\langle  \eta(-\tfrac1x;\varphi_{-1/x} \circ g_x) \, \psi(-\tfrac1y;\varphi_{-1/y} \circ g_y) 
\big\rangle 
\nonumber\\	
&\overset{\eqref{eq:change-coord-for-Vir}}{=} 
\big\langle  \tilde\eta_x(-\tfrac1x;\varphi_{-1/x}) \, \tilde\psi_y(-\tfrac1y;\varphi_{-1/y}) \big\rangle 
\nonumber\\	
&\overset{\eqref{eq:local-coord-example-aux2}}{=}  
x^{-2h-2} y^{-2h} \big\langle \eta(-\tfrac1x) \psi(-\tfrac1y) \big\rangle 
-2h x^{-2h-1}
y^{-2h} \big\langle \psi(-\tfrac1x) \psi(-\tfrac1y) \big\rangle  ~.
\end{align}
And while it is not immediately obvious, substituting \eqref{eq:local-coord-example1} for the insertion points $-\frac1x>-\frac1y$ does indeed give $-2h(x-y)^{-2h-1}$, as required.

{\small
\newcommand\arxiv[2]      {\href{http://arXiv.org/abs/#1}{#2}}
\newcommand\jdoi[2]        {\href{http://dx.doi.org/#1}{#2}}

}

\begin{thebibliography}{10}


\bibitem{Petkova:2000ip}
V.B.~Petkova and J.-B.~Zuber,
{\it Generalised twisted partition functions},
\jdoi{10.1016/S0370-2693(01)00276-3}{Phys.\ Lett.\ B {\bf 504} (2001) 157--164},
\arxiv{hep-th/0011021}{[hep-th/0011021]}.

\bibitem{Frohlich:2006ch} 
J.~Fr\"ohlich, J.~Fuchs, I.~Runkel and C.~Schweigert,
{\it Duality and defects in rational conformal field theory},
\jdoi{10.1016/j.nuclphysb.2006.11.017}{Nucl.\ Phys.\ B {\bf 763} (2007) 354--430},
\arxiv{hep-th/0607247}{[hep-th/0607247]}.

\bibitem{PhysRevLett.98.160409}
A.~Feiguin, S.~Trebst, A.W.W.~Ludwig, M.~Troyer, A.~Kitaev, Z.~Wang and M.H.~Freedman,
\textit{Interacting anyons in topological quantum liquids: The golden chain},
\jdoi{10.1103/PhysRevLett.98.160409}{Phys.\ Rev.\ Lett.\ \textbf{98} (2007) 160409},
[\arxiv{cond-mat/0612341}{cond-mat/0612341 [cond-mat.str-el]}].

\bibitem{Pfeifer2012}
R.N.C. Pfeifer, O.~Buerschaper, S.~Trebst, A.W.W.~Ludwig, M.~Troyer and G.~Vidal,
\textit{Translation invariance, topology, and protection of criticality in chains of interacting anyons},
\jdoi{10.1103/physrevb.86.155111}{Phys.\ Rev.\ B \textbf{86} (2012) 155111},
[\arxiv{1005.5486}{1005.5486 [cond-mat.str-el]}].

\bibitem{Buican:2017rxc}
M.~Buican and A.~Gromov,
\textit{Anyonic Chains, Topological Defects, and Conformal Field Theory},
\jdoi{10.1007/s00220-017-2995-6}{Commun.\ Math.\ Phys.\ \textbf{356} (2017) 1017--1056},
[\arxiv{1701.02800}{1701.02800 [hep-th]}].

\bibitem{belletete2020topological}
J.~Bellet\^ete, A.M.~Gainutdinov, J.L.~Jacobsen, H.~Saleur and T.S.~Tavares,
\textit{Topological defects in periodic RSOS models and anyonic chains},
\arxiv{2003.11293}{2003.11293 [math-ph]}.

\bibitem{Levin:2004mi}
M.A.~Levin and X.G.~Wen,
\textit{String net condensation: A Physical mechanism for topological phases},
\jdoi{10.1103/PhysRevB.71.045110}{Phys.\ Rev.\ B \textbf{71} (2005) 045110},
[\arxiv{cond-mat/0404617}{cond-mat/0404617 [cond-mat.str-el]}].

\bibitem{Kitaev:2011dxc}
A.~Kitaev and L.~Kong,
\textit{Models for Gapped Boundaries and Domain Walls},
\jdoi{10.1007/s00220-012-1500-5}{Commun.\ Math.\ Phys.\ \textbf{313} (2012) 351--373},
[\arxiv{1104.5047}{1104.5047 [cond-mat.str-el]}].

\bibitem{Vanhove:2018wlb}
R.~Vanhove, M.~Bal, D.J. Williamson, N.~Bultinck, J.~Haegeman and
  F.~Verstraete,
\textit{Mapping topological to conformal field theories through
  strange correlators},
\jdoi{10.1103/PhysRevLett.121.177203}{Phys.\ Rev.\ Lett.\ \textbf{121} (2018) 177203},
[\arxiv{1801.05959}{1801.05959 [quant-ph]}].

\bibitem{Lootens:2020mso}
L.~Lootens, J.~Fuchs, J.~Haegeman, C.~Schweigert and F.~Verstraete,
\textit{Matrix product operator symmetries and intertwiners in string-nets with domain walls},
\jdoi{10.21468/SciPostPhys.10.3.053}{SciPost Phys.\ \textbf{10} 053 (2021)},
[\arxiv{2008.11187}{2008.11187 [quant-ph]}].

\bibitem{Aasen:2016dop}
D.~Aasen, R.S.K.~Mong and P.~Fendley,
\textit{Topological Defects on the Lattice I: The Ising model},
\jdoi{10.1088/1751-8113/49/35/354001}{J.\ Phys.\ A \textbf{49} (2016) 354001},
[\arxiv{1601.07185}{1601.07185 [cond-mat.stat-mech]}].

\bibitem{Aasen:2020jwb}
D.~Aasen, P.~Fendley and R.S.K.~Mong,
\textit{Topological Defects on the Lattice: Dualities and
  Degeneracies}, 
\arxiv{2008.08598}{2008.08598 [cond-mat.stat-mech]}.

\bibitem{Cardy:1989ir}
J.~L. Cardy,
\textit{Boundary Conditions, Fusion Rules and the Verlinde Formula},
\jdoi{10.1016/0550-3213(89)90521-X}{Nucl.\ Phys.\ B \textbf{324} (1989) 581--596}.

\bibitem{PhysRevB.82.115126}
R.N.C.~Pfeifer, P.~Corboz, O.~Buerschaper, M.~Aguado, M.~Troyer and G.~Vidal,
\textit{Simulation of anyons with tensor network algorithms},
\jdoi{10.1103/PhysRevB.82.115126}{Phys.\ Rev.\ B \textbf{82} (2010) 115126},
[\arxiv{1006.3532}{1006.3532 [cond-mat.str-el]}].

\bibitem{Eguchi:1986sb}
T.~Eguchi and H.~Ooguri,
\textit{Conformal and Current Algebras on General Riemann Surface},
\jdoi{10.1016/0550-3213(87)90686-9}{Nucl.\ Phys.\ B \textbf{282} (1987) 308--328}.

\bibitem{Zhu:1996}
Y.~Zhu,
\textit{Modular invariance of characters of vertex operator algebras},
\href{https://www.jstor.org/stable/2152847}{J.\ American Math.\ Soc.\ \textbf{9} (1996) 237--302}.

\bibitem{Gaberdiel:2008pr}
M.R.~Gaberdiel and C.A.~Keller,
\textit{Modular differential equations and null vectors},
\jdoi{10.1088/1126-6708/2008/09/079}{JHEP \textbf{09} (2008) 079},
[\arxiv{0804.0489}{0804.0489 [hep-th]}].

\bibitem{Gaberdiel:2012yb}
M.R.~Gaberdiel, T.~Hartman and K.~Jin,
\textit{Higher Spin Black Holes from CFT},
\jdoi{10.1007/JHEP04(2012)103}{JHEP \textbf{04} (2012) 103},
[\arxiv{1203.0015}{1203.0015 [hep-th]}].

\bibitem{Kilford:2008}
L.J.P.~Kilford,
\textit{Modular Forms, A Classical and Computational Introduction},
\jdoi{10.1142/p564}{Imperial College Press}.

\bibitem{DiFrancesco:1987ez}
P.~Di Francesco, H.~Saleur and J.B.~Zuber,
\textit{Critical Ising correlation functions in the plane and on the torus},
\jdoi{10.1016/0550-3213(87)90202-1}{Nucl.\ Phys.\ B \textbf{290} (1987) 527--581}.

\bibitem{you2014wave}
Y.-Z. You, Z.~Bi, A.~Rasmussen, K.~Slagle and C.~Xu,
\textit{Wave function and strange correlator of short-range entangled states},
\jdoi{10.1103/PhysRevLett.112.247202}{Phys.\ Rev.\ Lett.\ \textbf{112} (2014) 247202},
[\arxiv{1312.0626}{1312.0626 [cond-mat.str-el]}].

\bibitem{Chang:2018iay}
C.M.~Chang, Y.H.~Lin, S.H.~Shao, Y.~Wang and X.~Yin,
\textit{Topological Defect Lines and Renormalization Group Flows in Two Dimensions},
\jdoi{10.1007/JHEP01(2019)026}{JHEP \textbf{01} (2019) 026},
[\arxiv{1802.04445}{1802.04445 [hep-th]}].

\bibitem{Kikuchi:2021qxz}
K.~Kikuchi,
\textit{Symmetry enhancement in RCFT},
\arxiv{2109.02672}{2109.02672 [hep-th]}.

\bibitem{Gaberdiel:2001xm}
M.R.~Gaberdiel, A.~Recknagel and G.~M.~T. Watts,
\textit{The Conformal boundary states for SU(2) at level 1},
\jdoi{10.1016/S0550-3213(02)00033-0}{Nucl. Phys. B \textbf{626} (2002) 344--362},
[\arxiv{hep-th/0108102}{hep-th/0108102}].

\bibitem{Fuchs:2007tx}
J.~Fuchs, M.R.~Gaberdiel, I.~Runkel and C.~Schweigert,
\textit{Topological defects for the free boson CFT},
\jdoi{10.1088/1751-8113/40/37/016}{J.\ Phys.\ A \textbf{40} (2007) 11403},
[\arxiv{0705.3129}{0705.3129 [hep-th]}].

\bibitem{Thorngren:2021yso}
R.~Thorngren and Y.~Wang,
\textit{Fusion Category Symmetry II: Categoriosities at $c$ = 1 and Beyond}, \arxiv{2106.12577}{2106.12577 [hep-th]}.

\bibitem{Runkel:2012rp}
I.~Runkel, M.R.~Gaberdiel and S.~Wood,
\textit{Logarithmic bulk and boundary conformal field theory and the
  full centre construction},
in: C.~Bai C. et al. (eds) \jdoi{10.1007/978-3-642-39383-9_4}{Conformal Field Theories and Tensor Categories}. Math.\ Lectures from Peking Univ., Springer (2012),
[\arxiv{1201.6273}{1201.6273 [hep-th}].

\bibitem{Creutzig:2016fms}
T.~Creutzig and T.~Gannon,
\textit{Logarithmic conformal field theory, log-modular tensor categories and modular forms},
\jdoi{10.1088/1751-8121/aa8538}{J.\ Phys.\ A \textbf{50} (2017) 404004},
[\arxiv{1605.04630}{1605.04630 [math.QA]}].

\bibitem{Fuchs:2017unc}
J.~Fuchs, T.~Gannon, G.~Schaumann and C.~Schweigert,
\textit{The logarithmic Cardy case: Boundary states and annuli},
\jdoi{10.1016/j.nuclphysb.2018.03.005}{Nucl.\ Phys.\ B \textbf{930} (2018) 287--327},
[\arxiv{1712.01922}{1712.01922 [math.QA]}].

\bibitem{Fuchs:2002cm}
J.~Fuchs, I.~Runkel and C.~Schweigert,
\textit{TFT construction of RCFT correlators 1. Partition functions},
\jdoi{10.1016/S0550-3213(02)00744-7}{Nucl.\ Phys.\ B \textbf{646} (2002) 353--497},
[\arxiv{hep-th/0204148}{hep-th/0204148}].

\bibitem{Runkel:2008gr}
I.~Runkel and R.R.~Suszek,
\textit{Gerbe-holonomy for surfaces with defect networks},
\jdoi{10.4310/ATMP.2009.v13.n4.a5}{Adv.\ Theor.\ Math.\ Phys.\ \textbf{13} (2009) 1137--1219},
[\arxiv{0808.1419}{0808.1419 [hep-th]}].

\bibitem{Runkel:2010ym}
I.~Runkel,
\textit{Non-local conserved charges from defects in perturbed conformal field theory},
\jdoi{10.1088/1751-8113/43/36/365206}{J.\ Phys.\ A \textbf{43} (2010) 365206},
[\arxiv{1004.1909}{1004.1909 [hep-th]}].

\bibitem{Kojita:2016jwe}
T.~Kojita, C.~Maccaferri, T.~Masuda and M.~Schnabl,
\textit{Topological defects in open string field theory},
\jdoi{10.1007/JHEP04(2018)057}{JHEP \textbf{04} (2018) 057},
[\arxiv{1612.01997}{1612.01997 [hep-th]}].

\bibitem{Bhardwaj:2017xup}
L.~Bhardwaj and Y.~Tachikawa,
\textit{On finite symmetries and their gauging in two dimensions},
\jdoi{10.1007/JHEP03(2018)189}{JHEP \textbf{03} (2018) 189},
[\arxiv{1704.02330}{1704.02330 [hep-th]}].

\bibitem{Konechny:2019wff}
A.~Konechny,
\textit{Open topological defects and boundary RG flows},
\jdoi{10.1088/1751-8121/ab7c8b}{J.\ Phys.\ A \textbf{53} (2020) 155401},
[\arxiv{1911.06041}{1911.06041 [hep-th]}].

\bibitem{Thorngren:2019iar}
R.~Thorngren and Y.~Wang,
\textit{Fusion Category Symmetry I: Anomaly In-Flow and Gapped Phases},
\arxiv{1912.02817}{1912.02817 [hep-th]}.

\bibitem{kirillov2011string}
A.~Kirillov~Jr,
\textit{String-net model of Turaev-Viro invariants},
\arxiv{1106.6033}{1106.6033 [math.AT]}.

\bibitem{Goosen2018Oriented1V}
G.~Goosen,
\textit{Oriented 123-TQFTs via string-nets and state-sums}, 
\href{http://hdl.handle.net/10019.1/103324}{PhD Thesis, Stellenbosch University, 2018}.

\bibitem{Fjelstad:2006aw}
J.~Fjelstad, J.~Fuchs, I.~Runkel and C.~Schweigert,
\textit{Uniqueness of open / closed rational CFT with given algebra of open states},
\jdoi{10.4310/ATMP.2008.v12.n6.a4}{Adv.\ Theor.\ Math.\ Phys.\ \textbf{12} (2008) 1283--1375},
[\arxiv{hep-th/0612306}{hep-th/0612306}].

\bibitem{Kong:2009inh}
L.~Kong and I.~Runkel,
\textit{Cardy algebras and sewing constraints. I.},
\jdoi{10.1007/s00220-009-0901-6}{Commun.\ Math.\ Phys.\ \textbf{292} (2009) 871--912},
[\arxiv{0807.3356}{0807.3356 [math.QA]}].

\bibitem{Frohlich:2009gb}
J.~Fr\"ohlich, J.~Fuchs, I.~Runkel and C.~Schweigert,
\textit{Defect lines, dualities, and generalised orbifolds},
In \jdoi{10.1142/9789814304634_0056}{16th International Congress on Mathematical Physics (2009)},
[\arxiv{0909.5013}{0909.5013 [math-ph]}].

\bibitem{Bachas:1992cr}
C.~Bachas and P.M.S.~Petropoulos,
\textit{Topological models on the lattice and a remark on string theory cloning},
\jdoi{10.1007/BF02097063}{Commun.\ Math.\ Phys.\ \textbf{152} (1993) 191--202},
[\arxiv{hep-th/9205031}{hep-th/9205031}].

\bibitem{Fukuma:1993hy}
M.~Fukuma, S.~Hosono and H.~Kawai,
\textit{Lattice topological field theory in two-dimensions},
\jdoi{10.1007/BF02099416}{Commun.\ Math.\ Phys.\ \textbf{161} (1994) 157--175},
[\arxiv{hep-th/9212154}{hep-th/9212154}].

\bibitem{Davydov:2011kb}
A.~Davydov, L.~Kong and I.~Runkel,
\textit{Field theories with defects and the centre functor},
\jdoi{10.1090/pspum/083}{Mathematical Foundations of Quantum Field Theory and Perturbative String Theory}, Proceedings of Symposia in Pure Mathematics, AMS, 2011, [\arxiv{1107.0495}{1107.0495 [math.QA]}].

\bibitem{Ishibashi:1988kg}
N.~Ishibashi,
\textit{The Boundary and Crosscap States in Conformal Field Theories},
\jdoi{10.1142/S0217732389000320}{Mod.\ Phys.\ Lett.\ A \textbf{4} (1989) 251--265 }.

\bibitem{Cardy:1988tk}
J.~L. Cardy and I.~Peschel,
\textit{Finite-size dependence of the free energy in two-dimensional critical systems},
\jdoi{10.1016/0550-3213(88)90604-9}{Nucl.\ Phys.\ B \textbf{300} (1988) 377--392}.

\bibitem{Gaberdiel:1994fs}
M. Gaberdiel,
\textit{A General transformation formula for conformal fields},
\jdoi{10.1016/0370-2693(94)90026-4}{Phys. Lett. B \textbf{325} (1994) 366--370},
[\arxiv{hep-th/9401166}{hep-th/9401166}].

\bibitem{Huang:2002mx}
Y.-Z. Huang,
\textit{Riemann surfaces with boundaries and the theory of vertex operator algebras},
\jdoi{10.1090/fic/039}{Fields Inst.\ Commun.\ \textbf{39} (2003) 109--126},
[\arxiv{math/0212308}{math/0212308 [math.QA]}].

\bibitem{Frenkel:2004jn}
E.~Frenkel and D.~Ben-Zvi,
\textit{Vertex algebras and algebraic curves} (2004).

\bibitem{Fuchs:2004xi}
J.~Fuchs, I.~Runkel and C.~Schweigert,
\textit{TFT construction of RCFT correlators IV: Structure constants and correlation functions},
\jdoi{10.1016/j.nuclphysb.2005.03.018}{Nucl.\ Phys.\ B \textbf{715} (2005) 539--638},
[\arxiv{hep-th/0412290}{hep-th/0412290}].

\bibitem{Brehm:2020zri}
E.M.~Brehm, M. Broccoli,
\textit{Correlation functions and quantum measures of descendant states},
\jdoi{10.1007/JHEP04(2021)227}{JHEP \textbf{04} (2021) 227},
[\arxiv{2012.11255}{2012.11255 [hep-th]}].

\bibitem{Cardy:1984bb}
J.~L. Cardy,
\textit{Conformal Invariance and Surface Critical Behavior},
\jdoi{10.1016/0550-3213(84)90241-4}{Nucl. Phys. B \textbf{240} (1984) 514--532}.

\bibitem{Blau:1987pn}
S.K.~Blau, M.~Clements, S.~Della Pietra, S.~Carlip and V.~Della Pietra,
\textit{The String Amplitude on Surfaces With Boundaries and Crosscaps},
\jdoi{10.1016/0550-3213(88)90346-X}{Nucl.\ Phys.\ B \textbf{301} (1988) 285--303}.

\bibitem{DiFrancesco:1997nk}
P.~Di Francesco, P.~Mathieu and D.~Senechal,
\textit{Conformal Field Theory},
\jdoi{10.1007/978-1-4612-2256-9}{Springer (1997)}.

\bibitem{Abramowitz:1970}
M.~Abramowitz and I.A.~Stegun,
\textit{Handbook of Mathematical Functions},
Dover Publications (1970).

\bibitem{Eichler:1982}
M.~Eichler and D.~Zagier
\textit{On the Zeros of the Weierstrass $\wp$-Function}
\jdoi{10.1007/BF01453974}{Math.\ Ann.\ 258 (1982) 399–407}.

\bibitem{nehari1975conformal}
Z.~Nehari,
\textit{Conformal Mapping}, Dover Publications (1975).

\bibitem{zwillinger2014table}
I.S.~Gradshteyn and I.M.~Ryzhik,
Editor: D.~Zwillinger and V.~Moll
\textit{Table of Integrals, Series, and Products},
\jdoi{10.1016/C2010-0-64839-5}{Academic Press (2014)}.

\bibitem{priestley2003introduction}
H.~Priestley,
\textit{Introduction to Complex Analysis},
OUP Oxford (2003).


\end{thebibliography}
\end{document}